%% file: Sample.tex
\documentclass[11pt,a4paper]{article}

\usepackage[a4paper,margin=1in]{geometry}
\usepackage{setspace}
\usepackage[T1]{fontenc}
\usepackage{newtxtext}

\usepackage{amsmath,amssymb,amsfonts}
\usepackage{amsthm}
\usepackage{mathrsfs}

\usepackage{graphicx}
\usepackage{multicol,multirow}
\usepackage{caption}
\usepackage{subcaption}
\usepackage{array}
\usepackage{rotating}

\usepackage{xcolor}
\usepackage{textcomp}
\usepackage{appendix}
\usepackage{ifpdf}
\usepackage{lipsum}
\usepackage{natbib}
\usepackage{makecell}

\usepackage{tikz}
\usetikzlibrary{
  arrows.meta,
  shapes,
  positioning,
  calc
}

\usepackage[colorlinks=true,
            linkcolor=blue,
            citecolor=blue,
            urlcolor=blue]{hyperref}

\theoremstyle{definition}

\numberwithin{equation}{section}

\title{Generative Unsupervised Downscaling of Climate Models via Domain Alignment:\\
Application to Wind Fields}

\author{
Julie Keisler\thanks{INRIA Paris} \and
Boutheina Oueslati\thanks{EDF R\&D} \and
Anastase Charantonis\thanks{INRIA Paris} \and
Yannig Goude\thanks{Université Paris-Saclay} \and
Claire Monteleoni\thanks{University of Colorado Boulder}
}

\date{} % vide = pas de date visible (classique arXiv)

\begin{document}
\maketitle

\abstract{
General Circulation Models (GCMs) are widely used for future climate projections, but their coarse spatial resolution and systematic biases limit their direct use for impact studies. This limitation is particularly critical for wind-related applications, such as wind energy, which require spatially coherent, multivariate, and physically plausible near-surface wind fields. Classical statistical downscaling and bias correction methods partly address this issue. Still, they struggle to preserve spatial structure, inter-variable consistency, and robustness under climate change, especially in high-dimensional settings.

Recent advances in generative machine learning offer new opportunities for downscaling and bias correction, eliminating the need for explicitly paired low- and high-resolution datasets. However, many existing approaches remain difficult to interpret and challenging to deploy in operational climate impact studies.

In this work, we apply SerpentFlow, an interpretable, generative, domain alignment framework, to the multivariate downscaling and bias correction of wind variables from GCM outputs. This is a method that generates low-resolution/high-resolution training data pairs by separating large-scale spatial patterns from small-scale variability.
Large-scale components are aligned across climate model and observational domains.
Conditional fine-scale variability is then learned using a flow-matching generative model. We apply the approach to multiple wind variables downscaling, including average and maximal wind speed, zonal and meridional components, and compare it with widely used multivariate bias correction methods. Results show improved spatial coherence, inter-variable consistency, and robustness under future climate conditions, highlighting the potential of interpretable generative models for wind and energy applications.}

\section{Introduction}

Anthropogenic climate change entails changes in atmospheric circulation and the statistical properties of climate variables, with important consequences for natural and socio-economic systems. General Circulation Models (GCMs) are the primary tools used to simulate these changes and produce climate projections under different emission scenarios. While GCMs provide a physically consistent representation of the climate system at the global scale, their coarse spatial resolution and systematic biases limit their direct use for regional and local impact studies \citep{maraun2010, teutschbein2012}.

Many climate impact applications require climate information that is spatially coherent, multivariate, and physically plausible at fine spatial scales. This is particularly important for wind-related applications, such as wind energy assessments, which depend on the joint behaviour of wind speed, direction, variability, and extremes. GCM outputs are therefore commonly post-processed using downscaling and bias correction methods \citep{bartok2019climate}.

Statistical downscaling and bias correction provide a computationally efficient alternative to dynamical downscaling by learning empirical relationships between climate simulations and observations. Classical approaches include distribution-based methods such as CDF-t \citep{cdft, vrac2015cdf} and multivariate frameworks such as MBCn \citep{cannon2018mbcn}, DOTC \citep{dotc}, R2D2 \citep{mehrotra2020r2d2}, and MRec \citep{zhang2021mrec}. While multivariate methods better preserve inter-variable dependencies, challenges remain for high-dimensional, spatially structured fields, particularly in maintaining spatial coherence under non-stationary climate conditions \citep{francois2020multivariate, allard2025assessing}.

Machine learning approaches, especially deep learning, have been increasingly explored for downscaling and bias correction. These models can capture complex nonlinear relationships and spatial structures, showing promising results across climate variables \citep{vandal2017deepsd, soares2024cmip6dl}. Most approaches rely on supervised learning with paired low- and high-resolution datasets, which are rarely available in realistic climate projection settings, making explicit pairing ill-defined \citep{bartok2019climate}. Consequently, many studies adopt perfect-model or synthetic training frameworks that do not fully reflect real-world applications.

These limitations have motivated interest in generative and probabilistic approaches that operate in unpaired or weakly supervised settings. Generative domain adaptation methods align climate model outputs and observations within a common distribution while preserving spatial structure and providing a measure of uncertainty. Recent studies explore adversarial learning \citep{groenke2020climalign}, diffusion bridge-based models \citep{bischoff2024unpaired, hess2025fast}, and hybrid approaches combining optimal transport and generative sampling \citep{wan2023debias}. While promising, these methods rely on implicit alignment mechanisms---cycle-consistency losses, Gaussian latent spaces, or diffusion paths---to bridge the gap between domains, without explicitly defining what should be shared and what should differ. As we show in the results, and as also reported in~\cite{serpentflow2024}, some of these approaches fail to preserve the large-scale dynamics of the driving GCM: the downscaled fields can drift away from the source signal even at scales that the GCM resolves well, making it difficult to trust their projections under future climate conditions. This lack of an interpretable, physically grounded separation of scales calls for methods that explicitly control which information is retained from the source and which is generated.

Frequency-based decompositions offer a natural framework for this purpose. Gaussian filtering and Fourier cutoffs have a long history in fluid dynamics, from subgrid-scale modelling in Large Eddy Simulations to more recent super-resolution and data reconstruction approaches for turbulent flows~\citep{buzzicotti2023data, suresh2026guided}. These methods explicitly separate resolved and unresolved scales, providing both physical interpretability and control over the scale-dependent behaviour of the reconstruction.

In this work, we investigate SerpentFlow \citep{serpentflow2024}, an interpretable generative domain adaptation framework for multivariate downscaling and bias correction. SerpentFlow operates in an unpaired setting by constructing pseudo low-/high-resolution pairs through an explicit separation of spatial scales: large-scale patterns are aligned between climate model outputs and observations, while small-scale variability is learned conditionally via a flow-matching generative model. This enables generation of fine-scale wind variability while maintaining physically meaningful large-scale climate information.

We apply SerpentFlow to multiple near-surface wind variables, including wind speed, zonal and meridional components, and maximum wind speed, evaluating its performance in both historical and future climate conditions. Comparisons with established multivariate bias correction methods focus on spatial coherence, inter-variable consistency, and robustness under climate change, highlighting the potential of interpretable generative domain adaptation for wind energy impact studies.\footnote{The code is available here: \url{https://github.com/JulieKeisler/serpentflow}}

The main contributions of this work are:
\begin{itemize}
\item Application of an interpretable generative domain adaptation framework, SerpentFlow, for multivariate downscaling and bias correction of wind variables in an unpaired setting.
\item Introduction of a Gaussian-blur-based decomposition to handle irregular observational domains, enabling flexible separation of large-scale and local features even when Fourier-based filtering is not applicable.
\item Comprehensive evaluation against established downscaling and bias correction methods, assessing distributions, spatial structure, inter-variable relationships, and robustness under future climate.
\end{itemize}

\section{Method}\label{method}

Our work uses SerpentFlow~\citep{serpentflow2024}, a generative domain adaptation framework designed for unpaired multivariate downscaling and bias correction. 
While we do not modify the original method, we provide a brief overview to clarify its main principles and how it applies to wind variables.

\subsection{Core idea}

SerpentFlow transfers samples from a source domain (coarse GCM outputs) to a target domain (observations or reanalyses) by representing each field in a latent space and separating it into a shared and a domain-specific component. 
The shared component captures structures common to both domains, while the domain-specific component encodes elements unique to each domain. 
During training, pseudo-pairs are constructed by combining the shared component of a target sample with stochastic realizations of its domain-specific part. 
A generative model then learns to reconstruct the target field conditioned on the shared component. 
At inference, source samples are projected into the shared-domain space, combined with stochastic noise in the domain-specific component, and mapped to the target domain through the learned generative model.

\subsection{Frequency-based latent space}

In practice, SerpentFlow implements the shared-domain hypothesis in the Fourier domain for unsupervised super-resolution. 
Fields are decomposed into spatial frequencies, with low-frequency modes representing large-scale shared structures and high-frequency modes representing domain-specific variability. 
This decomposition is conceptually analogous to the Reynolds decomposition used in atmospheric science~\citep{10.1098/rsta.1895.0004}, which separates a flow into its mean and fluctuating components---here, the cutoff frequency $\omega_{cut}$ plays the role of the scale separator between resolved large-scale dynamics and unresolved fine-scale variability. 
As stated in \cite{serpentflow2024}, the cutoff can be determined automatically by a classifier trained to distinguish between domains, providing a convenient strategy for hyperparameter tuning. 
Alternatively, for climate downscaling, a more interpretable choice consists in setting $\omega_{cut}$ to match the effective resolution of the driving GCM~\citep{abdalla2013effective}, directly linking the scale separation to the physical resolution gap between models. 
As we confirm in the results, the cutoff identified by the classifier indeed corresponds to the effective resolution of the GCM, bridging both approaches. 
This frequency-based latent space provides an interpretable and physically meaningful separation of scales, naturally suited for superresolution tasks such as wind downscaling.

\subsection{Generative mapping}

SerpentFlow is agnostic to the choice of the generative model $f_\theta$. 
In the original implementation, Flow Matching~\citep{flowmatching} is used to learn a continuous mapping from pseudo-inputs to target fields. 
This allows the model to generate physically plausible fine-scale details while respecting the shared large-scale patterns. 
In inference, coarse GCM wind fields are projected into the shared-frequency space, and the learned generator produces downscaled and bias-corrected wind variables consistent with the target distribution.

\subsection{Practical workflow}

\input{diagram}

Figure~\ref{fig:pipeline} illustrates the training and inference workflow. Pseudo-pairs generated from high-resolution observations/reanalyses are used to train $f_\theta$, and coarse GCM fields are subsequently downscaled through the same pipeline. 
The resulting outputs are physically coherent, statistically consistent, and preserve both large-scale climate dynamics and realistic small-scale variations.

1) Decompose target fields into low/high frequencies and generate pseudo-pairs,  
2) train the generative model to reconstruct high-resolution fields conditioned on low-frequency structure,  
3) project source fields into the shared space and apply the trained generator to obtain downscaled and bias-corrected outputs.  

For full technical details and theoretical background, we refer the reader to~\cite{serpentflow2024}.

\section{Experiments}\label{sec:exp}

\subsection{Data and study domain}

We evaluate SerpentFlow and baseline methods over the French territory using outputs from the ACCESS Earth System Model \citep{access} ($1.9^{\circ}\times1.2^{\circ}$, 145 km $\times$ 130 km) under the SSP2-4.5 scenario. ACCESS is a participating model in the Coupled Model Intercomparison Project Phase 6 (CMIP6, \cite{cmip6}), an international framework designed to coordinate and standardize climate model experiments. Our analysis focuses on four near-surface wind variables: wind speed (\texttt{sfcWind}), maximum wind speed (\texttt{sfcWindMax}), and the zonal (\texttt{uas}) and meridional (\texttt{vas}) components. ERA5 (\cite{era5}, $0.25^{\circ}\times0.25^{\circ}$, 25 km) reanalyses, averaged at a daily temporal resolution, serve as the observational reference.

The dataset is split to maximize training data while keeping independent evaluation periods. All GCM fields are bilinearly interpolated to the ERA5 grid prior to computing metrics and producing plots.
\begin{itemize}
\item \textbf{Training (1979--1999):} used to learn the statistical mapping between GCM outputs and ERA5 observations.
\item \textbf{Evaluation against ERA5 (2000--2020):} assesses historical bias correction and downscaling performance.
\item \textbf{Evaluation against GCM projections (2000--2100):} assesses the consistency of downscaling methods under future climate conditions. This evaluation is necessary because some statistical downscaling methods can alter the large-scale dynamics of the driving GCM, introducing spurious trends or distorting the climate change signal. By comparing downscaled outputs against the raw GCM projections, we verify that the downscaling preserves the broad physical consistency of the climate system over the full projection period.
\end{itemize}

GCM outputs are regridded to the ERA5 resolution before downscaling: linear interpolation is used for baseline methods, while SerpentFlow employs a spectral interpolation as described by \cite{serpentflow2024}.

\subsection{Baselines}

We compare the following approaches:

\begin{itemize}
    \item \textbf{CDF-t}~\citep{cdft, vrac2015cdf}: a univariate bias correction method that maps quantiles between distributions, implemented as in~\cite{francois2020multivariate}. Unlike simpler quantile mapping approaches, CDF-t explicitly accounts for the evolution of the GCM distribution between the historical and future periods, allowing it to preserve the climate change signal during correction. Moreover, wind variables—unlike temperature—do not exhibit a strong systematic shift in their overall distribution under future climate scenarios, which further limits the risk of degraded performance over the projection period.
    \item \textbf{R2D2}~\citep{mehrotra2020r2d2}: a multivariate stochastic bias correction method that preserves inter-variable dependencies. Due to computational constraints, we process the domain using $5 \times 5$ pixel patches.
    \item \textbf{Dual FM}: a domain adaptation method implemented in \cite{serpentflow2024}, inspired by ClimAlign \citep{groenke2020climalign}, but replacing normalizing flows with flow matching.
    \item \textbf{SerpentFlow}~\citep{serpentflow2024}: the full method, sharing the UNet generator architecture with Dual FM. SerpentFlow decomposes fields into three Fourier-based frequency bands (cutoff scales: 1200, 750, 300,km) to separate large-scale shared structures from finer domain-specific patterns, enabling control over scale-dependent downscaling performance. The primary cutoff frequency of 1200,km was identified by the classifier described in \cite{serpentflow2024}, and also corresponds to the effective resolution of the GCM \citep{abdalla2013effective}. To explore the sensitivity to this cutoff, we additionally tested two lower values (750 and 300,km), which allow more fine-scale information from the GCM to pass into the shared representation.
\end{itemize}

For deep learning–based methods (SerpentFlow and Dual FM), we generate 10 ensemble members per simulation. We report both the ensemble mean and the output of a single member (for SerpentFlow at 1200~km cutoff), the latter corresponding to the baseline in \cite{serpentflow2024}. While Dual FM has been described as a generative model, our analysis shows that its integration of an ODE makes the outputs effectively deterministic, so the ensemble mean and a single member are identical (discussed Figure~\ref{fig:ts_ERA5}). Inputs are normalized using the spatial mean and standard deviation of the training period (GCM statistics for source fields, ERA5 statistics for target fields), and outputs are denormalized using ERA5 mean and standard deviation at each grid point.

\subsection{Results}

We evaluate the downscaling and bias correction methods from two complementary perspectives. First, we assess their ability to reproduce the statistical and spatial characteristics of the observational reference (ERA5), which is critical for impact and energy applications. Second, we examine whether the methods preserve the large-scale climate dynamics and temporal evolution of the original GCM projections, a key requirement for climate change studies.
Regarding computational cost, training SerpentFlow for a given cutoff frequency $\omega_{cut}$ requires approximately 4 hours on a single NVIDIA H100 GPU using 20 years of training data. Once trained, generating a single ensemble member over a 100-year projection takes roughly 10 minutes. Dual FM, which requires training two separate Flow Matching models (one per domain), approximately doubles both training and inference time. Importantly, the GCM only intervenes in the SerpentFlow training pipeline through the choice of $\omega_{cut}$: the generator $f_\theta$ is trained exclusively on the target domain. This means that a trained model can be directly applied to any GCM sharing the same effective resolution, without retraining---substantially reducing the computational burden when downscaling multi-model ensembles.

Evaluation metrics and protocols are provided in Appendix~\ref{sec:metrics}, while detailed spatial maps, distributional diagnostics, and time series are available in Appendix~\ref{app:plots}.

\paragraph{Overview}
Figure~\ref{fig:radar_metrics} summarizes the performance of all methods using radar plots, where each axis represents a different metric and values closer to the outer edge indicate better performance. The left panel shows comparison against ERA5, highlighting the methods’ ability to reproduce observational statistics, whereas the right panel evaluates consistency with the raw GCM projections, emphasizing preservation of large-scale dynamics.

\begin{figure}[htbp]
    \centering
    \begin{subfigure}{0.45\textwidth}
        \centering
        \includegraphics[width=\textwidth]{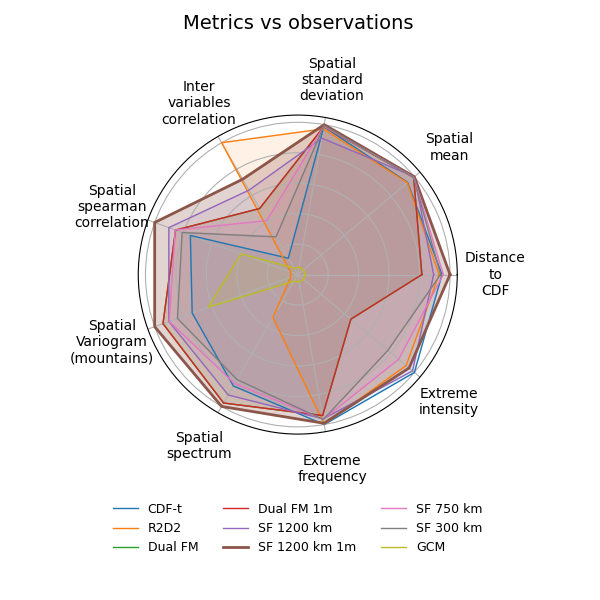}
        \caption{Metrics evaluated against ERA5. Results are detailed Table~\ref{tab:metrics_era5}.}
        \label{fig:radar_era5}
    \end{subfigure}
    \hfill
    \begin{subfigure}{0.45\textwidth}
        \centering
        \includegraphics[width=\textwidth]{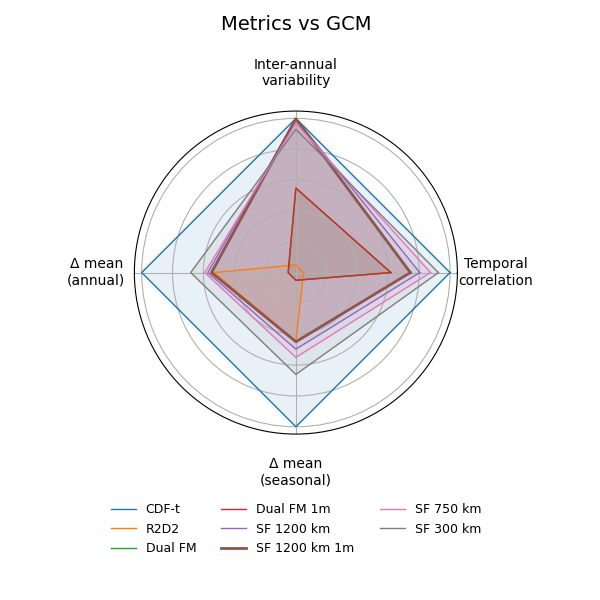}
        \caption{Metrics evaluated against the GCM. Results are detailed Table~\ref{tab:metrics_gcm}.}
        \label{fig:radar_gcm}
    \end{subfigure}
    \caption{Radar plots summarizing the performance of all methods averaged over all climate variables (results on average wind speed only are given Figure~\ref{fig:radar_metrics_sfcwind}). Higher values (closer to the outer circle) indicate better agreement with the reference. SF stands for SerpentFlow in all the plots, and ``mbr" indicates ``one member" for a generative method, the average of the members being shown otherwise}
    \label{fig:radar_metrics}
\end{figure}

\paragraph{Evaluation against ERA5}

The left panel of Figure~\ref{fig:radar_era5} highlights clear differences between classical statistical methods and deep learning based approaches. 
CDF-t performs well for distributional similarity (Figures~\ref{fig:cdf_global_ERA5}, \ref{fig:cdf_alps_ERA5}, \ref{fig:cdf_med_ERA5}) and extreme-event metrics, but its performance is weaker for inter-variable relationships (Figures~\ref{fig:sfcwind_era5_corr}, \ref{fig:sfcwindmax_era5_corr}, \ref{fig:uas_era5_corr}, \ref{fig:vas_era5_corr}) and spatial correlations (Figures~\ref{fig:cdf_spearman_corr_ERA5}, \ref{fig:spectrum_era5}), especially in mountainous regions (Figure~\ref{fig:orography}), reflecting its univariate nature. 

R2D2 improves inter-variable dependencies compared to CDF-t, but remains limited in reproducing spatial coherence and extremes (Figures~\ref{fig:sfcwind_era5_corr}, \ref{fig:sfcwindmax_era5_corr}, \ref{fig:uas_era5_corr}, \ref{fig:vas_era5_corr}).

Deep learning methods consistently outperform the statistical baselines in reconstructing ERA5's spatial dynamics (Figure~\ref{fig:cdf_spearman_corr_ERA5}). While Dual FM improves the spatial spectrum (Figure~\ref{fig:spectrum_era5}), it struggles to capture extreme-value distributions and is less effective than certain SerpentFlow configurations across most metrics. Thanks to its generative nature, SerpentFlow can represent a range of possible values at a given grid point and time (see Figure~\ref{fig:ts_ERA5}), effectively capturing uncertainty in the reconstructed fields. A calibration study for SF 1200~km is given section~\ref{app:calibration}. The cutoff scale strongly influences performance: smaller cutoffs (e.g., 300~km) better preserve large-scale coherence but may reduce spatial consistency (Figures~\ref{fig:cdf_variogram_ERA5},\ref{fig:spectrum_era5}), whereas larger cutoffs (e.g., 1200~km), especially the one-member version (SF 1200~kmmbr), achieves high values across all ERA5 metrics. Overall, SF 1200~kmmbr provides the best balance between distributional accuracy, spatial coherence, and inter-variable consistency.

\paragraph{Consistency with GCM projections}

The right panel of Figure~\ref{fig:radar_gcm} evaluates how well the downscaled outputs preserve the large-scale temporal dynamics of the GCM. As expected, CDF-t closely follows the GCM for mean changes and temporal correlations (Figures~\ref{fig:temp_corr_era5}, \ref{fig:full_delta_era5}, \ref{fig:full_seasonal_delta_sfcwind_era5}, \ref{fig:full_seasonal_delta_sfcwindmax_era5}). R2D2 performs worse for temporal metrics, reflecting partial loss of large-scale climate signals, partly due to the $5 \times 5$ patching, which reduces spatial coherence (Figure~\ref{fig:temp_corr_era5}).

Dual FM struggles to follow the GCM signal and is consistently outperformed by SerpentFlow across all metrics, performing worst on the delta metrics (Figures~\ref{fig:full_delta_era5}, \ref{fig:full_seasonal_delta_sfcwind_era5}, \ref{fig:full_seasonal_delta_sfcwindmax_era5}). SerpentFlow, while slightly below CDF-t, better preserves inter-annual variability and temporal correlations than R2D2 and Dual FM. The performance trade-off is controlled by the cutoff frequency: higher cutoffs retain more GCM dynamics but reduce similarity to observations, whereas lower cutoffs improve agreement with observations at the cost of some large-scale consistency. This allows users to select an appropriate balance depending on the application.

Taken together, these results demonstrate that SerpentFlow provides a favorable trade-off, substantially improving agreement with observations while preserving the essential large-scale dynamics of the driving GCM.

\subsection{Univariate application}\label{app:cnrm_safran}

To further illustrate SerpentFlow’s flexibility, we applied it to a univariate downscaling task for near-surface wind speed, using outputs from the CNRM-CM6-1 climate model \citep{cnrm} under the SSP3-7.0 scenario ($1.4^{\circ}\times1.4^{\circ}$, 130 km) and the high-resolution SAFRAN reanalysis (\cite{safran}, 8 km), at daily temporal resolution. This application serves several purposes. First, it demonstrates that SerpentFlow can operate without relying on a standard Fourier-based low-pass filter. Instead, we introduce a Gaussian-blur-based decomposition to separate a large-scale “shared-domain” component from a domain-specific residual. This approach enables the method to handle irregular domains, such as land-only observations from SAFRAN, where a Fourier decomposition would produce \texttt{NaNs} over the ocean. Second, it shows that SerpentFlow can work with higher-resolution observational data: whereas previous experiments used ERA5 (25 km), this setup leverages SAFRAN at 8 km, illustrating the method’s scalability and adaptability to finer spatial grids. Third, it generalizes the evaluation to a different GCM and scenario, confirming that SerpentFlow is not tied to a single climate model or pathway. Finally, for this particular setup with CNRM-CM6-1 and SAFRAN, we also have a dynamically downscaled reference provided by the CNRM-Aladin regional climate model \citep{rcm} (RCM, 12.5 km resolution) corrected using CDF-t. This allows us to compare statistical (SerpentFlow) and dynamical (RCM) downscaling approaches, highlighting their respective limitations and strengths, and illustrating how they could be used complementarily to improve local-scale climate projections. To prepare for the subsequent mathematical description, it is important to emphasize the goal of the Gaussian-blur-based decomposition: it provides a simple, flexible way to separate large-scale patterns from local-scale variability, while respecting irregular observational domains.

\paragraph{Gaussian-blur-based decomposition}

Let \(x\) denote the input wind-speed field and \(\sigma\) the standard deviation of the Gaussian kernel controlling the blur. The Gaussian-blur-based decomposition separates \(x\) into a large-scale shared component 
\(\mu = \text{GaussianBlur}(x, \sigma)\), ignoring missing values, and a high-frequency residual 
\(\epsilon_{\rm HF} = \epsilon - \text{GaussianBlur}(\epsilon, \sigma)\) with stochastic noise 
\(\epsilon \sim \mathcal{N}(0, \mathbf{I})\). Pseudo-training pairs for the target domain are then constructed as 
\(\tilde{x}_B = \mu_B + \epsilon_{\rm HF}\), where \(\mu_B\) denotes the shared component of the target field. The blur intensity \(\sigma\) plays a role analogous to the cutoff frequency \(\omega_{\rm cut}\) in the Fourier decomposition: larger \(\sigma\) retains only very large-scale structures, while smaller \(\sigma\) preserves finer-scale details.

\paragraph{Implementation details}

\begin{itemize}
    \item Only wind speed (\texttt{sfcWind}) is considered, making this a univariate application.
    \item The same UNet generator architecture from the main SerpentFlow experiments is used.
    \item Pseudo-pairs are constructed using the Gaussian-blur decomposition, enabling training despite missing oceanic values.
    \item Baselines include SerpentFlow at different cutoffs (300, 500, 600, and 750,km, the latter identified by the classifier as the optimal separation scale and matches the GCM effective resolution), CDF-t as in the main experiments, and the Regional Climate Model (RCM) CNRM-Aladin \citep{rcm} (12.5 km resolution), bias-corrected using CDF-t on SAFRAN.
    \item Data is split between the training data (1980--2000), evaluation against SAFRAN (2000--2020) and evaluation against GCM projections (2000--2100).
\end{itemize}

\paragraph{Results}

\begin{figure}[htbp]
    \centering
    \begin{subfigure}{0.45\textwidth}
        \centering
        \includegraphics[width=\textwidth]{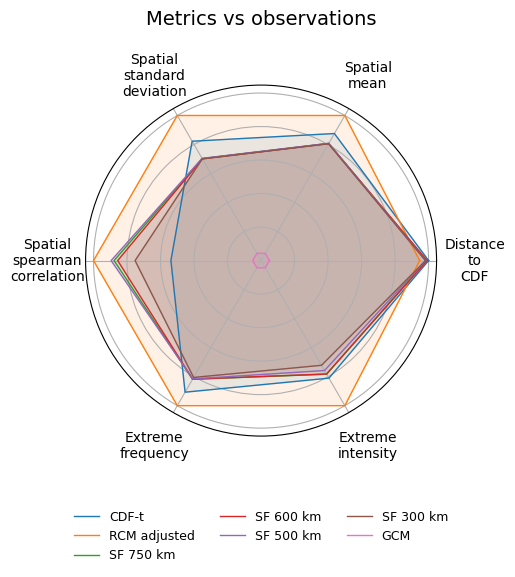}
        \caption{Metrics evaluated against SAFRAN.  Results are detailed Table~\ref{tab:metrics_safran}.}
        \label{fig:radar_safran}
    \end{subfigure}
    \hfill
    \begin{subfigure}{0.45\textwidth}
        \centering
        \includegraphics[width=\textwidth]{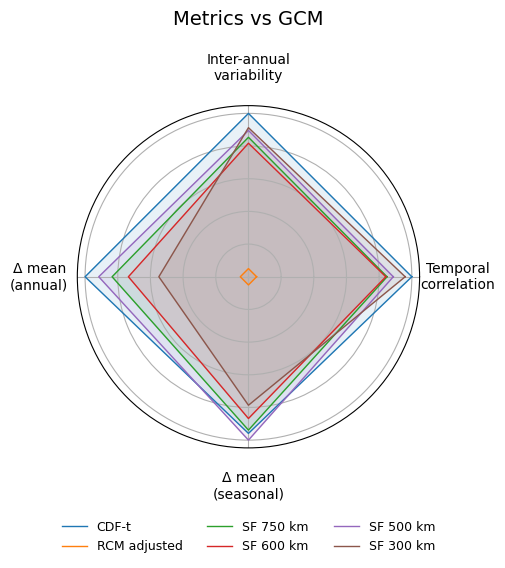}
        \caption{Metrics evaluated against the GCM. Results are detailed Table~\ref{tab:metrics_gcm_safran}}
        \label{fig:radar_safran_gcm}
    \end{subfigure}
    \caption{Radar plots summarizing the performance of all methods. Higher values (closer to the outer circle) indicate better agreement with the reference}
    \label{fig:radar_metrics_safran}
\end{figure}

The results (plots in Appendix~\ref{app:plots_cnrm}) are broadly consistent with the ERA5 experiment, confirming the robustness of the findings across datasets, resolutions, GCMs, and scenarios.

As expected, the RCM outperforms all statistical baselines on SAFRAN metrics, as it explicitly represents regional-scale processes and boundary conditions inaccessible to statistical methods like SerpentFlow. Its fine-scale dynamics related to topography and surface heterogeneity explain why it does not necessarily follow the temporal evolution of the driving GCM (see Figure~\ref{fig:temp_corr_safran}). However, Figure~\ref{fig:cdf_spearman_corr_SAFRAN} shows that SerpentFlow can better capture spatial interactions in some regions, reconstructing physically consistent structures directly from large-scale GCM information without additional dynamical assumptions. This suggests SerpentFlow could be used as a bias-correction method for RCM outputs, rather than relying on CDF-t.

More generally, SerpentFlow substantially outperforms CDF-t in reproducing SAFRAN spatial interactions, while achieving comparable or slightly lower performance on distributional metrics (Figure~\ref{fig:cdf_global_SAFRAN}), except for extreme local quantiles (Figures~\ref{fig:cdf_alps_SAFRAN}, \ref{fig:cdf_med_SAFRAN}). The cutoff scale provides control over reconstruction: lower cutoffs yield fields closer to SAFRAN, while higher cutoffs preserve more large-scale GCM characteristics, highlighting a trade-off between local realism and large-scale consistency.

Comparison with the RCM also highlights differences in projected climate change: Figures~\ref{fig:full_delta_safran} and~\ref{fig:full_seasonal_delta_safran} show that even large-scale future deltas are not preserved by the RCM compared to the driving GCM. Assessing which model provides a more realistic representation of future climate change remains challenging. 
Local phenomena are not resolved by GCMs, and several studies have shown that some processes and assumptions are not fully represented in RCMs for certain variables \citep{boe, nabat}, including near-surface wind \citep{wohland}.

Overall, these results emphasize the importance of maintaining diverse downscaling approaches. Dynamical (RCM) and statistical methods, such as SerpentFlow, are complementary and offer different perspectives on regional climate change. In this context, SerpentFlow provides a computationally efficient framework to explore uncertainties and to post-process both GCM and RCM outputs, correcting biases and enhancing spatial resolution.

\section{Conclusion}\label{sec:conclusion}

Our evaluation shows that SerpentFlow reproduces ERA5 spatial, temporal, and multivariate characteristics over France while largely preserving large-scale GCM dynamics. Compared to classical statistical methods, it improves spatial coherence, inter-variable consistency, and extreme-event representation, and its scale-aware decomposition allows control over the trade-off between local realism and large-scale consistency. Experiments with CNRM-CM6-1 and SAFRAN confirm that these results hold across resolutions, domains, and climate models, including irregular observational grids. Limitations include the short training period and focus on near-surface wind, leaving other variables and longer-term trends for future study. Overall, SerpentFlow demonstrates strong performance as a bias-correction and downscaling method under realistic climate scenarios.

%\paragraph{Funding Statement}

%\paragraph{Competing Interests}
%A statement about any financial, professional, %contractual or personal relationships or situations that could be perceived to impact the presentation of the work --- or `None' if none exist

%\paragraph{Data Availability Statement}
%A statement about how to access data, code and other materials allowing users to understand, verify and replicate findings --- e.g. Replication data and code can be found in Harvard Dataverse: \verb+\url{https://doi.org/link}+.

%\paragraph{Author Contributions}
%Please provide an author contributions statement using the CRediT taxonomy roles as a guide {\verb+\url{https://www.casrai.org/credit.html}+}. Conceptualization: A.A; A.B. Methodology: A.A; A.B. Data curation: A.C. Data visualisation: A.C. Writing original draft: A.A; A.B. All authors approved the final submitted draft.

%\renewcommand\bibpreamble{By default, this template uses \texttt{bibtex} and adopts the AMS referencing style. However, the journal you’re submitting to may require a different reference style; specify the journal you're using with the class' \texttt{journal} option --- see lines 1--19 of \emph{sample.tex} for a list of options and instructions for selecting the journal.}

% If using any of the following journal options:
%   wet, dap, dce, eds, prm, flw, jdm, psy, rsm
% then use the \printbibliography line instead of:
%\bibliographystyle{plain}
%\bibliography{example}

\bibliographystyle{plainnat}  % ou apa, alpha, etc.
\bibliography{example}        % fichier example.bib

\appendix

\section{Metrics}
\label{sec:metrics}

This section describes the evaluation protocol and the set of metrics used to compare the performance of SerpentFlow against the baseline. Most of the metrics are inspired by the work produced in \citep{francois2020multivariate}. The evaluation is performed separately against the reanalysis and the raw GCM (interpolated) output. All metrics are computed per climate variable and then averaged across the set of variables $\mathcal{V}$ considered in the study.

\subsection{Notations}

Let $X^{m,v}_{t,i,j}$ denote the value of climate variable $v \in \mathcal{V}$ produced by method $m$ at time $t$ and grid point $(i,j)$. The following notations are used:

\begin{itemize}
    \item $X^{\mathrm{obs},v}_{t,i,j}$ : reanalysis for variable $v$,
    \item $X^{\mathrm{GCM},v}_{t,i,j}$ : raw (interpolated) GCM output for variable $v$,
    \item $m \in \mathcal{M}$ : downscaling method (CDF-t, R2D2, Dual FM, SerpentFlow variants),
    \item $T$ : set of time steps,
    \item $S$ : set of spatial grid points,
    \item $\mathcal{V}$ : set of variables (e.g., wind speed, maximum wind speed, and the zonal and meridional wind components).
\end{itemize}

Spatial averages are taken over all grid points $(i,j) \in S$, and temporal averages over the evaluation period $T$. Metrics are then averaged across all climate variables $v \in \mathcal{V}$ unless otherwise specified.

\subsection{Metrics relative to the observations}

\subsubsection{Mean and standard deviation differences}

For each climate variable $v$, the mean difference between method $m$ and observations is:
\begin{equation}
\Delta \mu^{m,v} =
\frac{1}{|S|}\sum_{(i,j)\in S}
\left|\mu^{m,v}_{i,j} - \mu^{\mathrm{obs},v}_{i,j}\right|,
\quad
\mu^{m,v}_{i,j} = \frac{1}{|T|}\sum_{t\in T} X^{m,v}_{t,i,j}
\label{eq:delta_mu}
\end{equation}

The standard deviation difference is:
\begin{equation}
\Delta \sigma^{m,v} =
\frac{1}{|S|}\sum_{(i,j)\in S}
\left|\sigma^{m,v}_{i,j} - \sigma^{\mathrm{obs},v}_{i,j}\right|,
\quad
\sigma^{m,v}_{i,j} = \sqrt{\frac{1}{|T|}\sum_{t\in T} \left(X^{m,v}_{t,i,j}-\mu^{m,v}_{i,j}\right)^2}
\label{eq:delta_sigma}
\end{equation}

Finally, metrics are averaged across variables:
\begin{equation}
\Delta \mu^{m} = \frac{1}{|\mathcal{V}|}\sum_{v\in \mathcal{V}} \Delta \mu^{m,v}, \quad
\Delta \sigma^{m} = \frac{1}{|\mathcal{V}|}\sum_{v\in \mathcal{V}} \Delta \sigma^{m,v}.
\end{equation}

\subsubsection{Inter-variable correlation consistency}

For each downscaling method $m$, climate variable pair $(v_1,v_2)$ and grid point $(i,j)$, the temporal Pearson correlation is defined as
\begin{equation}
\rho^{m,v_1,v_2}_{i,j} = 
\frac{
\sum_{t \in T} \left( X^{m,v_1}_{t,i,j} - \mu^{m,v_1}_{i,j} \right) \left( X^{m,v_2}_{t,i,j} - \mu^{m,v_2}_{i,j} \right)
}{
\sqrt{
\sum_{t \in T} \left( X^{m,v_1}_{t,i,j} - \mu^{m,v_1}_{i,j} \right)^2
\sum_{t \in T} \left( X^{m,v_2}_{t,i,j} - \mu^{m,v_2}_{i,j} \right)^2
}
},
\end{equation}
where $\mu^{m,v}_{i,j} = \frac{1}{|T|} \sum_{t \in T} X^{m,v}_{t,i,j}$ is the temporal mean of variable $v$ at grid point $(i,j)$.

The mean absolute difference of Pearson correlations between method $m$ and the observations is then computed as
\begin{equation}
\Delta \rho^{m,v_1,v_2} = \frac{1}{|S|}\sum_{(i,j)\in S} 
\left|\rho^{m,v_1,v_2}_{i,j} - \rho^{\mathrm{obs},v_1,v_2}_{i,j}\right|.
\end{equation}

Finally, the metric is averaged over all distinct variable pairs to produce a single inter-variable correlation consistency score:
\begin{equation}
\Delta \rho^{m} = \frac{2}{|\mathcal{V}|(|\mathcal{V}|-1)} \sum_{v_1 < v_2} \frac{1}{|S|}\sum_{(i,j)\in S} 
\left|\rho^{m,v_1,v_2}_{i,j} - \rho^{\mathrm{obs},v_1,v_2}_{i,j}\right|
\label{eq:delta_rho}
\end{equation}

\subsubsection{Spatial Spearman correlation structure}

For each variable $v$, we first define spatial anomalies by removing the spatial mean at each time step:
\begin{equation}
X'^{m,v}_{t,i,j} = X^{m,v}_{t,i,j} - \frac{1}{|S|}\sum_{(i,j)\in S} X^{m,v}_{t,i,j}.
\end{equation}

Let $r^{m,v}_{t,i,j}$ denote the rank of $X'^{m,v}_{t,i,j}$ among all grid points $(i,j)$ at fixed time $t$.  
The Spearman correlation matrix $\mathbf{R}^{m,v}$ is then defined by computing the Pearson correlation of the ranks across time for every pair of grid points $(p,q)$:
\begin{equation}
\mathbf{R}^{m,v}_{p,q} = 
\frac{
\sum_{t\in T} \left(r^{m,v}_{t,p} - \overline{r}^{m,v}_{p}\right)\left(r^{m,v}_{t,q} - \overline{r}^{m,v}_{q}\right)
}{
\sqrt{
\sum_{t\in T} \left(r^{m,v}_{t,p} - \overline{r}^{m,v}_{p}\right)^2
\sum_{t\in T} \left(r^{m,v}_{t,q} - \overline{r}^{m,v}_{q}\right)^2
}
}, \quad
\overline{r}^{m,v}_{p} = \frac{1}{|T|}\sum_{t\in T} r^{m,v}_{t,p}.
\end{equation}

The spatial correlation error for variable $v$ is then computed as the mean absolute difference with the observations:
\begin{equation}
\Delta R^{m,v} = \frac{1}{|S|^2}\sum_{p,q \in S} \left| \mathbf{R}^{m,v}_{p,q} - \mathbf{R}^{\mathrm{obs},v}_{p,q} \right|.
\end{equation}

Finally, this metric is averaged across all variables to produce a single score:
\begin{equation}
\Delta \mathbf{R}^{m} = \frac{1}{|\mathcal{V}|} \sum_{v\in \mathcal{V}} \frac{1}{|S|^2}\sum_{p,q \in S} \left| \mathbf{R}^{m,v}_{p,q} - \mathbf{R}^{\mathrm{obs},v}_{p,q} \right|
\label{eq:delta_R}
\end{equation}

\subsubsection{Spatial power spectrum metric}

For each climate variable $v$, the isotropic spatial power spectrum is computed over the grid, and averaged over time.  
A scalar metric quantifies the relative large-scale energy difference between the downscaled field and observations:

\begin{equation}
\mathrm{SSM}^{m,v} = 
\frac{
\sqrt{\frac{1}{|k \le k_{\max}|}\sum_{k \le k_{\max}} (P^{m,v}_k - P^{\mathrm{obs},v}_k)^2}
}{
\frac{1}{|k \le k_{\max}|}\sum_{k \le k_{\max}} P^{\mathrm{obs},v}_k
}, \quad
\mathrm{SSM}^{m} = \frac{1}{|\mathcal{V}|} \sum_{v \in \mathcal{V}} \mathrm{SSM}^{m,v}
\end{equation}

where $P^{m,v}_k$ denotes the mean power spectrum in wavenumber bin $k$, and $k_{\max}$ is a cutoff separating large-scale features.

\subsubsection{Spatial variogram metric (mountains)}

For mountainous regions (elevation $> \theta_\text{alt}$), the semi-variogram $\gamma(h)$ is computed for each variable $v$, and compared to the observations:

\begin{equation}
\gamma^{m,v}(h) = \frac{1}{2 |(i,j) \in \text{bin } h|} \sum_{(i,j) \in \text{bin } h} \left( X^{m,v}_i - X^{m,v}_j \right)^2,
\end{equation}
\begin{equation}
\mathrm{OVM}^{m} = \frac{1}{|\mathcal{V}|} \sum_{v\in\mathcal{V}} 
\frac{\sqrt{\frac{1}{h \le h_{\max}} \sum_{h \le h_{\max}} (\gamma^{m,v}(h) - \gamma^{\mathrm{obs},v}(h))^2}}
{\frac{1}{h \le h_{\max}}\sum_{h \le h_{\max}} \gamma^{\mathrm{obs},v}(h)}
\end{equation}

where $h$ denotes distance bins and $h_{\max}$ a cutoff distance.

\subsubsection{Kolmogorov--Smirnov statistic}

For each variable $v$, let $X^{m,v}_{t,i,j}$ denote the values produced by method $m$ at all times $t \in T$ and all spatial points $(i,j) \in S$.  
The empirical cumulative distribution function (CDF) of method $m$ is defined as
\begin{equation}
F^{m,v}(x) = \frac{1}{|T||S|} \sum_{t\in T} \sum_{(i,j)\in S} \mathbb{I}\left(X^{m,v}_{t,i,j} \le x\right),
\end{equation}
where $\mathbb{I}$ is the indicator function.  
Similarly, $F^{\mathrm{obs},v}(x)$ is the empirical CDF of the observations' values for variable $v$.

The two-sample Kolmogorov--Smirnov (KS) statistic between method $m$ and the observations for variable $v$ is then defined as
\begin{equation}
\mathrm{KS}^{m,v} = \sup_x \left| F^{m,v}(x) - F^{\mathrm{obs},v}(x) \right|.
\end{equation}

Finally, the KS statistic is averaged over all variables:
\begin{equation}
\mathrm{KS}^{m} = \frac{1}{|\mathcal{V}|} \sum_{v \in \mathcal{V}} \sup_x \left| F^{m,v}(x) - F^{\mathrm{obs},v}(x) \right|
\label{eq:KS}
\end{equation}

\subsubsection{Extreme value metrics}

Using variable-specific observations thresholds $\theta^{v}_{i,j} = Q_{0.95}(X^{\mathrm{obs},v}_{\cdot,i,j})$:

\paragraph{Extreme frequency}
\begin{equation}
f^{m,v}_{i,j} = \frac{1}{|T|}\sum_{t\in T} \mathbb{I}(X^{m,v}_{t,i,j}>\theta^{v}_{i,j}), \quad
f^{m} = \frac{1}{|\mathcal{V}|}\sum_{v\in \mathcal{V}} \frac{1}{|S|}\sum_{(i,j)\in S} |f^{m,v}_{i,j}-f^{\mathrm{obs},v}_{i,j}|
\label{eq:ext_freq}
\end{equation}

\paragraph{Extreme intensity}
\begin{equation}
I^{m,v}_{i,j} = \frac{1}{N^{m,v}_{i,j}} \sum_{t:X^{m,v}_{t,i,j}>\theta^{v}_{i,j}} X^{m,v}_{t,i,j}, \quad
I^{m} = \frac{1}{|\mathcal{V}|}\sum_{v\in \mathcal{V}} \frac{1}{|S|} \sum_{(i,j)\in S} |I^{m,v}_{i,j}-I^{\mathrm{obs},v}_{i,j}|
\label{eq:ext_intensity}
\end{equation}

\subsection{Metrics relative to the GCM}

\subsubsection{Temporal correlation}

For each variable $v$, the temporal Pearson correlation with the GCM is:
\begin{equation}
\rho^{m,v}_{\mathrm{GCM},i,j} = \mathrm{corr}_t(X^{m,v}_{\cdot,i,j}, X^{\mathrm{GCM},v}_{\cdot,i,j}), \quad
\rho^{m}_{\mathrm{GCM}} = \frac{1}{|\mathcal{V}|}\sum_{v\in \mathcal{V}}\frac{1}{|S|}\sum_{(i,j)\in S} \mathrm{corr}_t(X^{m,v}_{\cdot,i,j}, X^{\mathrm{GCM},v}_{\cdot,i,j})
\label{eq:rho_GCM}
\end{equation}

\subsubsection{Global annual anomalies}

Let $\overline{X}^{m,v}_{y} = \frac{1}{|S|}\sum_{(i,j)\in S} X^{m,v}_{\text{year } y,i,j}$. Annual anomaly deviations:
\begin{equation}
A^{m,v}_y = \overline{X}^{m,v}_y - \frac{1}{N_y}\sum_y \overline{X}^{m,v}_y, \quad
A^m = \frac{1}{|\mathcal{V}|} \sum_{v\in \mathcal{V}} \frac{1}{N_y}\sum_y |A^{m,v}_y - A^{\mathrm{GCM},v}_y|
\label{eq:annual_anom}
\end{equation}

\subsubsection{Relative climate change signal}

For each variable $v$, the relative change between a historical period and a future period is computed as
\begin{equation}
\Delta^{m,v}_{\mathrm{full}} = \frac{\overline{X}^{m,v}_{\mathrm{fut}} - \overline{X}^{m,v}_{\mathrm{hist}}}{\overline{X}^{m,v}_{\mathrm{hist}}} \times 100,
\label{eq:delta_full}
\end{equation}
where $\overline{X}^{m,v}_{\mathrm{hist}}$ and $\overline{X}^{m,v}_{\mathrm{fut}}$ are spatial averages over the historical and future periods, respectively.

For the seasonal variant, the same metric is computed for each meteorological season $s$:
\begin{itemize}
    \item DJF: December–January–February,
    \item MAM: March–April–May,
    \item JJA: June–July–August,
    \item SON: September–October–November.
\end{itemize}

The seasonal relative change is then
\begin{equation}
\Delta^{m,v}_{\mathrm{season},s} = \frac{\overline{X}^{m,v}_{\mathrm{fut},s} - \overline{X}^{m,v}_{\mathrm{hist},s}}{\overline{X}^{m,v}_{\mathrm{hist},s}} \times 100,
\end{equation}
with averages computed over the months corresponding to season $s$.  

Finally, metrics are averaged over all variables and, for the seasonal metric, over all seasons:
\begin{equation}
\Delta^m_{\mathrm{full}} = \frac{1}{|\mathcal{V}|}\sum_{v\in \mathcal{V}} |\Delta^{m,v}_{\mathrm{full}} - \Delta^{\mathrm{GCM},v}_{\mathrm{full}}|, 
\end{equation}
\begin{equation}    
\Delta^m_{\mathrm{season}} = \frac{1}{|\mathcal{V}|}\sum_{v\in \mathcal{V}} \frac{1}{|S|} \sum_{(i,j)\in S} \frac{1}{4} \sum_{s \in \{\text{DJF,MAM,JJA,SON}\}} |\Delta^{m,v}_{\mathrm{season},s} - \Delta^{\mathrm{GCM},v}_{\mathrm{season},s}|.
\label{eq:delta_season}
\end{equation}

\subsection{Summary}

All metrics are computed per variable, averaged over space, and then aggregated across variables to produce method-level scores.

\begin{table}[htbp]
\centering
\small
\setlength{\tabcolsep}{6pt}
\renewcommand{\arraystretch}{1.2}
\caption{Summary of evaluation metrics}
\label{tab:metrics_summary_extended}
\begin{tabular}{>{\centering\arraybackslash}m{2.2cm} >{\raggedright\arraybackslash}m{10cm}}
\hline
\textbf{Metric} & \textbf{Description / Equation} \\
\hline
\multicolumn{2}{l}{\textit{Metrics relative to the observations}} \\
$\Delta \mu$ & Mean difference (Eq.~\ref{eq:delta_mu}) \\
$\Delta \sigma$ & Standard deviation difference (Eq.~\ref{eq:delta_sigma}) \\
$\Delta \rho$ & Inter-variable temporal correlation consistency (Eq.~\ref{eq:delta_rho}) \\
$\Delta R$ & Spatial Spearman correlation error (Eq.~\ref{eq:delta_R}) \\
SSM & Spatial power spectrum metric (large-scale energy) \\
OVM & Orographic variogram metric (spatial variability in mountains) \\
KS & Kolmogorov--Smirnov statistic (CDF error, Eq.~\ref{eq:KS}) \\
$f$ & Extreme frequency (Eq.~\ref{eq:ext_freq}) \\
$I$ & Extreme intensity (Eq.~\ref{eq:ext_intensity}) \\

\hline
\multicolumn{2}{l}{\textit{Metrics relative to GCM}} \\
$\rho_{\mathrm{GCM}}$ & Temporal correlation with GCM (Eq.~\ref{eq:rho_GCM}) \\
$A$ & Annual anomaly deviations (inter-annual variability, Eq.~\ref{eq:annual_anom}) \\
$\Delta_{\mathrm{full}}$ & Full-period relative climate change signal (Eq.~\ref{eq:delta_full}) \\
$\Delta_{\mathrm{season}}$ & Seasonal relative climate change signal (Eq.~\ref{eq:delta_season}) \\
\hline
\end{tabular}
\end{table}

Below are the values for each metric for the two experiments detailed Section~\ref{sec:exp}.

%\begin{landscape}
\begin{table}[htbp]
\centering
\small
\setlength{\tabcolsep}{4pt}
\renewcommand{\arraystretch}{1.2}
\caption{Performance metrics with respect to ERA5 observations. For ensemble methods, values are reported as mean $\pm$ standard deviation across members.}
\label{tab:metrics_era5}
\resizebox{\textwidth}{!}{
\begin{tabular}{lccccccccc}
\hline
Metric & CDF-t & R2D2 & DFM & DFMm & SF12 & SF12m & SF7 & SF3 & GCM \\ \hline
\makecell[l]{Distance \\ to \\ CDF} & 0.016 & 0.016 & 0.024 & 0.024 & 0.018 & \textbf{0.011} & 0.013 & 0.012 & 0.062 \\ 
\makecell[l]{Spatial \\ mean} & 0.109 & 0.098 & 0.064\tiny{$\pm$0.000} & 0.064 & \textbf{0.060\tiny{$\pm$0.001}} & 0.061 & 0.061\tiny{$\pm$0.000} & 0.063\tiny{$\pm$0.000} & 0.639 \\ 
\makecell[l]{Spatial \\ standard \\ deviation} & 0.062 & 0.074 & 0.068 & 0.068 & 0.089\tiny{$\pm$0.000} & 0.057 & 0.056\tiny{$\pm$0.000} & \textbf{0.055\tiny{$\pm$0.000}} & 0.452 \\ 
\makecell[l]{Inter \\ variables \\ correlation} & 0.121 & \textbf{0.024} & 0.108 & 0.108 & 0.069\tiny{$\pm$0.000} & 0.056 & 0.091\tiny{$\pm$0.000} & 0.103\tiny{$\pm$0.000} & 0.128 \\ 
\makecell[l]{Spatial \\ spearman \\ correlation} & 19350 & 42407 & 14504\tiny{$\pm$0} & 14504 & 14931\tiny{$\pm$33} & \textbf{10055} & 15733\tiny{$\pm$20} & 17513\tiny{$\pm$7} & 30929 \\ 
\makecell[l]{Spatial \\ Variogram \\ (mountains)} & 0.227 & 0.705 & 0.067 & 0.067 & 0.129\tiny{$\pm$0.000} & \textbf{0.033} & 0.131\tiny{$\pm$0.000} & 0.162\tiny{$\pm$0.000} & 0.305 \\ 
\makecell[l]{Spatial \\ spectrum} & 0.188 & 0.540 & \textbf{0.078\tiny{$\pm$0.000}} & 0.078 & 0.173\tiny{$\pm$0.001} & 0.089 & 0.203\tiny{$\pm$0.001} & 0.215\tiny{$\pm$0.000} & 0.719 \\ 
\makecell[l]{Extreme \\ frequency} & 0.006 & 0.007 & 0.011 & 0.011 & 0.007\tiny{$\pm$0.000} & \textbf{0.006} & 0.007\tiny{$\pm$0.000} & 0.007\tiny{$\pm$0.000} & 0.052 \\ 
\makecell[l]{Extreme \\ intensity} & \textbf{0.103} & 0.118 & 0.200 & 0.200 & 0.116\tiny{$\pm$0.001} & 0.118 & 0.137\tiny{$\pm$0.000} & 0.158\tiny{$\pm$0.001} & 0.349 \\ 
\hline
\end{tabular}
}
\vspace{2pt}\\\footnotesize{DFM = Dual FM, DFMm = Dual FM mbr, SF12 = SF 1200 km, SF12m = SF 1200 km mbr, SF7 = SF 750 km, SF3 = SF 300 km}
\end{table}
%\end{landscape}

\begin{table}[htbp]
\centering
\small
\setlength{\tabcolsep}{4pt}
\renewcommand{\arraystretch}{1.2}
\caption{Performance metrics with respect to GCM. For ensemble methods, values are reported as mean $\pm$ standard deviation across members.}
\label{tab:metrics_gcm}
\resizebox{\textwidth}{!}{
\begin{tabular}{lccccccccc}
\hline
Metric & CDF-t & R2D2 & DFM & DFMm & SF12 & SF12m & SF7 & SF3 & GCM \\ \hline
\makecell[l]{Temporal \\ correlation} & \textbf{0.996} & 0.473 & 0.715 & 0.715 & 0.893\tiny{$\pm$0.000} & 0.853 & 0.932\tiny{$\pm$0.000} & 0.956\tiny{$\pm$0.000} & -- \\ 
\makecell[l]{Inter-annual \\ variability} & \textbf{0.012} & 0.088 & 0.066\tiny{$\pm$0.000} & 0.066 & 0.013\tiny{$\pm$0.000} & 0.013 & 0.014\tiny{$\pm$0.000} & 0.017\tiny{$\pm$0.000} & -- \\ 
\makecell[l]{Delta mean \\ (annual)} & \textbf{0.167} & 0.684 & 0.880\tiny{$\pm$0.000} & 0.880 & 0.625\tiny{$\pm$0.012} & 0.636 & 0.620\tiny{$\pm$0.008} & 0.514\tiny{$\pm$0.006} & -- \\ 
\makecell[l]{Delta mean \\ (seasonal)} & \textbf{0.266} & 1.097 & 1.516 & 1.516 & 1.012\tiny{$\pm$0.018} & 1.098 & 0.921\tiny{$\pm$0.010} & 0.757\tiny{$\pm$0.006} & -- \\ 
\hline
\end{tabular}
}
\vspace{2pt}\\\footnotesize{DFM = Dual FM, DFMm = Dual FM mbr, SF12 = SF 1200 km, SF12m = SF 1200 km mbr, SF7 = SF 750 km, SF3 = SF 300 km}
\end{table}

%\begin{landscape}
\begin{table}[htbp]
\centering
\small
\setlength{\tabcolsep}{4pt}
\renewcommand{\arraystretch}{1.2}
\caption{Performance metrics with respect to ERA5 observations (sfcWind variable only). For ensemble methods, values are reported as mean $\pm$ standard deviation across members.}
\label{tab:metrics_era5_sfcWind}
\resizebox{\textwidth}{!}{
\begin{tabular}{lccccccccc}
\hline
Metric & CDF-t & R2D2 & DFM & DFMm & SF12 & SF12m & SF7 & SF3 & GCM \\ \hline
\makecell[l]{Distance \\ to \\ CDF} & 0.014 & 0.014 & 0.032 & 0.032 & 0.013 & \textbf{0.005} & 0.010 & 0.009 & 0.061 \\ 
\makecell[l]{Spatial \\ Variogram \\ (mountains)} & 0.170 & 0.190 & 0.026 & 0.026 & 0.083\tiny{$\pm$0.000} & \textbf{0.024} & 0.084\tiny{$\pm$0.000} & 0.111\tiny{$\pm$0.000} & 0.206 \\ 
\makecell[l]{Spatial \\ spectrum} & 0.124 & 0.139 & 0.054 & 0.054 & 0.090\tiny{$\pm$0.003} & \textbf{0.038} & 0.090\tiny{$\pm$0.001} & 0.104\tiny{$\pm$0.000} & 0.761 \\ 
\makecell[l]{Extreme \\ frequency} & 0.006 & 0.006 & 0.008 & 0.008 & \textbf{0.005} & 0.005 & 0.005 & 0.005 & 0.061 \\ 
\makecell[l]{Extreme \\ intensity} & 0.103 & 0.103 & 0.172 & 0.172 & \textbf{0.087} & 0.092 & 0.089 & 0.102 & 0.333 \\ 
\hline
\end{tabular}
}
\vspace{2pt}\\\footnotesize{DFM = Dual FM, DFMm = Dual FM mbr, SF12 = SF 1200 km, SF12m = SF 1200 km mbr, SF7 = SF 750 km, SF3 = SF 300 km}
\end{table}
%\end{landscape}

\begin{table}[htbp]
\centering
\small
\setlength{\tabcolsep}{4pt}
\renewcommand{\arraystretch}{1.2}
\caption{Performance metrics with respect to GCM (sfcWind variable only). For ensemble methods, values are reported as mean $\pm$ standard deviation across members.}
\label{tab:metrics_gcm_sfcWind}
\resizebox{\textwidth}{!}{
\begin{tabular}{lccccccccc}
\hline
Metric & CDF-t & R2D2 & DFM & DFMm & SF12 & SF12m & SF7 & SF3 & GCM \\ \hline
\makecell[l]{Temporal \\ correlation} & \textbf{0.997} & 0.835 & 0.646 & 0.646 & 0.879 & 0.837 & 0.922 & 0.949 & -- \\ 
\makecell[l]{Inter-annual \\ variability} & 0.006 & 0.012 & 0.077 & 0.077 & 0.005 & \textbf{0.005} & 0.007 & 0.010 & -- \\ 
\makecell[l]{Delta mean \\ (annual)} & \textbf{0.138} & 0.511 & 0.895 & 0.895 & 0.608 & 0.583 & 0.543 & 0.443 & -- \\ 
\makecell[l]{Delta mean \\ (seasonal)} & \textbf{0.175} & 0.802 & 1.057 & 1.057 & 1.037 & 1.085 & 0.712 & 0.608 & -- \\ 
\hline
\end{tabular}
}
\vspace{2pt}\\\footnotesize{DFM = Dual FM, DFMm = Dual FM mbr, SF12 = SF 1200 km, SF12m = SF 1200 km mbr, SF7 = SF 750 km, SF3 = SF 300 km}
\end{table}

\begin{table}[htbp]
\centering
\small
\setlength{\tabcolsep}{5pt}
\renewcommand{\arraystretch}{1.2}
\caption{Performance metrics with respect to SAFRAN}
\label{tab:metrics_safran}
\resizebox{\textwidth}{!}{%
\begin{tabular}{p{2.5cm}ccccccc}
\hline
Metric & CDF-t & RCM adjusted & SF 750 km & SF 600 km & SF 500 km & SF 300 km & GCM \\ \hline
Distance to CDF & \textbf{0.032} & 0.042 & 0.035 & 0.034 & 0.036 & 0.036 & 0.212 \\ 
Spatial mean & 0.320 & \textbf{0.196} & 0.388 & 0.389 & 0.388 & 0.391 & 1.147 \\ 
Spatial standard deviation & 0.180 & \textbf{0.099} & 0.235 & 0.236 & 0.234 & 0.235 & 0.532 \\ 
Spatial spearman correlation & 0.166 & \textbf{0.091} & 0.111 & 0.114 & 0.108 & 0.131 & 0.245 \\ 
Extreme frequency & 0.014 & \textbf{0.008} & 0.020 & 0.020 & 0.020 & 0.021 & 0.070 \\ 
Extreme intensity & 0.223 & \textbf{0.128} & 0.237 & 0.237 & 0.249 & 0.267 & 0.603 \\ 
\hline
\end{tabular}
}
\end{table}

\begin{table}[htbp]
\centering
\small
\setlength{\tabcolsep}{5pt}
\renewcommand{\arraystretch}{1.2}
\caption{Performance metrics with respect to GCM (SAFRAN experimentation)}
\label{tab:metrics_gcm_safran}
\resizebox{\textwidth}{!}{%
\begin{tabular}{p{2.5cm}cccccc}
\hline
Metric & CDF-t & RCM adjusted & SF 750 km & SF 600 km & SF 500 km & SF 300 km \\ \hline
Temporal correlation & \textbf{0.957} & 0.085 & 0.818 & 0.811 & 0.853 & 0.920\\ 
Inter-annual variability & \textbf{0.032} & 0.200 & 0.058 & 0.064 & 0.050 & 0.047\\ 
Delta mean (annual) & \textbf{0.423} & 2.077 & 0.711 & 0.884 & 0.569 & 1.209\\ 
Delta mean (seasonal) & 1.068 & 3.544 & 1.122 & 1.316 & \textbf{0.955} & 1.537\\ 
\hline
\end{tabular}
}
\end{table}
\newpage
\section{Ensemble calibration}
\label{app:calibration}
 
As described in Section~\ref{method}, SerpentFlow generates downscaled fields by sampling stochastic noise in the domain-specific (high-frequency) component and passing it through the learned generator $f_\theta$. At inference, this noise is drawn as $\boldsymbol{z} \sim \mathcal{N}(\mathbf{0},\, a^2 \mathbf{I})$, where the scaling factor $a$ controls the amplitude of the injected variability. When $a=1$, the inference noise matches the unit-variance distribution seen during training. Adjusting $a$ allows post-hoc control over ensemble spread without retraining: lower values produce tighter ensembles, while higher values increase diversity among members.
 
We generate $N=10$ ensemble members for each time step on ERA5 low-passed values (perfect model setup) and evaluate calibration against the ground truth over the 2000--2020 period using four standard diagnostics: the mean Continuous Ranked Probability Score (CRPS), the spread--skill ratio, rank histograms, and reliability diagrams. All diagnostics are shown for the four wind channels (\texttt{sfcWind}, \texttt{uas}, \texttt{vas}, \texttt{sfcWindmax}), with the rank histogram and reliability diagram displayed for \texttt{sfcWind}.
 
\begin{figure}[h]
    \centering
    \includegraphics[width=\textwidth]{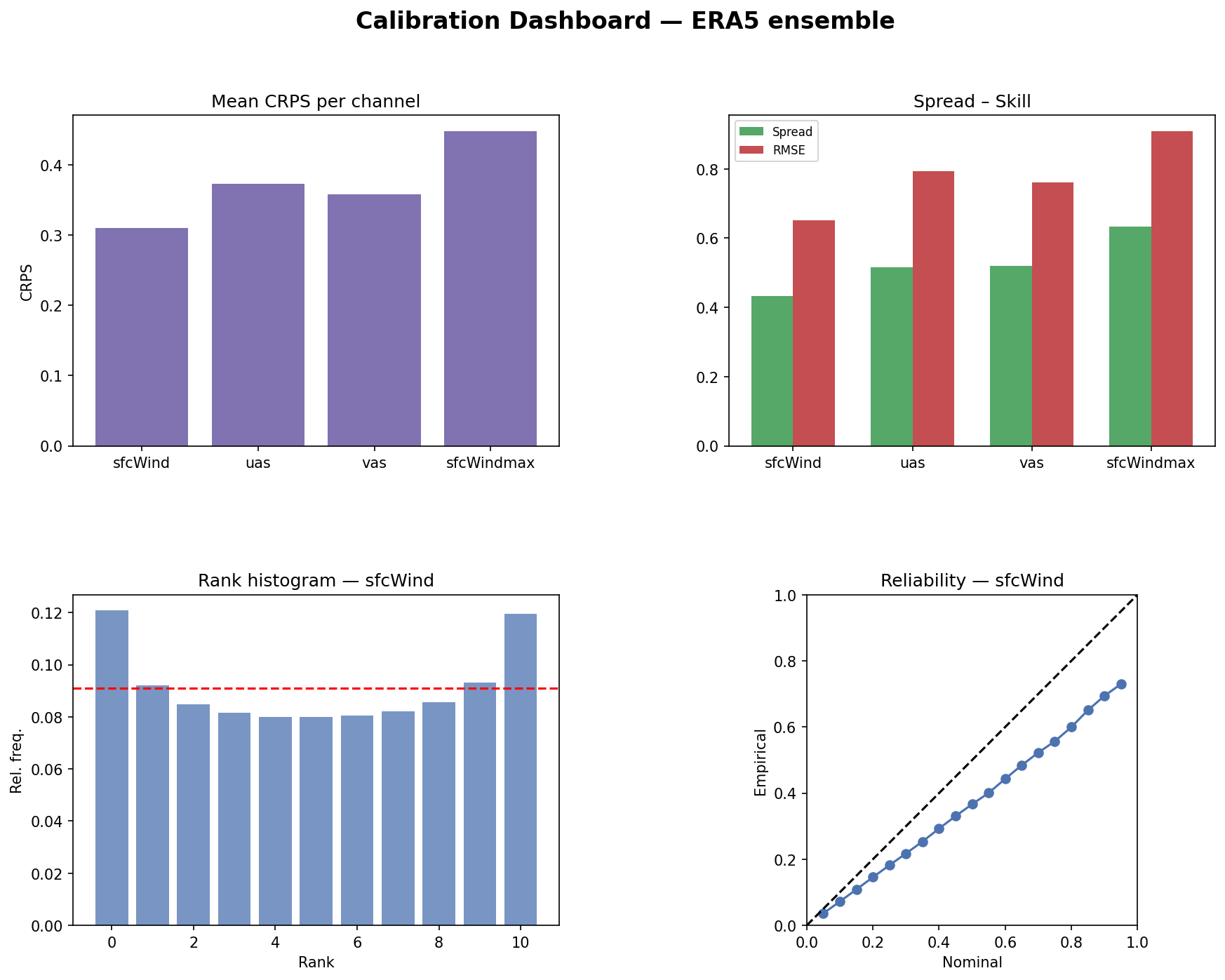}
    \caption{Calibration dashboard for the ERA5 ensemble with noise scaling $a=1.0$. The U-shaped rank histogram and the spread--skill gap indicate underdispersion.}
    \label{fig:calib_a1}
\end{figure}

\paragraph{$a=1.0$ (Figure~\ref{fig:calib_a1}).} The rank histogram exhibits a clear U-shape, indicating that the ensemble is underdispersive: ERA5 values frequently fall outside the ensemble range. Consistently, the ensemble spread is noticeably smaller than the RMSE across all channels, and the reliability curve lies above the diagonal at high nominal levels, confirming that prediction intervals are too narrow. This suggests that the unit-variance noise used during training does not fully capture the variability needed at inference, likely because the generator learns a partial mapping from noise to fine-scale structure, reducing the effective stochasticity of the outputs.
 
\begin{figure}[h]
    \centering
    \includegraphics[width=\textwidth]{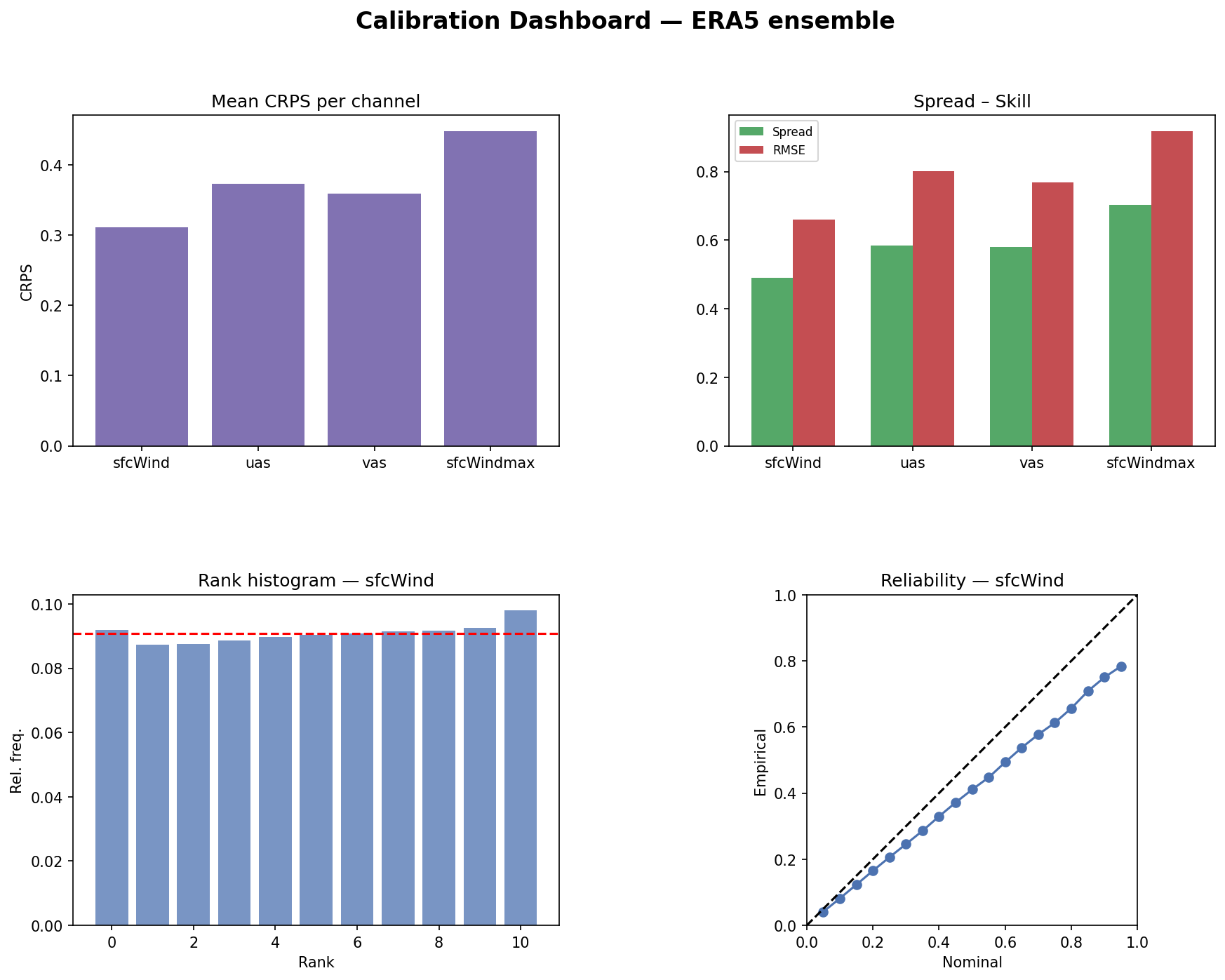}
    \caption{Calibration dashboard for $a=1.1$. The near-flat rank histogram, balanced spread--skill ratio, and diagonal reliability curve indicate well-calibrated ensembles.}
    \label{fig:calib_a11}
\end{figure}

\paragraph{$a=1.1$ (Figure~\ref{fig:calib_a11}).} A moderate increase in noise amplitude substantially improves calibration. The rank histogram becomes nearly flat, the spread--skill gap narrows across all variables, and the reliability curve closely follows the diagonal. This configuration yields the best overall calibration among the three tested values, indicating that a slight amplification of the domain-specific noise compensates for the variance reduction introduced by the deterministic component of the generator.
 
\begin{figure}[h]
    \centering
    \includegraphics[width=\textwidth]{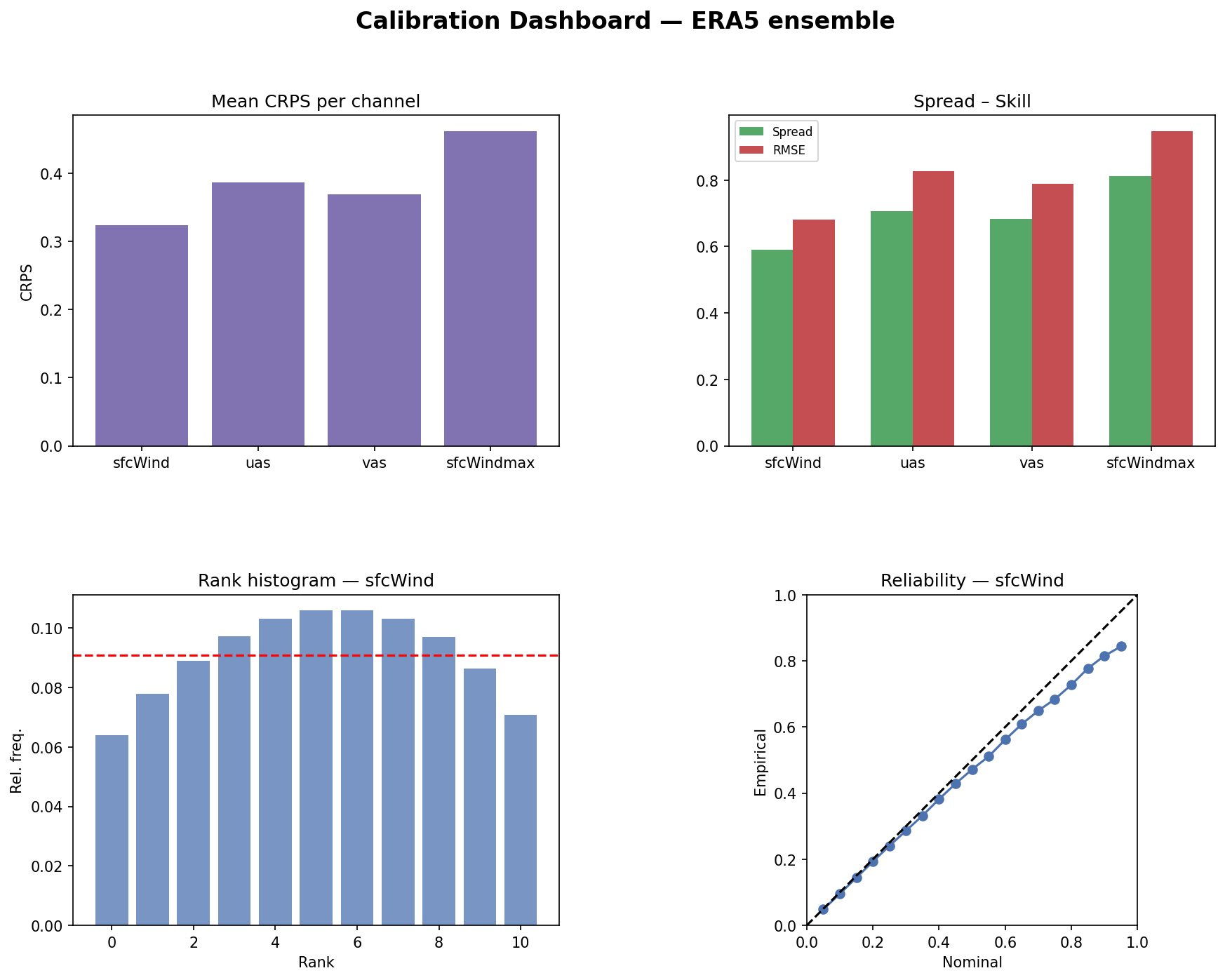}
    \caption{Calibration dashboard for $a=1.2$. The hump-shaped rank histogram and the reliability curve below the diagonal indicate overdispersion.}
    \label{fig:calib_a12}
\end{figure}

\paragraph{$a=1.2$ (Figure~\ref{fig:calib_a12}).} Further increasing the noise amplitude leads to overdispersion. The rank histogram develops a hump shape with depleted tails, indicating that ensemble members spread too widely. The spread now approaches or exceeds the RMSE for several channels, and the reliability curve falls below the diagonal, meaning that prediction intervals are wider than necessary. At this level, the injected noise dominates over the learned fine-scale structure, degrading the physical plausibility of individual members.
 
Based on this analysis, the value $a=1.1$ should be used as the default noise scaling factor. This simple post-hoc tuning of the noise amplitude provides an effective and computationally inexpensive way to calibrate probabilistic predictions from SerpentFlow without any retraining.
 
\section{Additional Plots ACCESS/ERA5}\label{app:plots}

\begin{figure}[htbp]
    \centering
    \begin{subfigure}{0.45\textwidth}
        \centering
        \includegraphics[width=\textwidth]{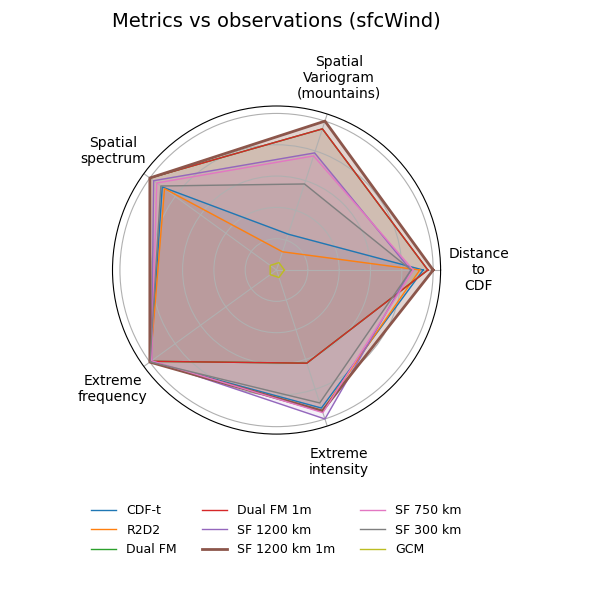}
        \caption{Metrics evaluated against ERA5. Results are detailed Table~\ref{tab:metrics_era5_sfcWind}.}
        \label{fig:radar_era5_sfc_wind}
    \end{subfigure}
    \hfill
    \begin{subfigure}{0.45\textwidth}
        \centering
        \includegraphics[width=\textwidth]{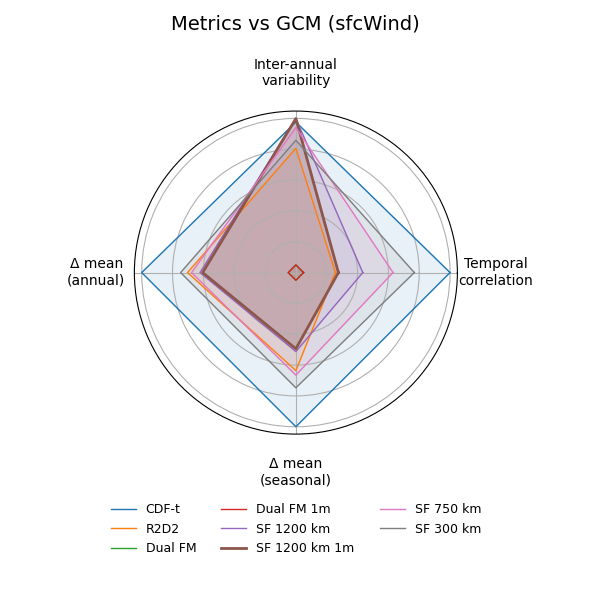}
        \caption{Metrics evaluated against the GCM. Results are detailed Table~\ref{tab:metrics_gcm_sfcWind}.}
        \label{fig:radar_gcm_sfcwind}
    \end{subfigure}
    \caption{Radar plots summarizing the performance of all methods on the mean wind speed only. Higher values (closer to the outer circle) indicate better agreement with the reference. SF stands for SerpentFlow in all the plots, and ``mbr" indicates ``one member" for a generative method, the average of the members being shown otherwise}
    \label{fig:radar_metrics_sfcwind}
\end{figure}

\subsection{Plots vs observations}

% Wind Maps
\begin{figure}[htbp]
    \centering
    \includegraphics[width=\linewidth]{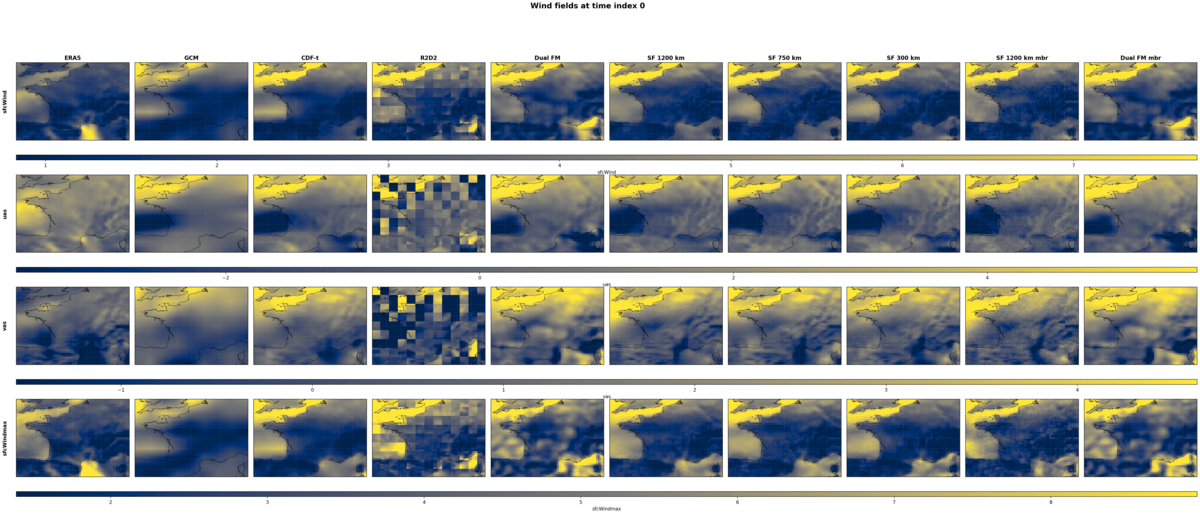}
    \caption{Wind maps for the first time step, for each climate variables and for each method. The data and methods in the columns are, respectively: ERA5, GCM, CDF-t, R2D2, Dual FM, SerpentFlow (SF) 1200 km, SF 750 km, SF 300 km, SF 1200 km (1 member only), and Dual FM (1 member only). The rows show sfcWind, uas, vas, and sfcWindmax, respectively. The squares appearing on the R2D2 plot for each climate variable are due to the $5\times5$ spatial patches that had to be used due to limitations of the method. ERA5 and GCM are not aligned in time, which explains the differences between the first two columns. For the different downscaling methods, we can see that the pattern given by the GCM is generally followed. Dual FM seems to have created some artifacts, particularly in Brittany (in the west)}
    \label{fig:wind_maps_era5}
\end{figure}
% Mean / std
\begin{figure}[htbp]
    \centering
    \begin{subfigure}{0.49\textwidth}
        \centering
        \includegraphics[width=\linewidth]{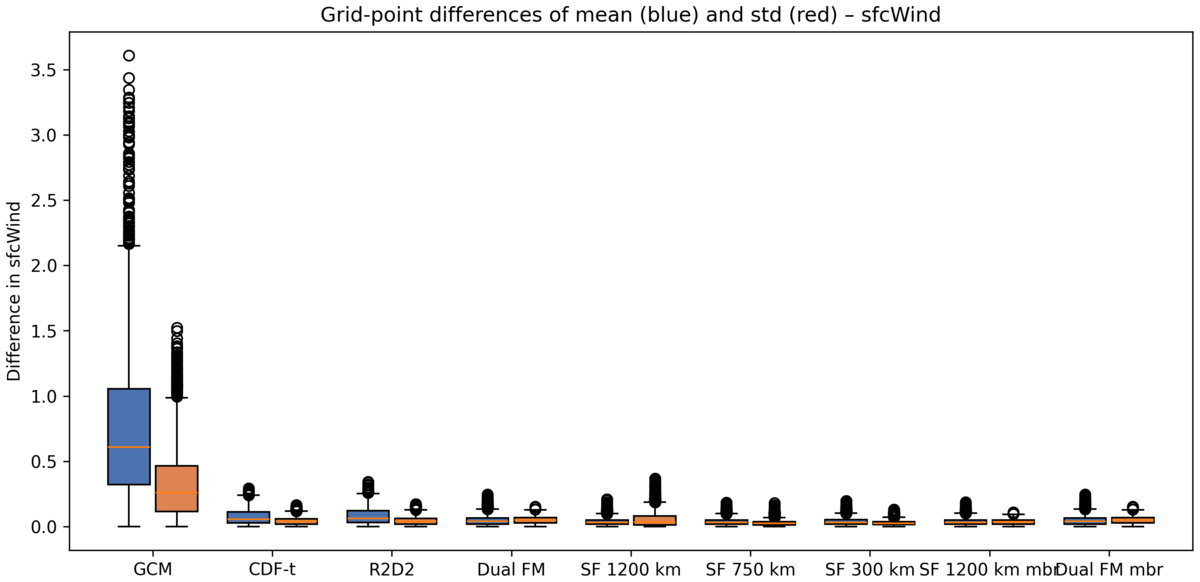}
        \caption{Wind Speed}
    \end{subfigure}
    \hfill
    \begin{subfigure}{0.49\textwidth}
        \centering
        \includegraphics[width=\linewidth]{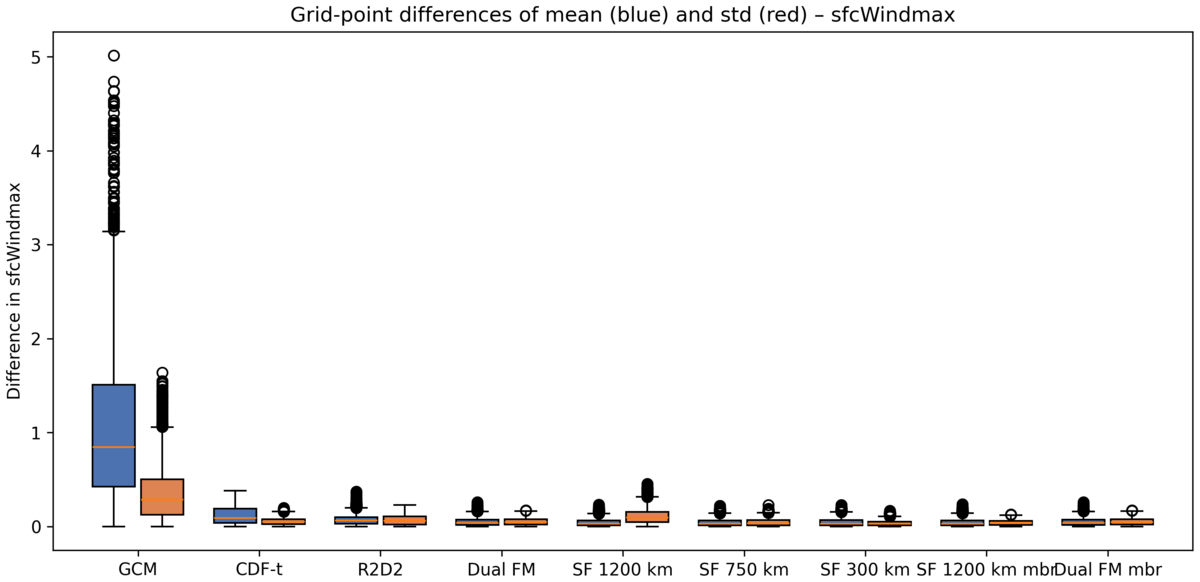}
        \caption{Maximum wind speed}
    \end{subfigure}

    \medskip

    \begin{subfigure}{0.49\textwidth}
        \centering
        \includegraphics[width=\linewidth]{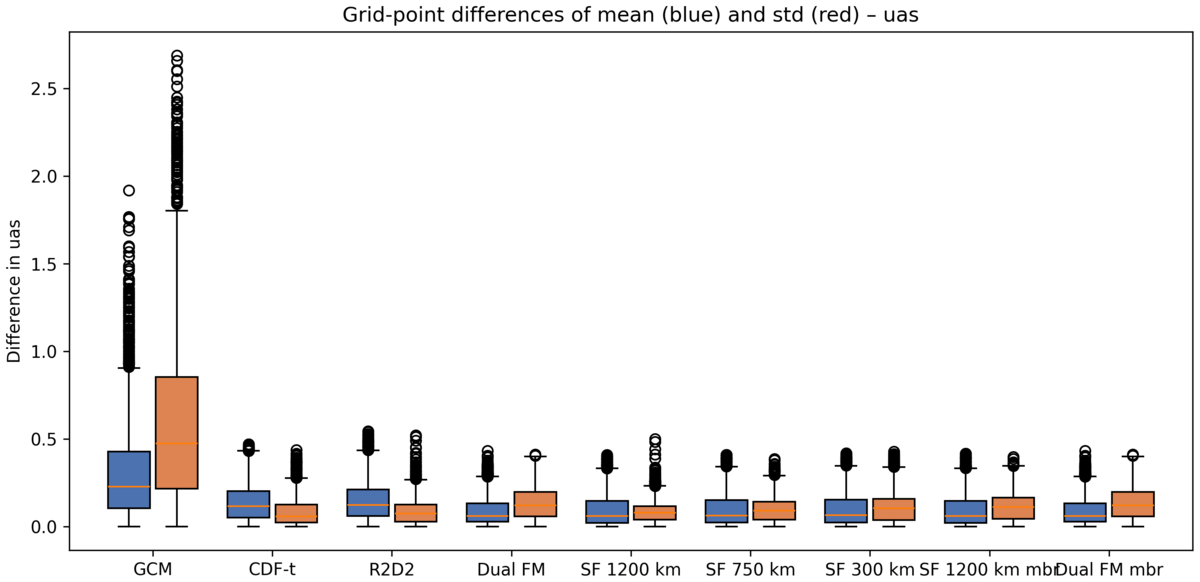}
        \caption{Zonal wind}
    \end{subfigure}
    \hfill
    \begin{subfigure}{0.49\textwidth}
        \centering
        \includegraphics[width=\linewidth]{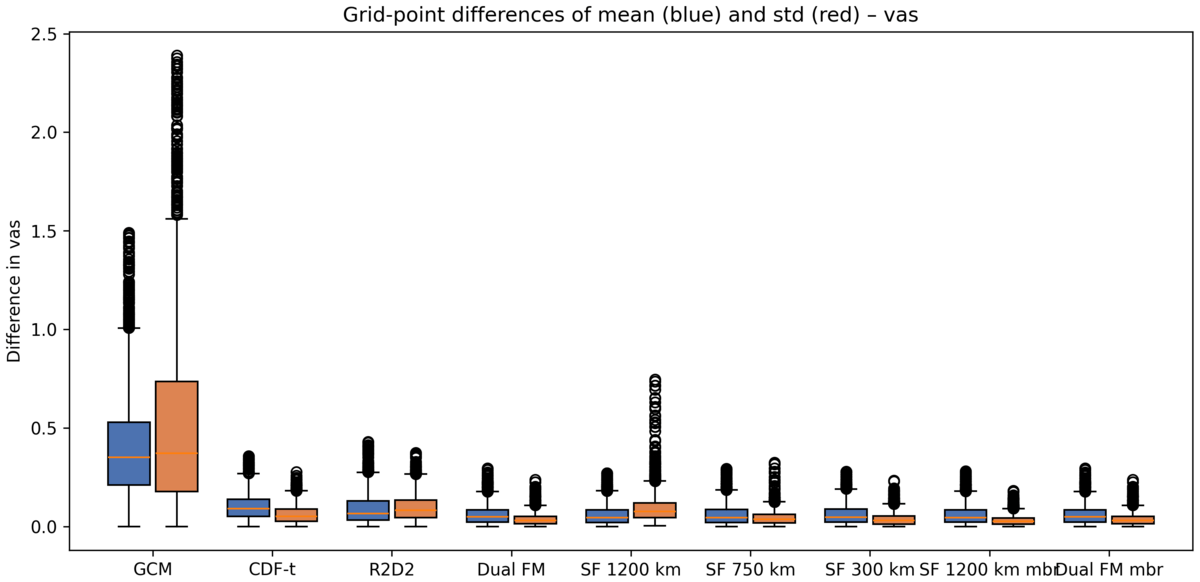}
        \caption{Meridional wind}
    \end{subfigure}
    \caption{Distribution of mean and standard deviations differences w.r.t ERA5 per grid point. All methods achieved spatially consistent bias correction, obtaining deviations from means and standard deviations well below those of the GCM. This is particularly true for wind speed variables. It is also noted that R2D2 deviates quite significantly from the mean for the meridional wind, which may explain the performance shown on the radar plot. Similarly, SF 1200 km (the mean) shows some deviations for the standard deviation of the meridional wind}
    \label{fig:mean_std_ERA5}
\end{figure}
% CDFs Global
\begin{figure}[htbp]
    \centering
    \begin{subfigure}{0.49\textwidth}
        \centering
        \includegraphics[width=\linewidth]{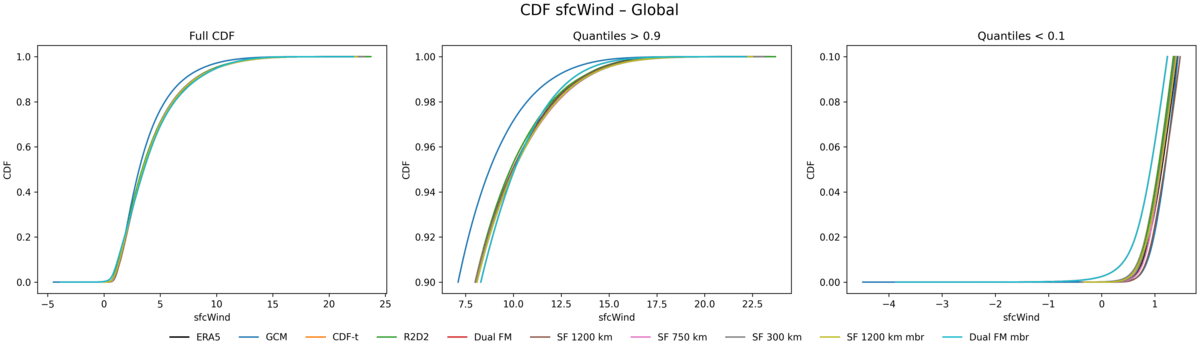}
        \caption{Wind Speed}
    \end{subfigure}
    \hfill
    \begin{subfigure}{0.49\textwidth}
        \centering
        \includegraphics[width=\linewidth]{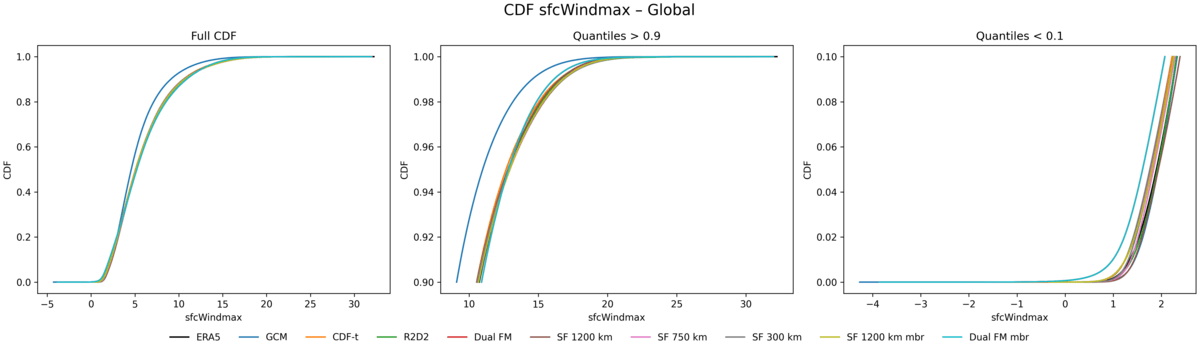}
        \caption{Maximum wind speed}
    \end{subfigure}

    \medskip

    \begin{subfigure}{0.49\textwidth}
        \centering
        \includegraphics[width=\linewidth]{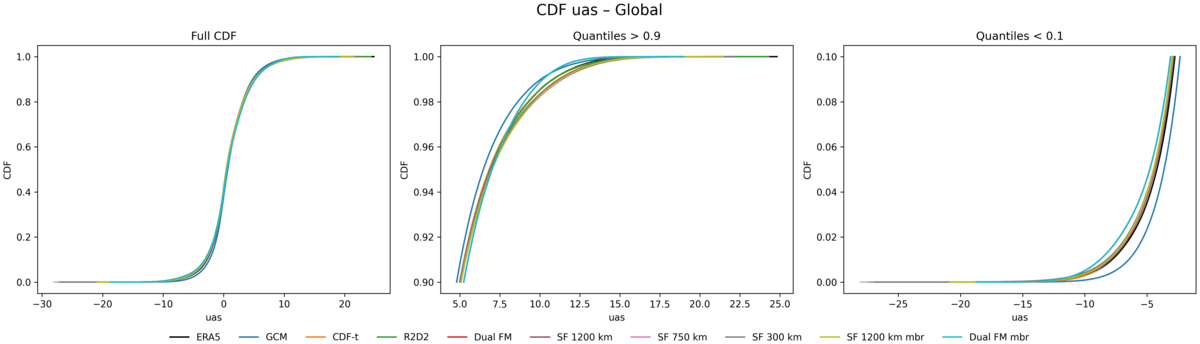}
        \caption{Zonal wind}
    \end{subfigure}
    \hfill
    \begin{subfigure}{0.49\textwidth}
        \centering
        \includegraphics[width=\linewidth]{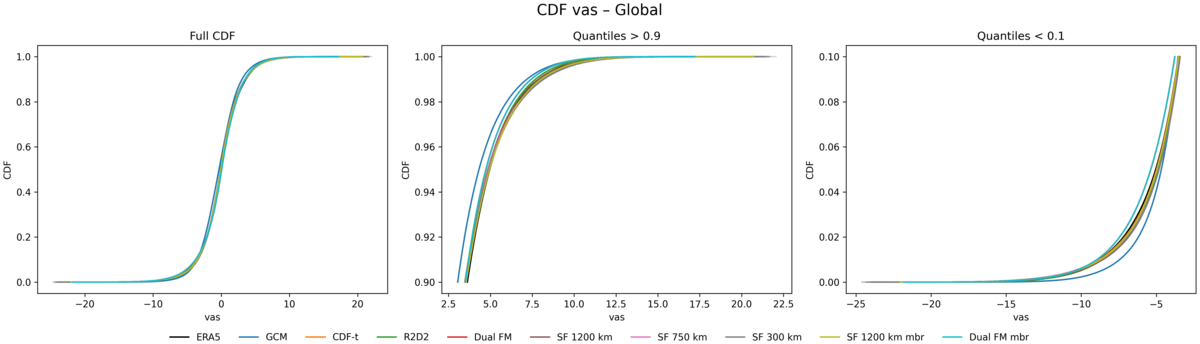}
        \caption{Meridional wind}
    \end{subfigure}
    \caption{Cumulative Distribution Functions (CDFs) for each method over the validation period. For each sublot a zoom is done on the extreme quantiles ($q\leq 0.1$ and $q\geq 0.9$). The bias correction is also clearly visible on these CDF plots. While the GCM tends to slightly underestimate wind speeds, downscaling methods correct this bias even in extreme quantiles. As in the paper \citep{serpentflow2024}, we note that Dual FM produces negative values. The correction of u and v is more complicated in the high and low quantiles. Dual FM in particular fails}
    \label{fig:cdf_global_ERA5}
\end{figure}

% CDFs Alps
\begin{figure}[htbp]
    \centering
    \begin{subfigure}{0.49\textwidth}
        \centering
        \includegraphics[width=\linewidth]{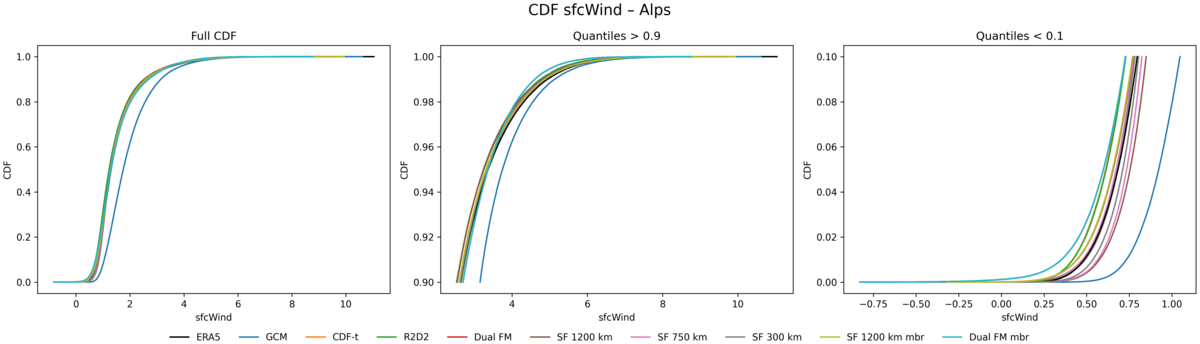}
        \caption{Wind Speed}
    \end{subfigure}
    \hfill
    \begin{subfigure}{0.49\textwidth}
        \centering
        \includegraphics[width=\linewidth]{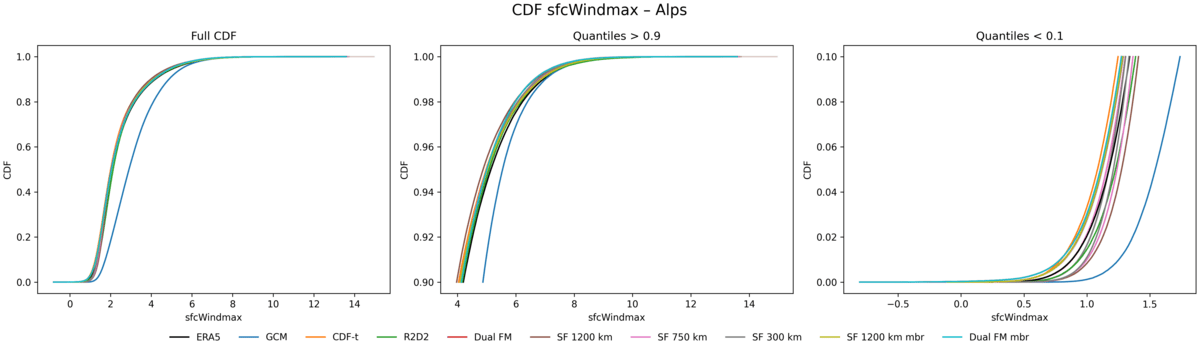}
        \caption{Maximum wind speed}
    \end{subfigure}

    \medskip

    \begin{subfigure}{0.49\textwidth}
        \centering
        \includegraphics[width=\linewidth]{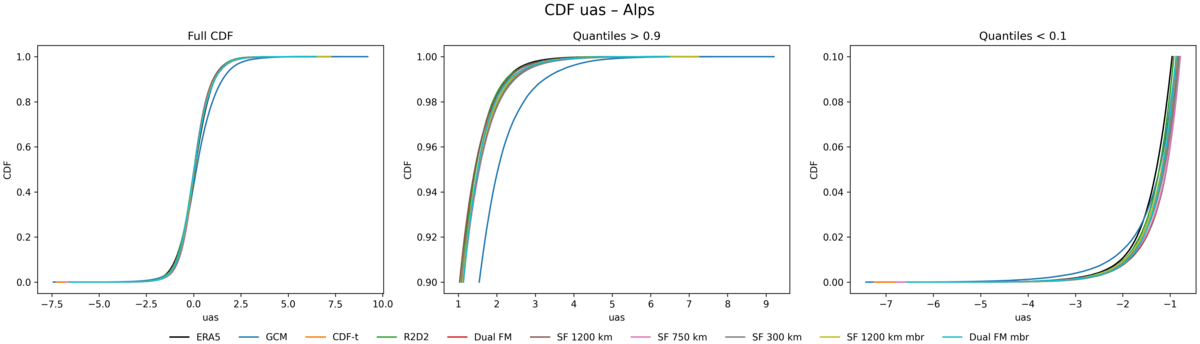}
        \caption{Zonal wind}
    \end{subfigure}
    \hfill
    \begin{subfigure}{0.49\textwidth}
        \centering
        \includegraphics[width=\linewidth]{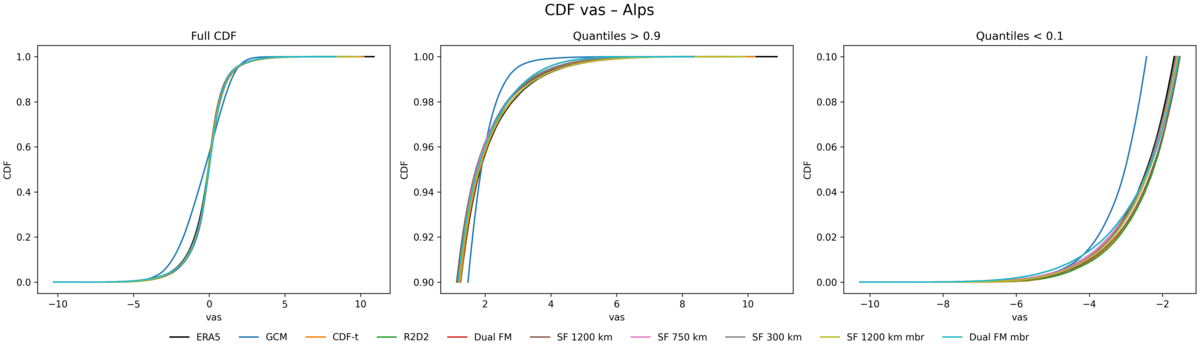}
        \caption{Meridional wind}
    \end{subfigure}
    \caption{Cumulative distribution functions (CDFs) for each method over a small region in the Alps ($\mathrm{latitude} \in [44,47]$, $\mathrm{longitude} \in [5,8]$). For each subplot, a zoom is applied to the extreme quantiles ($q \leq 0.1$ and $q \geq 0.9$). In this small region, the corrections remain very good, particularly for u and v. All methods are therefore successful in modeling marginal distributions. However, for wind speed variables, we note that SF 1200 km (single-member version) is significantly more accurate for the extreme quantiles (high and low) of ERA5}
    \label{fig:cdf_alps_ERA5}
\end{figure}

% CDFs Med
\begin{figure}[htbp]
    \centering
    \begin{subfigure}{0.49\textwidth}
        \centering
        \includegraphics[width=\linewidth]{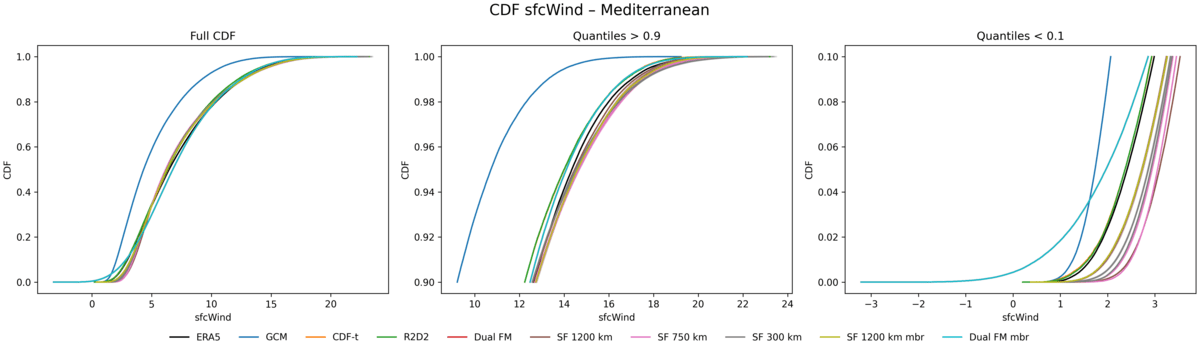}
        \caption{Wind Speed}
    \end{subfigure}
    \hfill
    \begin{subfigure}{0.49\textwidth}
        \centering
        \includegraphics[width=\linewidth]{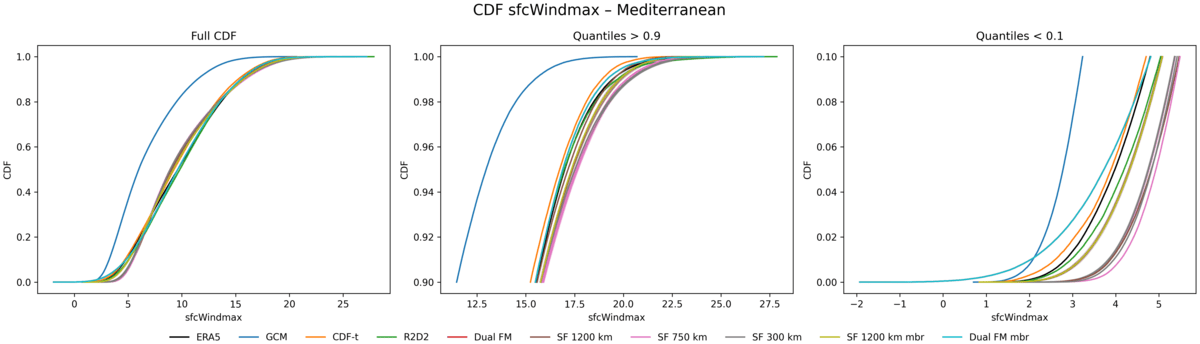}
        \caption{Maximum wind speed}
    \end{subfigure}

    \medskip

    \begin{subfigure}{0.49\textwidth}
        \centering
        \includegraphics[width=\linewidth]{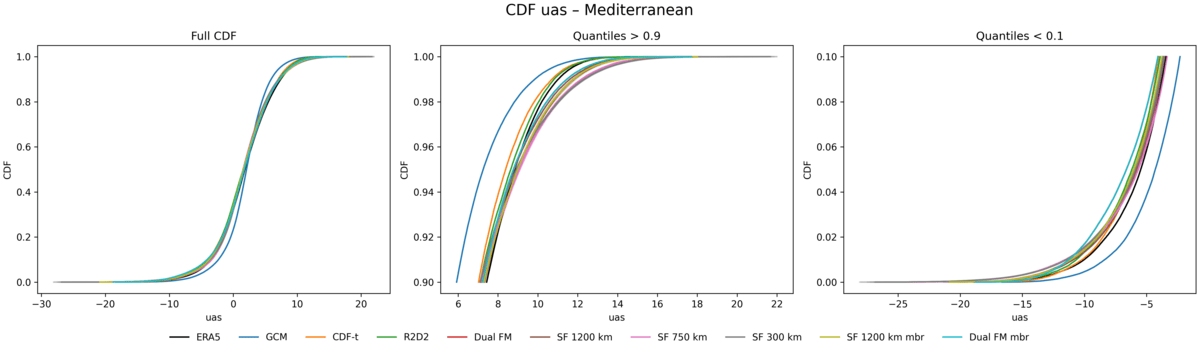}
        \caption{Zonal wind}
    \end{subfigure}
    \hfill
    \begin{subfigure}{0.49\textwidth}
        \centering
        \includegraphics[width=\linewidth]{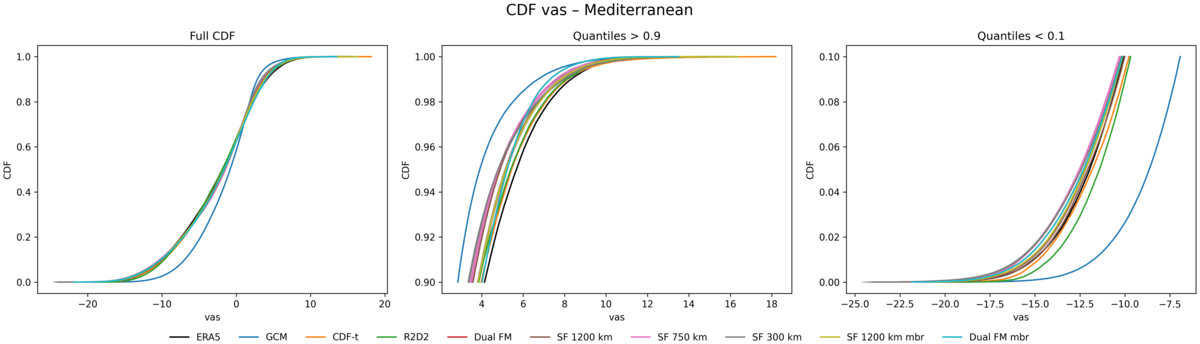}
        \caption{Meridional wind}
    \end{subfigure}
    \caption{Cumulative distribution functions (CDFs) for each method over a small region in the Mediterranean ($\mathrm{latitude} \in [41,43]$, $\mathrm{longitude} \in [3,6]$). For each subplot, a zoom is applied to the extreme quantiles ($q \leq 0.1$ and $q \geq 0.9$). In this region, R2D2 performs best in the high quantiles of all variables, while Dual FM performs best in the low quantiles. Nevertheless, in general, the distributions have been well corrected compared to the GCM}
    \label{fig:cdf_med_ERA5}
\end{figure}

% Intervar corr
\begin{figure}[htbp]
    \centering
    \includegraphics[height=0.8\textheight]{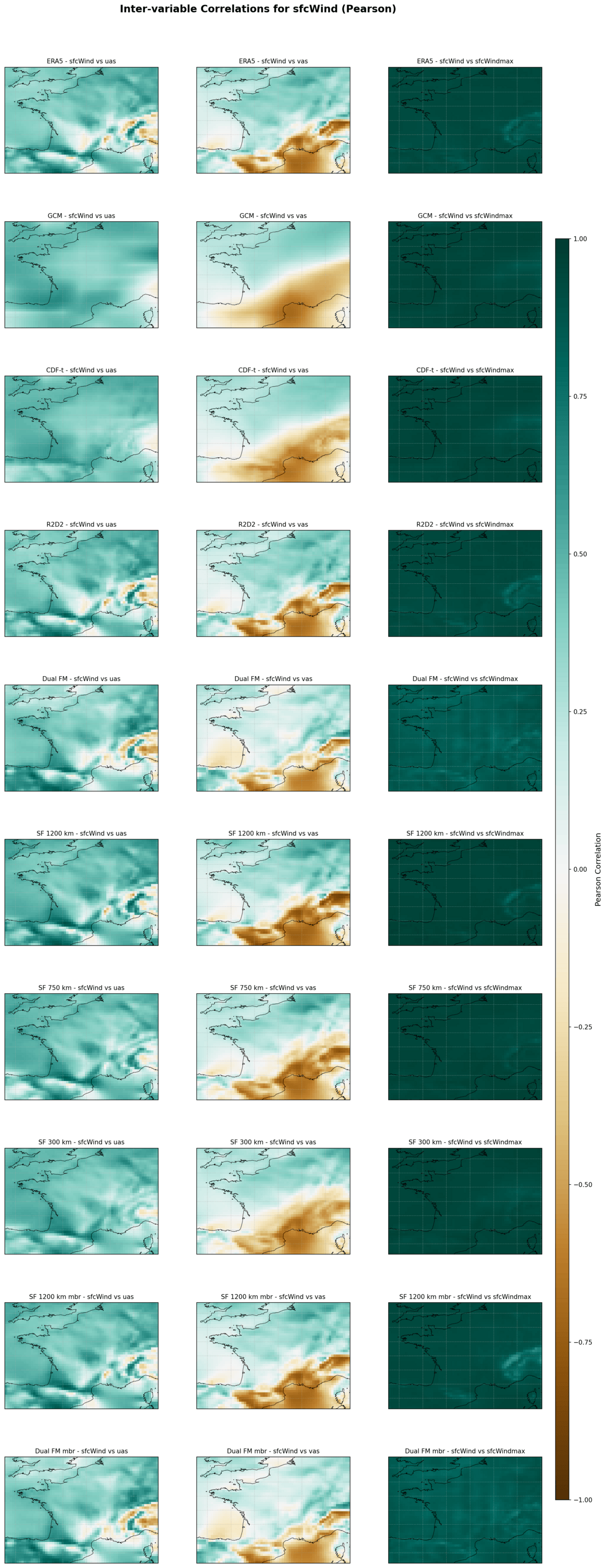}
    \caption{Pearson correlation coefficients between wind speed and all other climate variables, by grid point and method. The columns show correlations with zonal wind, meridional wind, and maximum wind speed, respectively. The CDF-t correlations are very close to those of the GCM. The maps lean heavily toward green, indicating strong correlation between variables. However, for wind speed versus meridional wind in the ERA5 data, relief features and coastlines tend toward anti-correlation (dark brown). We also verified the physical consistency constraint sfcWind $\leq$ sfcWindmax: this holds for nearly all grid points and time steps, with violation rates comparable to those found in the raw GCM output.}
    \label{fig:sfcwind_era5_corr}
\end{figure}
\begin{figure}[htbp]
    \centering
    \includegraphics[height=0.9\textheight]{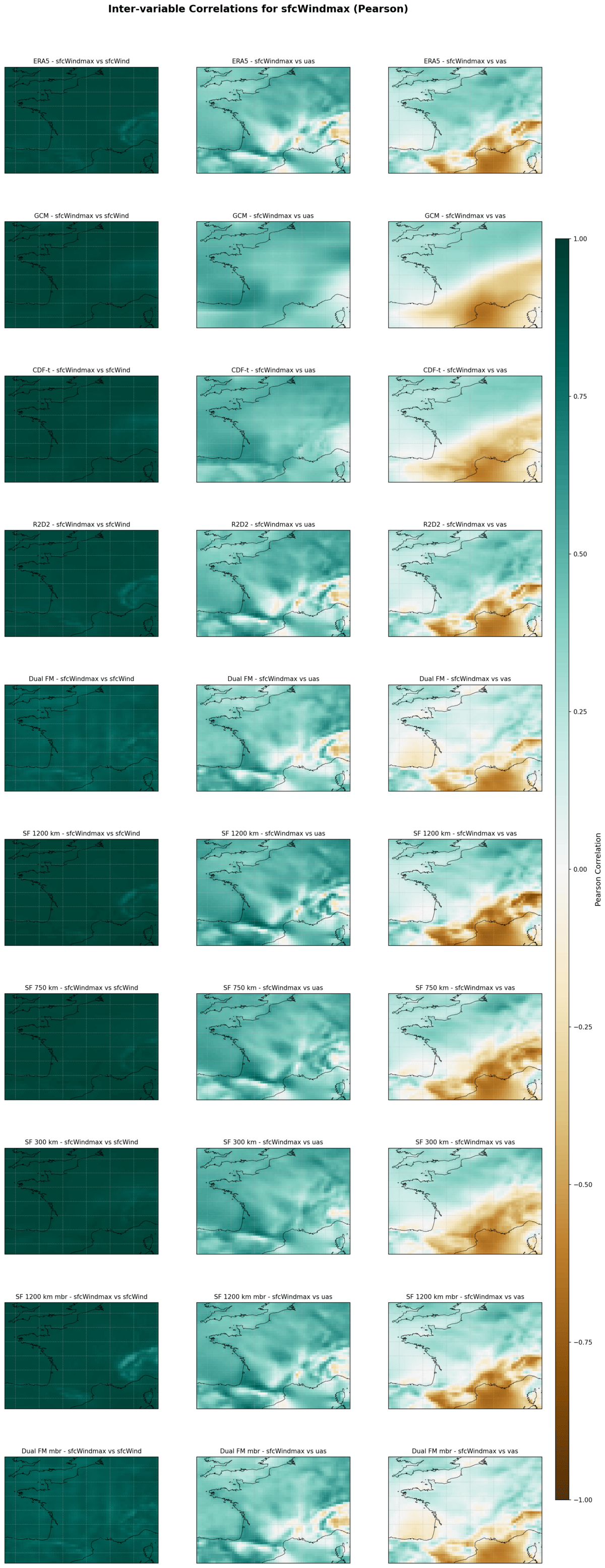}
    \caption{Pearson correlations coefficient between maximum wind speed and all the other climate variables, by grid point and method. Respectively, the columns are correlations vs wind speed, zonal and meridional wind. Dual FM appears to be significantly less effective than SerpentFlow or R2D2 at correctly correcting inter-variable correlations. Between average wind speed and zonal wind, excessive correlations appear in flat areas. Between average and maximum speed, Dual FM tends to accentuate the loss of correlation over relief areas}
    \label{fig:sfcwindmax_era5_corr}
\end{figure}
\begin{figure}[htbp]
    \centering
    \includegraphics[height=0.9\textheight]{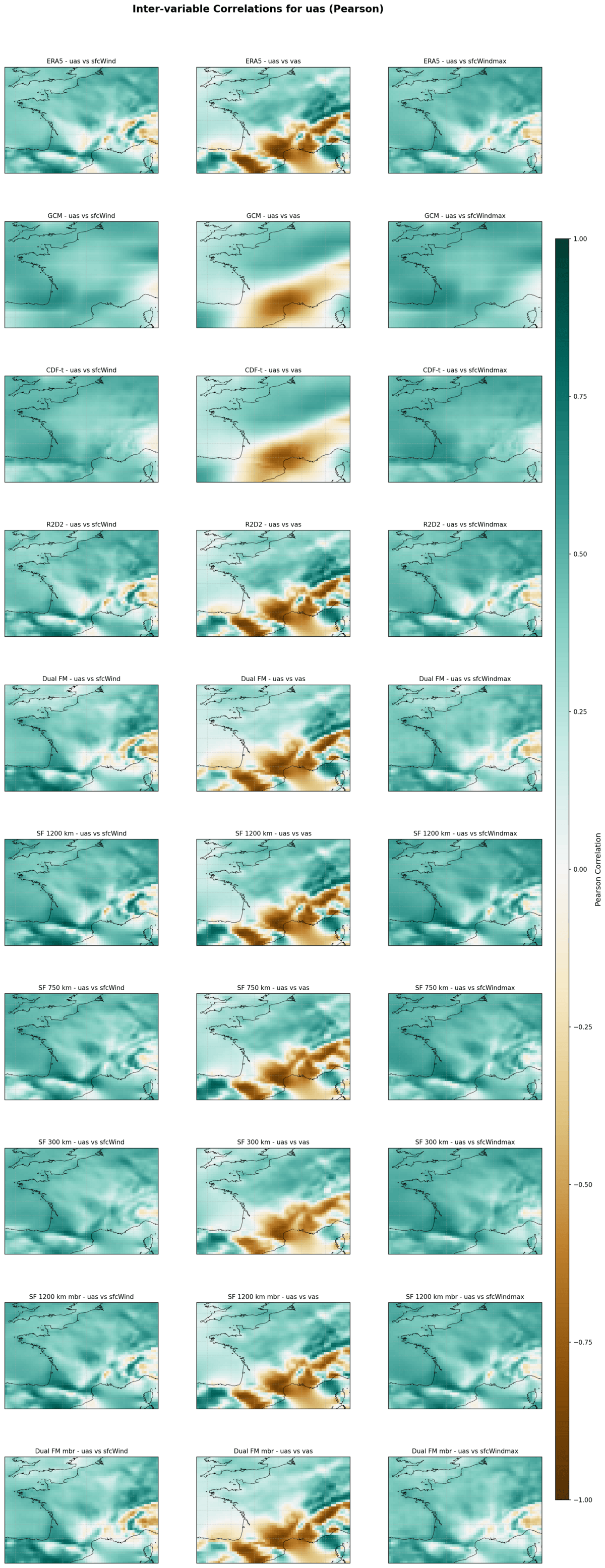}
    \caption{Pearson correlations coefficient between zonal wind and all the other climate variables, by grid point and method. Respectively, the columns are correlations vs wind speed, meridional wind and max wind. This figure clearly shows, particularly in the middle column (u vs. v), that the lower the SerpentFlow cutoff frequency (i.e., the more small-scale elements are removed), the closer the correlation with ERA5. SF 300 km is much less accurate here than SF 1200 km}
    \label{fig:uas_era5_corr}
\end{figure}
\begin{figure}[htbp]
    \centering
    \includegraphics[height=0.9\textheight]{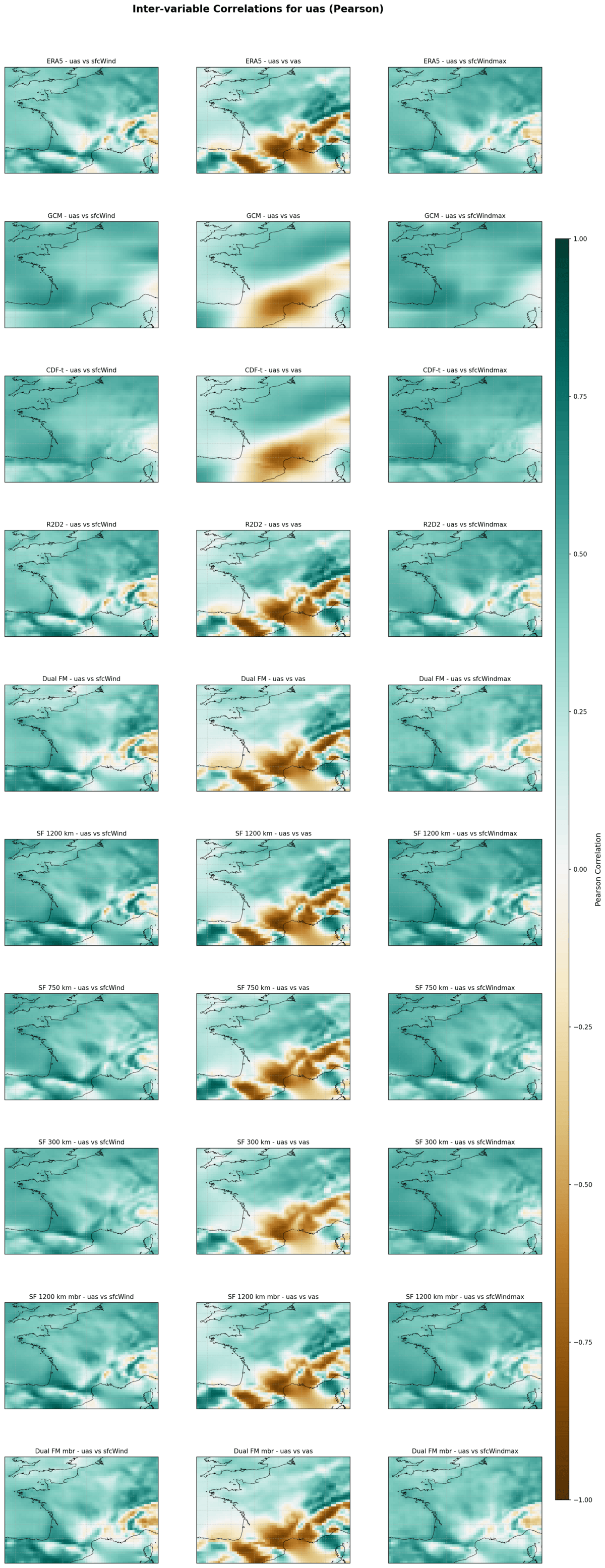}
    \caption{Pearson correlations coefficient between meridional wind and all the other climate variables, by grid point and method. Respectively, the columns are correlations vs wind speed, zonal wind and max wind}
    \label{fig:vas_era5_corr}
\end{figure}

% Spatial Spearman correlation
\begin{figure}[htbp]
    \centering
    \begin{subfigure}{0.49\textwidth}
        \centering
        \includegraphics[width=\linewidth]{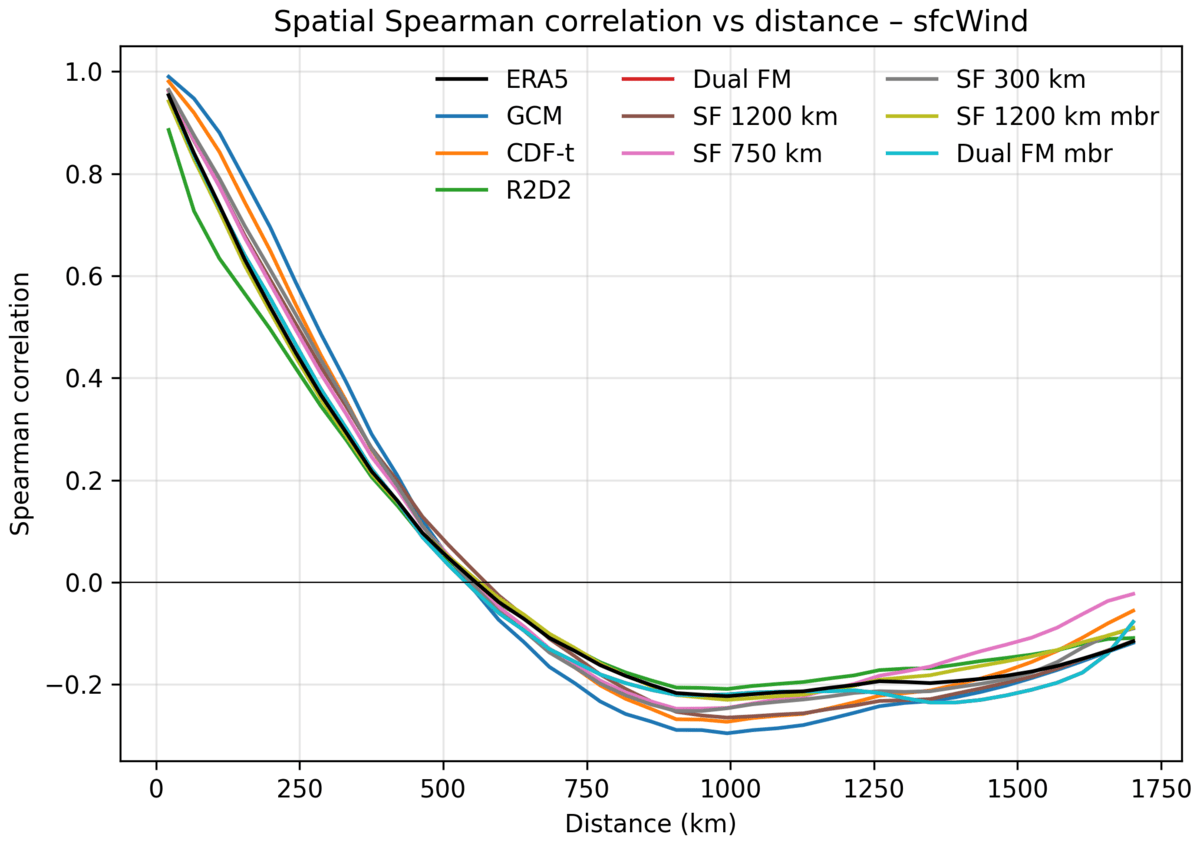}
        \caption{Wind Speed}
    \end{subfigure}
    \hfill
    \begin{subfigure}{0.49\textwidth}
        \centering
        \includegraphics[width=\linewidth]{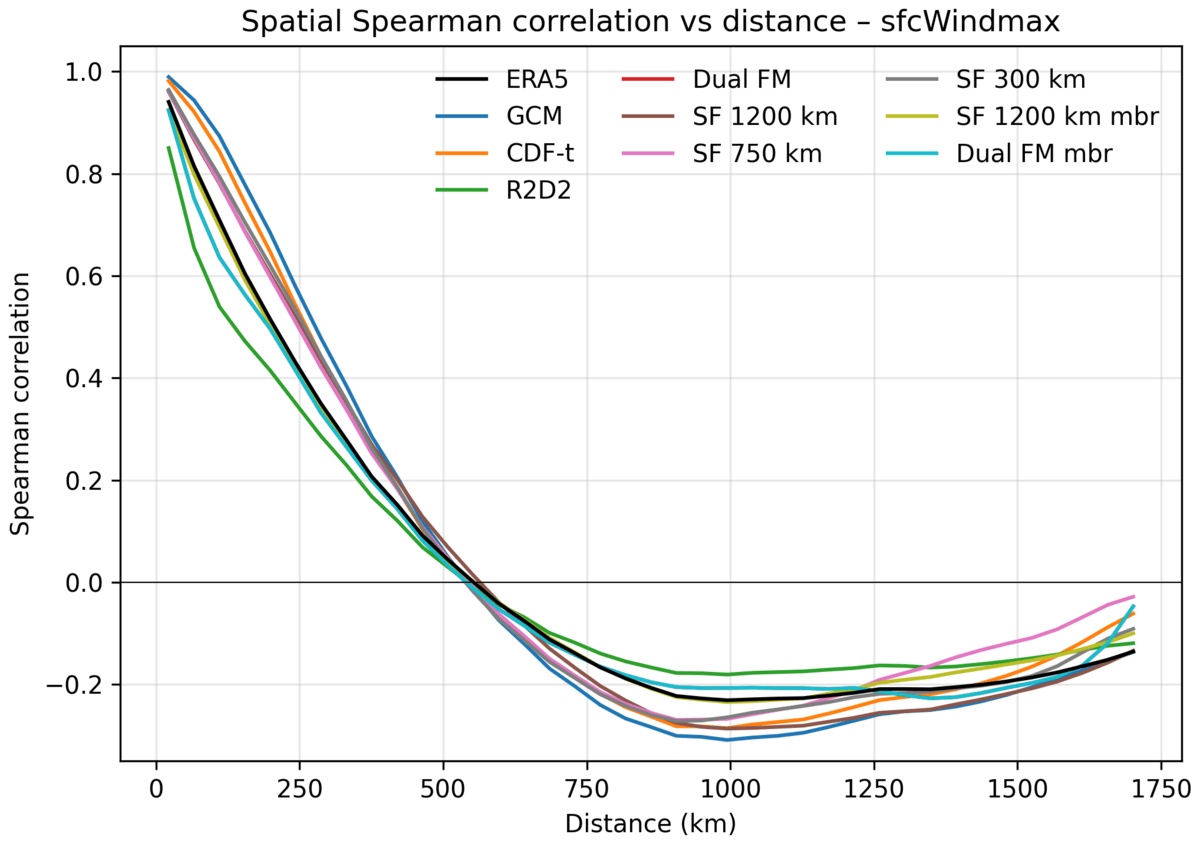}
        \caption{Maximum wind speed}
    \end{subfigure}

    \medskip

    \begin{subfigure}{0.49\textwidth}
        \centering
        \includegraphics[width=\linewidth]{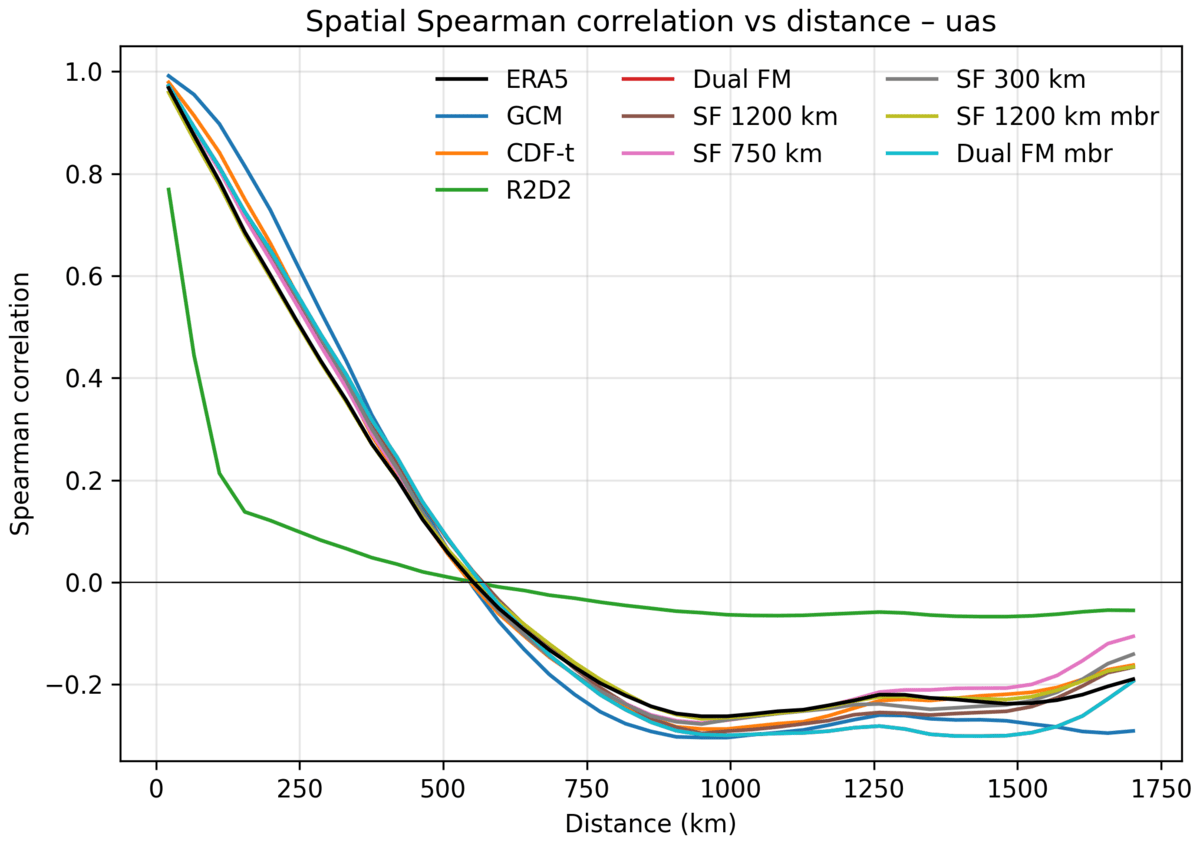}
        \caption{Zonal wind}
    \end{subfigure}
    \hfill
    \begin{subfigure}{0.49\textwidth}
        \centering
        \includegraphics[width=\linewidth]{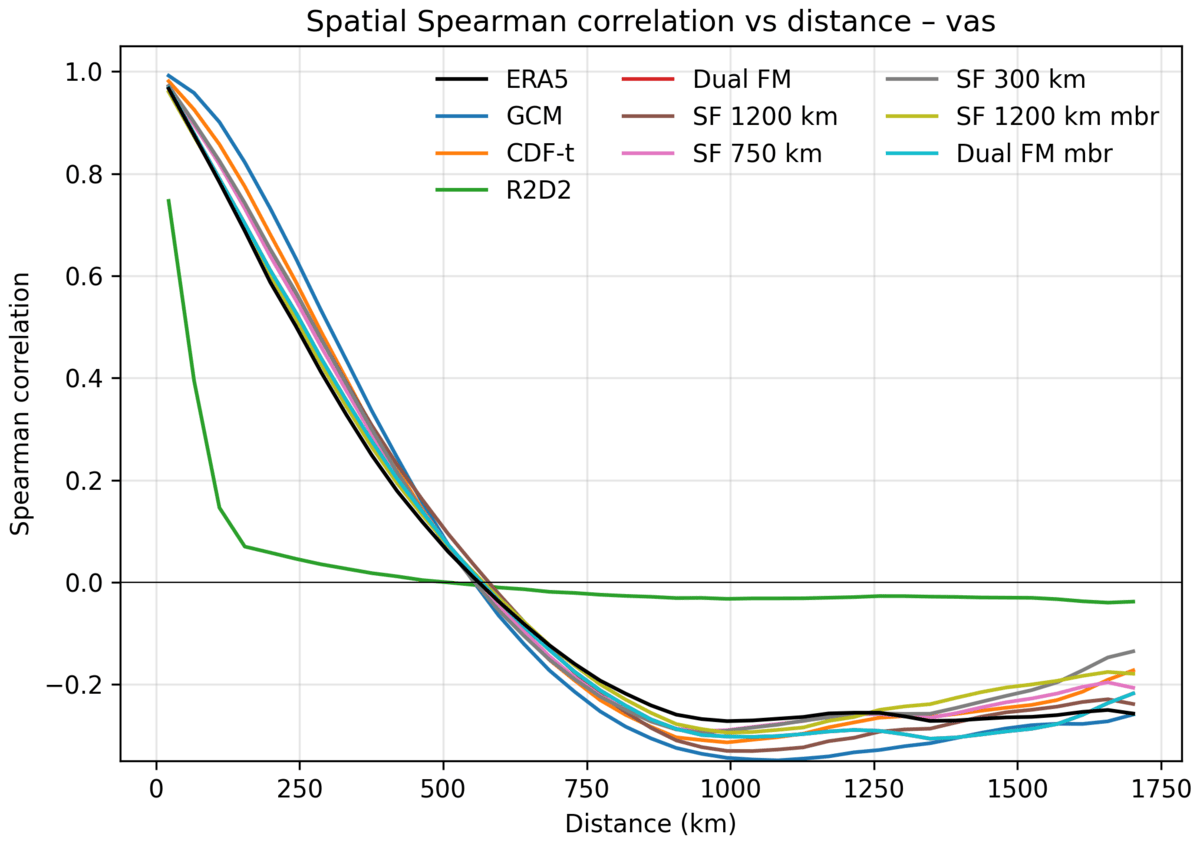}
        \caption{Meridional wind}
    \end{subfigure}
    \caption{Mean spatial Spearman correlation between pairs of locations as a function of their separation distance (distance bins in km) for each climate variable. Correlations are spatially averaged within each distance bin. On this plot, we can already see that the GCM grid points are much more correlated than those of ERA5. This is also the case for CDF-t, even though it is closer to the correlations of ERA5. While R2D2 is not too bad at reproducing wind speed correlations, it fails completely on u and v. Dual FM tends to decorrelate a little too strongly, and always ends up below the ERA5 curve. As for SerpentFlow, the curves are just above it. The lower the cutoff frequency, the closer we get to the ERA5 curve. The 1200 km single-member version fits the ERA5 curve almost perfectly for all variables}
    \label{fig:cdf_spearman_corr_ERA5}
\end{figure}
% Spatial Variogram
\begin{figure}[htbp]
    \centering
    \begin{subfigure}{0.99\textwidth}
        \centering
        \includegraphics[width=\linewidth]{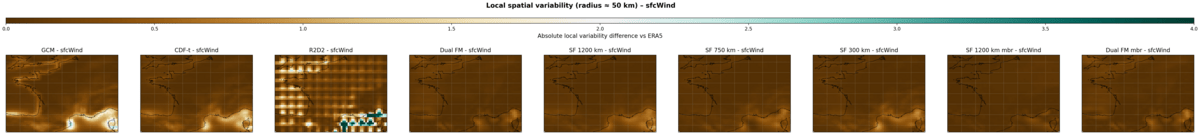}
        \caption{Wind Speed}
    \end{subfigure}
    \hfill
    \begin{subfigure}{0.99\textwidth}
        \centering
        \includegraphics[width=\linewidth]{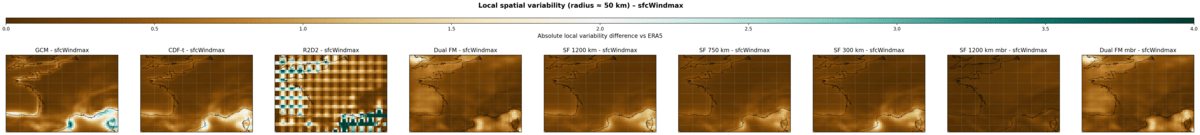}
        \caption{Maximum wind speed}
    \end{subfigure}

    \medskip

    \begin{subfigure}{0.99\textwidth}
        \centering
        \includegraphics[width=\linewidth]{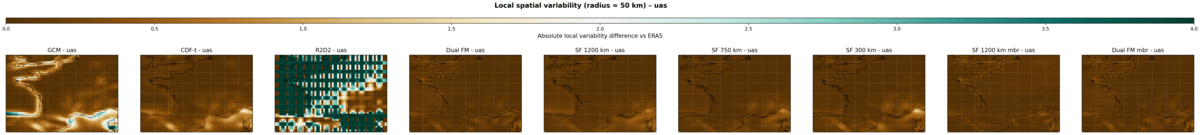}
        \caption{Zonal wind}
    \end{subfigure}
    \hfill
    \begin{subfigure}{0.99\textwidth}
        \centering
        \includegraphics[width=\linewidth]{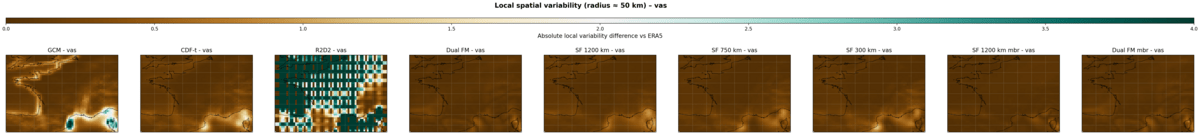}
        \caption{Meridional wind}
    \end{subfigure}
    \caption{Mean absolute difference in local spatial variability between each method and the ERA5 reference, computed over a neighborhood of approximately 50~km radius. Each panel corresponds to one method for a given climate variable. Values are averaged over all available times and grid points within the specified radius. The color scale indicates the magnitude of deviation from ERA5, with higher values representing stronger local differences. All maps show significantly higher biases on the sides and in mountainous areas (Alps and Pyrenees). However, the GCM shows much greater deviations over Corsica and the Mediterranean coast. For maximum wind speed, Dual FM significantly underperforms on the English coast and around Corsica. The single-member version of SerpentFlow 1200 km has much darker maps than the other methods, demonstrating its superior performance}
    \label{fig:cdf_variogram_ERA5}
\end{figure}
% Spatial power spectra
\begin{figure}[htbp]
    \centering
    \includegraphics[width=\linewidth]{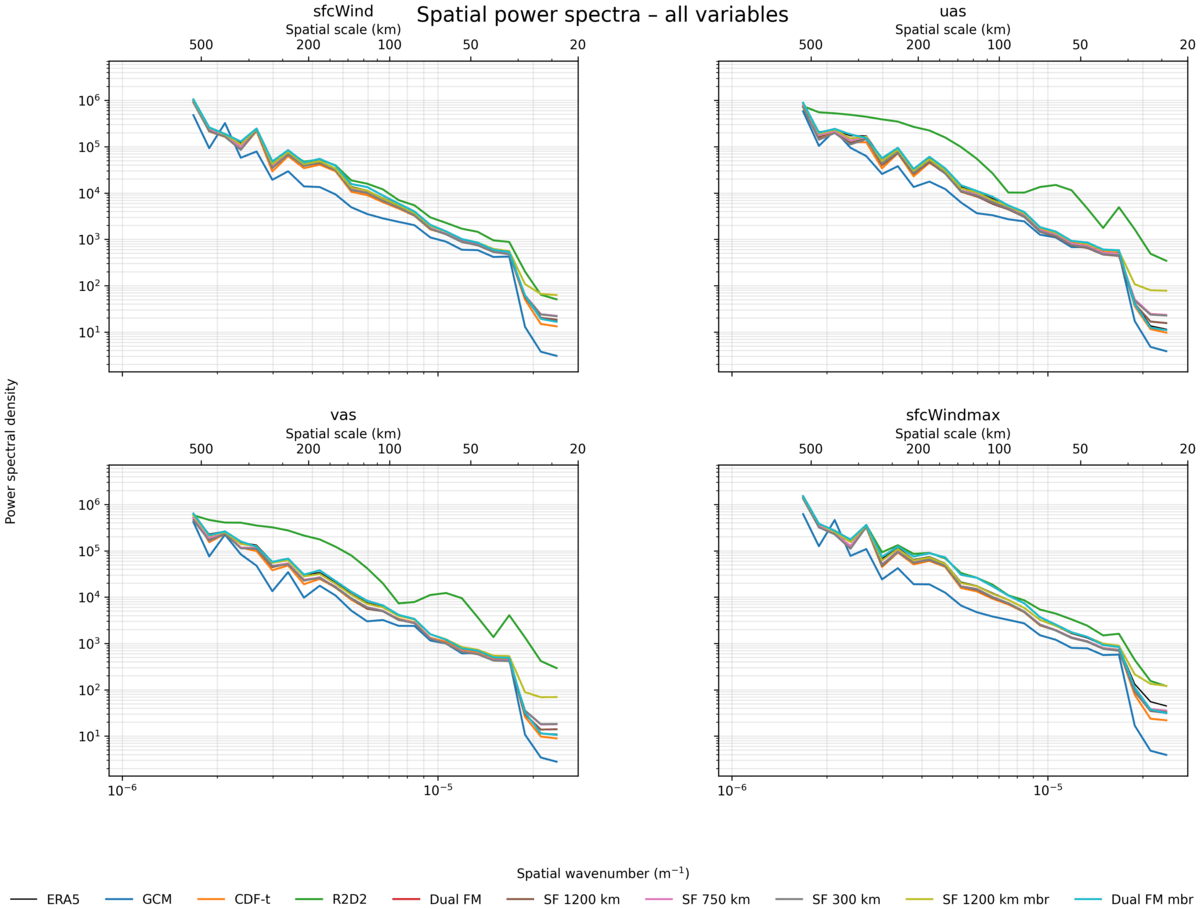}
    \caption{Spatial power spectra for all variables. Each subplot shows the power spectral density as a function of spatial wavenumber for one variable. The bottom x-axis displays the wavenumber (m$^{-1}$), while the top x-axis indicates the corresponding spatial scale in kilometers. Spectra are computed over the full domain, and differences among methods reflect deviations from the reference across spatial scales. This figure shows that the GCM has much less energy at high frequencies (small-scale effects) compared to ERA5 and downscaling methods. All plots show a significant decrease in spectral energy for effects at less than 40 km. This decrease is much stronger for CDF-t and SerpentFlow cut at 300 and 750 km, particularly for sfcWindmax. Dual FM performs slightly better, but still falls short of SerpentFlow 1200 km}
    \label{fig:spectrum_era5}
\end{figure}

% Correlations mountains
\begin{figure}[htbp]
    \centering
    \includegraphics[width=\linewidth]{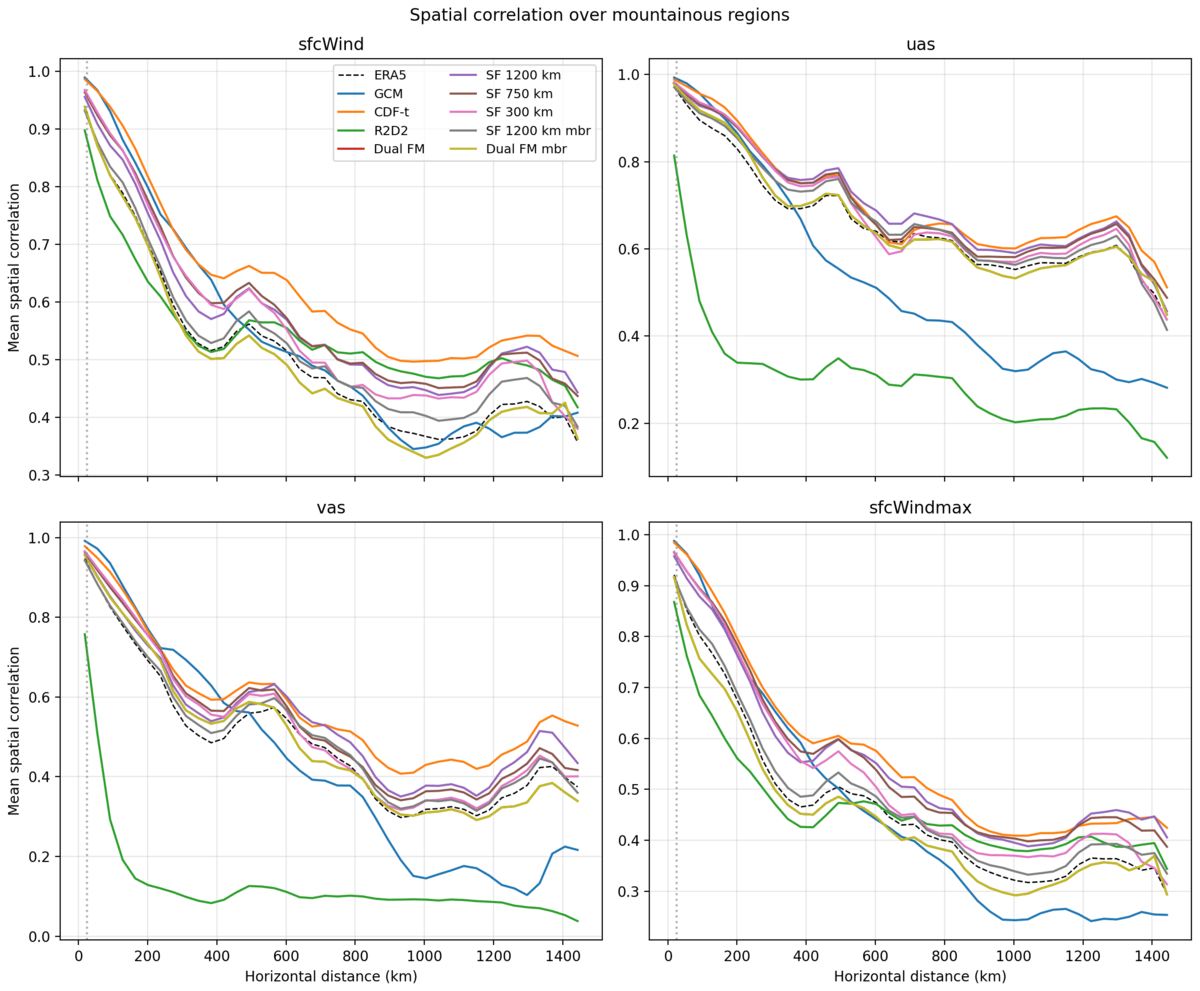}
    \caption{Mean spatial correlations over mountainous regions (altitude $>$ 800~m) as a function of horizontal distance. Each subplot corresponds to one variable. A vertical line at 25~km is added for reference. Correlations are computed only for grid points above 800~m to focus on high-elevation areas, highlighting method performance in orographic regions. The spatial over-correlations of the CDF-t and the GCM are much more visible in the mountains than in the plains (Figure~\ref{fig:cdf_spearman_corr_ERA5}).  Here we can see that the lower the GCM frequencies are cut for SerpentFlow, the better we are at reconstructing the dynamics of ERA5. The SerpentFlow 1200 km average cuts out certain effects, making it slightly less accurate than the single-member version}

    \label{fig:orography}
\end{figure}

\subsection{Plots vs GCM}

% Global Annual anomalies
\begin{figure}[htbp]
    \centering
    \includegraphics[width=\linewidth]{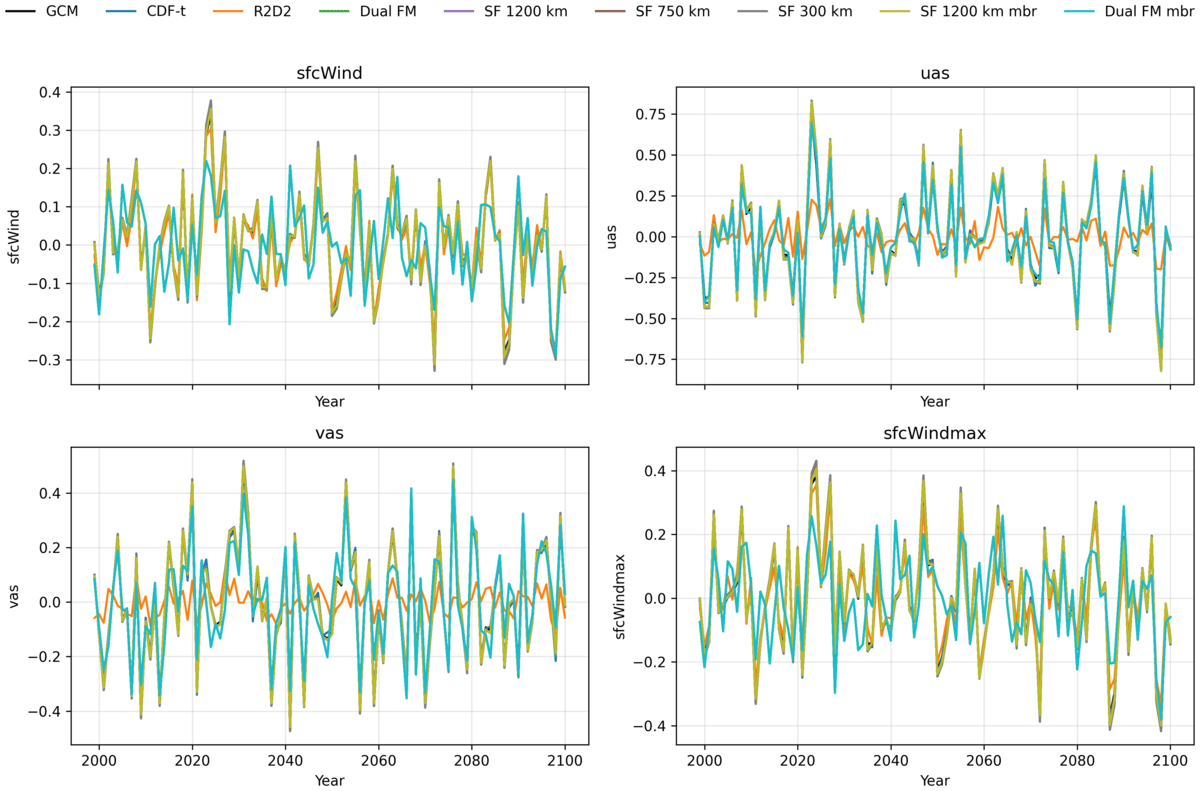}
    \caption{Global annual anomalies for all variables. Each subplot shows one climate variable. With the exception of R2D2, which deviates for u and v, all methods follow the inter-annual variability of the GCM almost perfectly. As noted in \citep{serpentflow2024}, Dual FM deviates from the GCM variations at certain peaks, predicting values that are too extreme}
    \label{fig:anomaly_era5}
\end{figure}

% Temporal correlation
\begin{figure}[htbp]
    \centering
    \includegraphics[width=\linewidth]{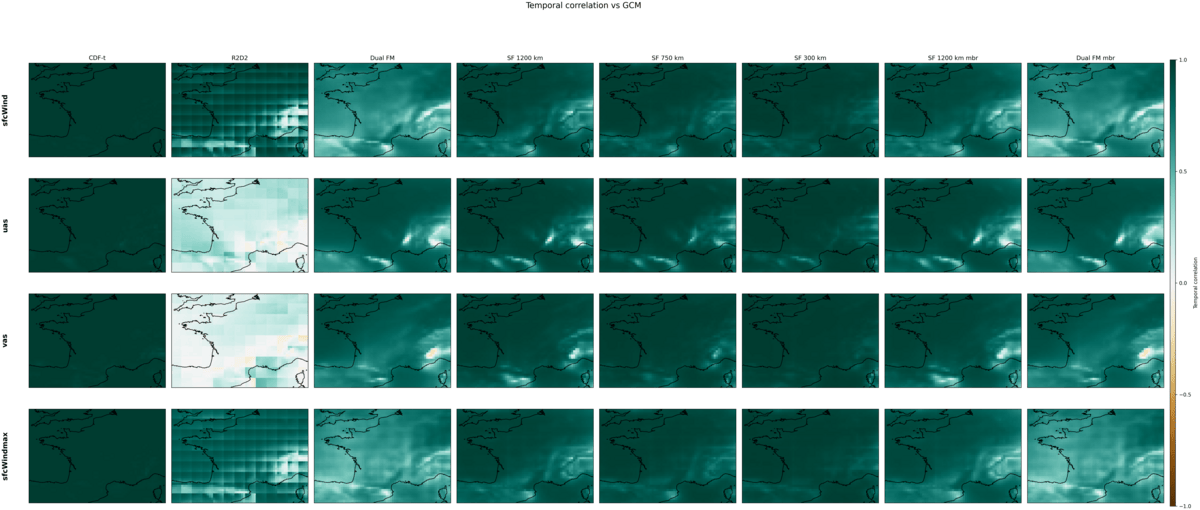}
    \caption{Temporal correlation between each method and the reference GCM for all variables. Each row corresponds to one variable, and each column shows one method. The color scale indicates the correlation coefficient at each grid point, ranging from ranging from $-1$ (perfect anti-correlation) to $1$ (perfect correlation). The CDF-t follows the temporal dynamics of the GCM grid point by grid point almost perfectly. Among the other methods, R2D2 falls completely behind for u and v, and Dual FM deviates more from the dynamics of the GCM than the various versions of SerpentFlow. The deviations are greater in mountainous areas (the Alps and Pyrenees) for deep learning methods. This is not surprising, as these are the regions where grid point values need to be modified the most to take into account local climatic effects and spatial correlations. It can be seen that the lower the SerpentFlow cut, the further it deviates from the GCM dynamics (and the closer it gets to the ERA5 dynamics)}
    \label{fig:temp_corr_era5}
\end{figure}

% Full delta
\begin{figure}[htbp]
    \centering
    \begin{subfigure}{0.99\textwidth}
        \centering
        \includegraphics[width=\linewidth]{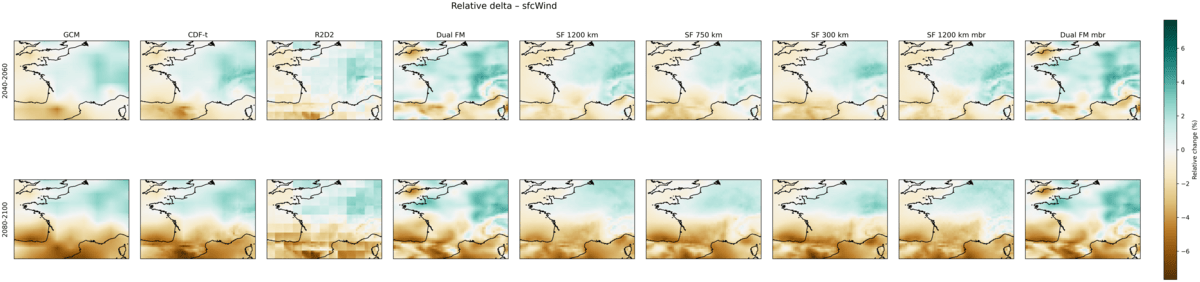}
        \caption{Wind Speed}
    \end{subfigure}
    \hfill
    \begin{subfigure}{0.99\textwidth}
        \centering
        \includegraphics[width=\linewidth]{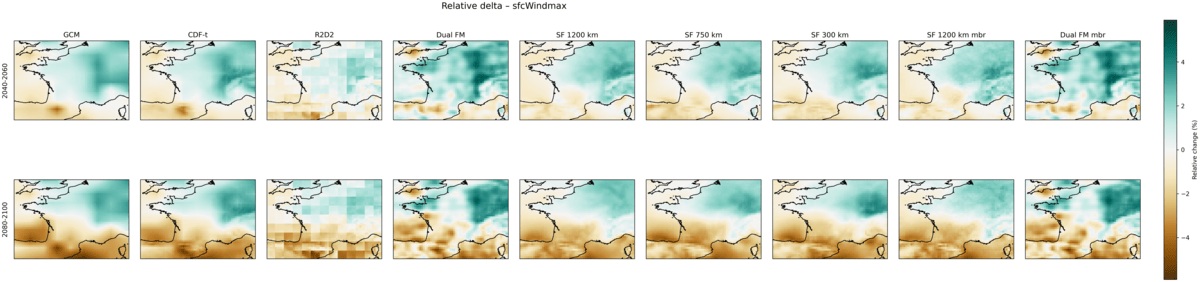}
        \caption{Maximum wind speed}
    \end{subfigure}
    \caption{Relative change (delta, in \%) of each climate variable for multiple methods and future periods. Each row corresponds to a future period, and each column shows one method, including the reference GCM in the first column. Values represent the relative change with respect to the historical baseline, computed as \% difference. Dual FM shows deltas that are quite significantly different from those of the GCM, with some increases in wind speed in Spain, whereas the GCM indicates a decrease. With regard to SerpentFlow, the lower the frequencies (e.g., for 1,200 km), the more local effects of increases or decreases emerge that are not present in the GCM. It is difficult to know whether these local effects are plausible or not, as they cannot, by design, be modeled by the GCM, which only represents effects on a larger scale}

    \label{fig:full_delta_era5}
\end{figure}
% Seasonal delta
\begin{figure}[htbp]
    \centering
    \includegraphics[width=\linewidth]{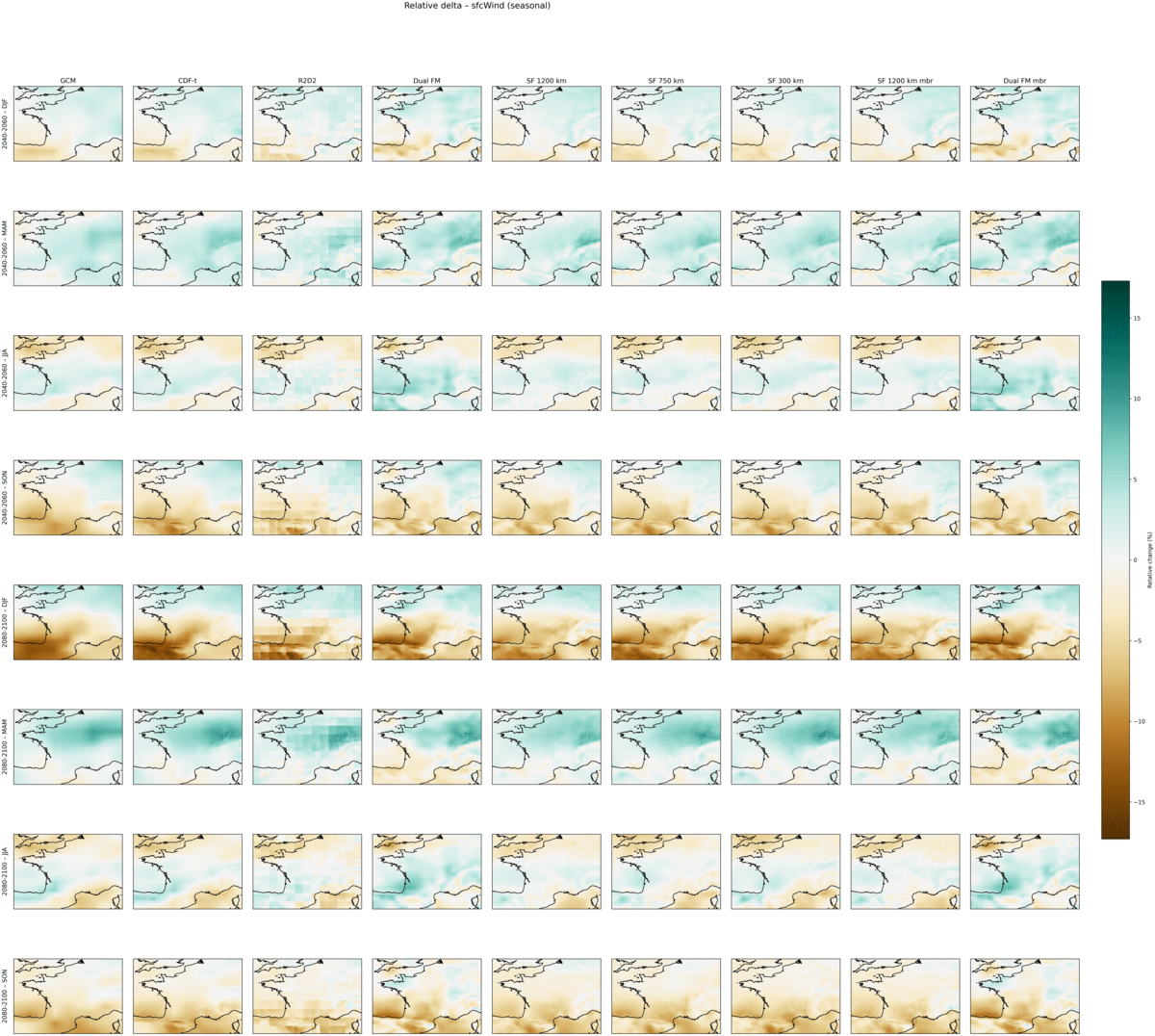}
        \caption{Relative seasonal change (delta, in \%) of mean wind speed for multiple methods and future periods. Each row corresponds to a future period and a season, and each column shows one method, including the reference GCM in the first column. Values represent the relative change with respect to the historical baseline, computed as \% difference. Results are similar to the ones from Figure~\ref{fig:full_delta_era5}}
    \label{fig:full_seasonal_delta_sfcwind_era5}
\end{figure}
\begin{figure}[htbp]
    \centering
    \includegraphics[width=\linewidth]{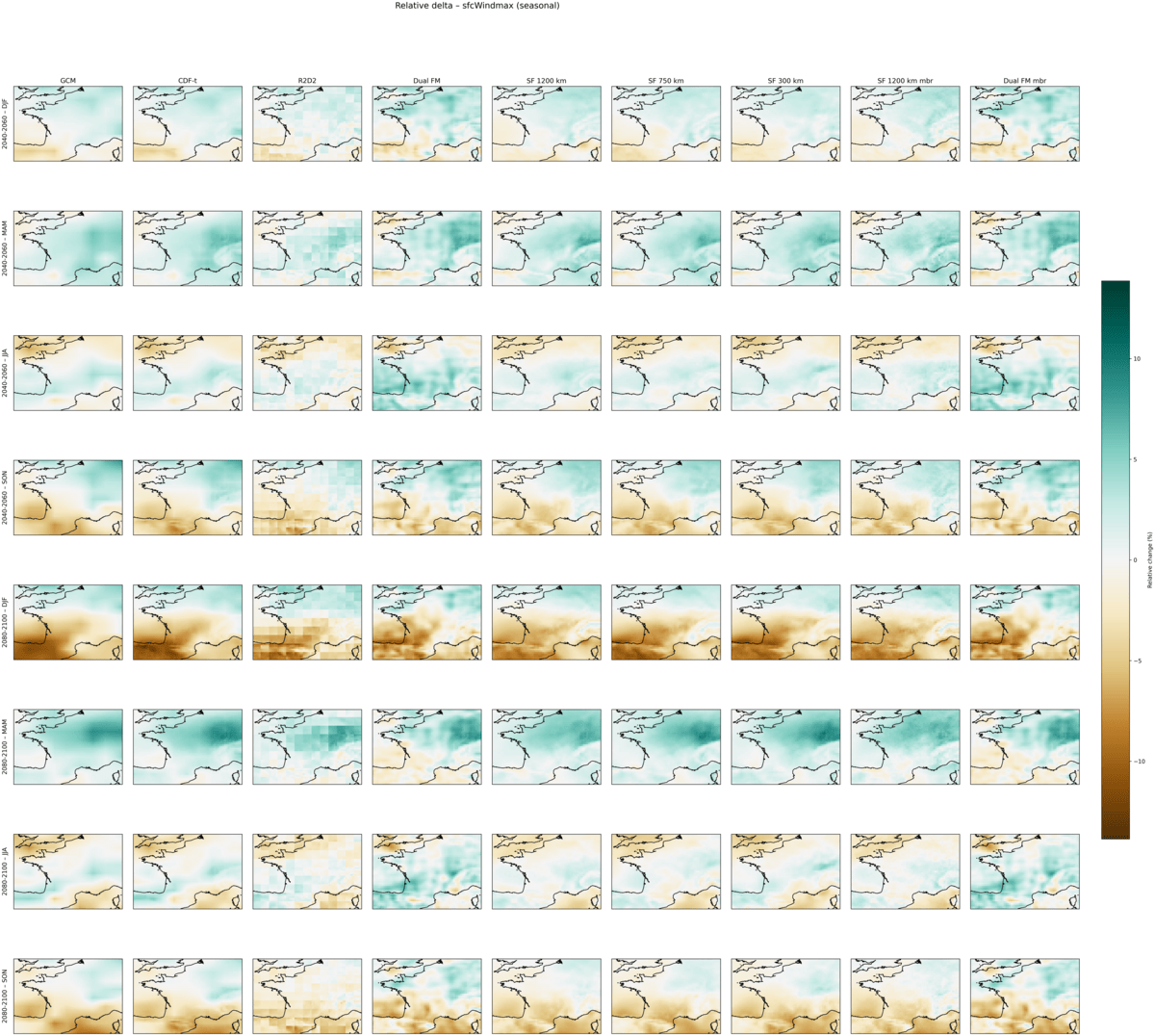}
        \caption{Relative seasonal change (delta, in \%) of maximal wind speed for multiple methods and future periods. Each row corresponds to a future period and a season, and each column shows one method, including the reference GCM in the first column. Values represent the relative change with respect to the historical baseline, computed as \% difference. Results are similar to the ones from Figure~\ref{fig:full_delta_era5}}
    \label{fig:full_seasonal_delta_sfcwindmax_era5}
\end{figure}

\subsection{Deep learning based methods plots}
% Time series
\begin{figure}[htbp]
    \centering
    \begin{subfigure}{0.45\textwidth}
        \centering
        \includegraphics[width=\linewidth]{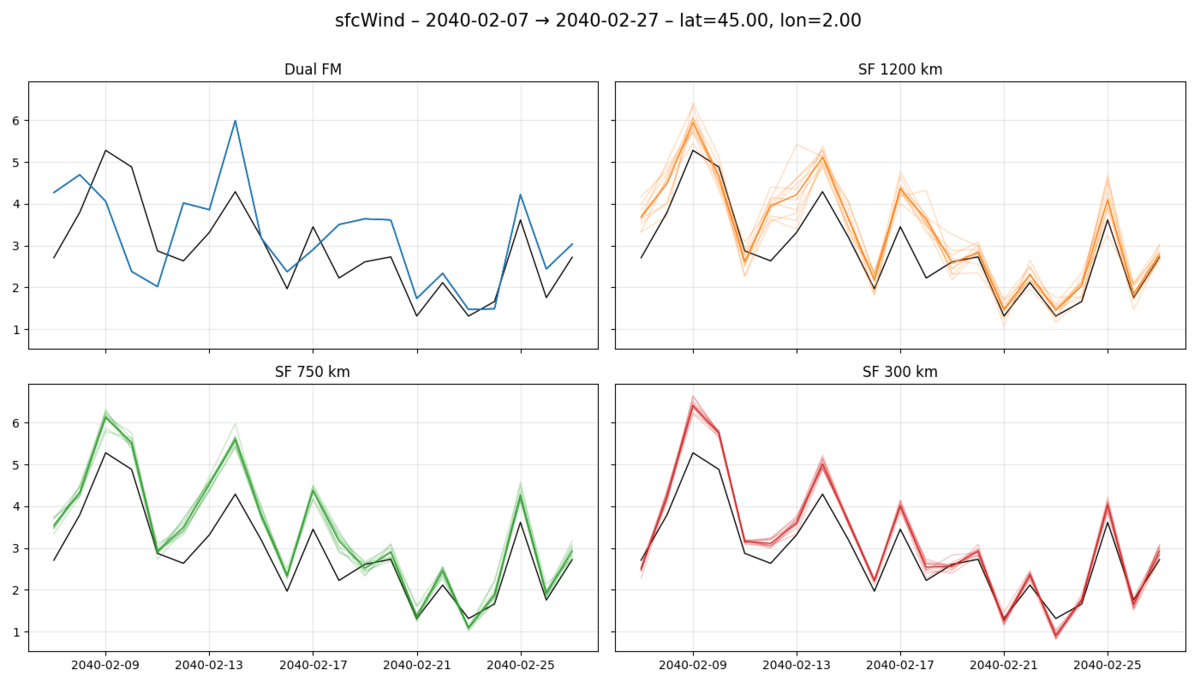}
        \caption{Wind Speed}
    \end{subfigure}
    \hfill
    \begin{subfigure}{0.45\textwidth}
        \centering
        \includegraphics[width=\linewidth]{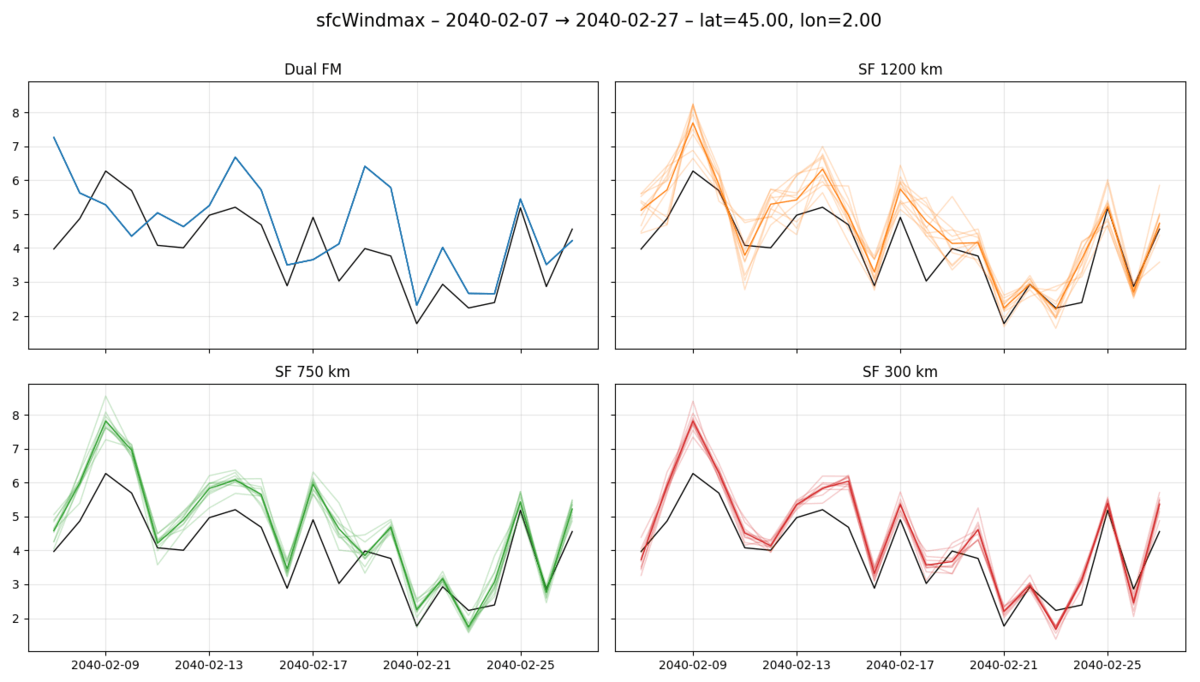}
        \caption{Maximum wind speed}
    \end{subfigure}
    \hfill
    \begin{subfigure}{0.45\textwidth}
        \centering
        \includegraphics[width=\linewidth]{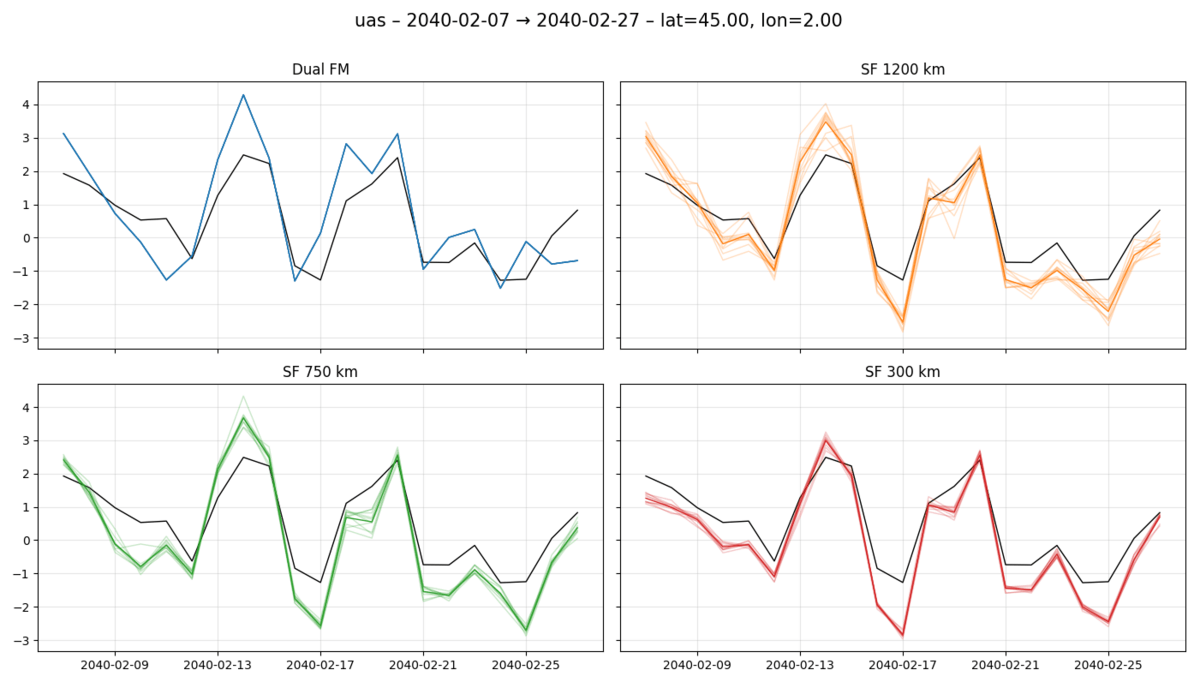}
        \caption{Zonal wind}
    \end{subfigure}
    \hfill
    \begin{subfigure}{0.45\textwidth}
        \centering
        \includegraphics[width=\linewidth]{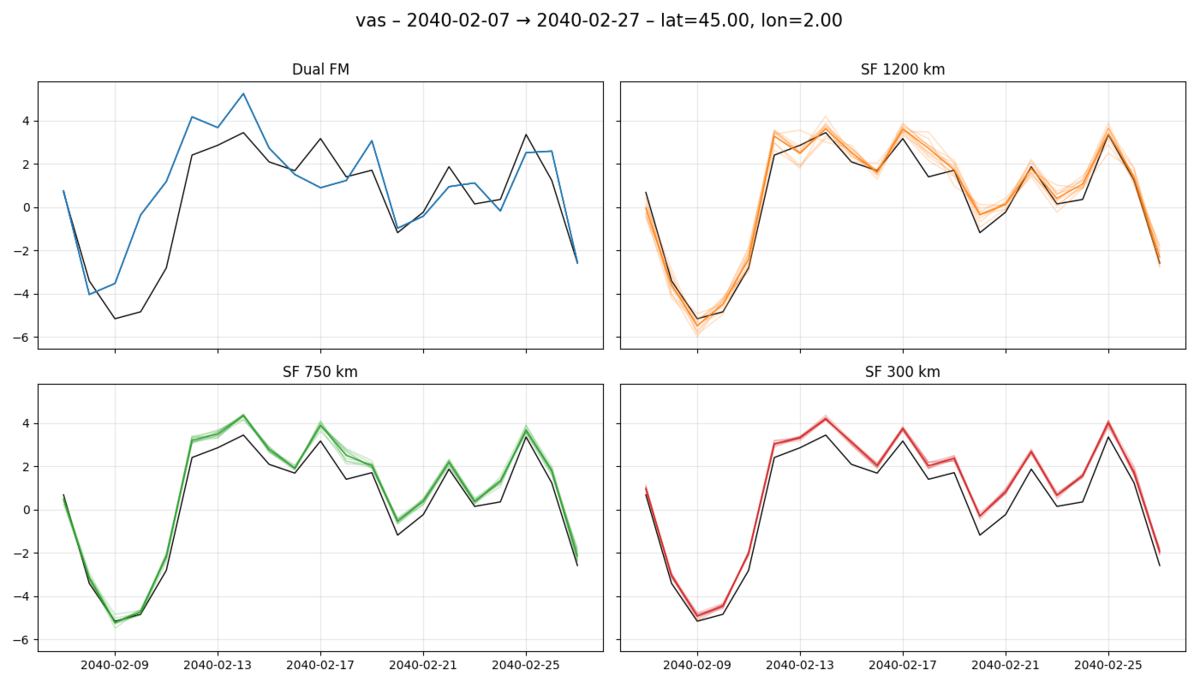}
        \caption{Meridional wind}
    \end{subfigure}
    \caption{Time series of each climate variable at a specific location (lat=45, lon=2) over a 20-day window starting 2040-02-07. Each panel shows one method: the colored line is the ensemble mean, faint colored lines show individual ensemble members, and the black line represents the GCM reference. For Dual FM, all ensemble members are identical, resulting in a single trajectory that does not closely follow the GCM dynamics; this demonstrates that, despite its generative design, the method is effectively deterministic and does not capture uncertainty. In contrast, SerpentFlow’s generative framework allows for variability: a smaller frequency cutoff (1200 km) produces a wider spread that captures more local variability but may deviate further from the GCM, whereas larger cutoffs (300 km) constrain the model more, leading to a tighter spread and closer agreement with the reference. From a climate perspective, this reflects the trade-off between representing local variability and reproducing the large-scale climate signal: a wider spread better represents the range of possible values at each point while maintaining the observed spatial and temporal structure, whereas a tighter spread prioritizes fidelity to the reference climatology.}
    \label{fig:ts_ERA5}
\end{figure}

\section{Additional Plots CNRM/SAFRAN}\label{app:plots_cnrm}

\subsection{Plots vs observations}
% Wind Maps
\begin{figure}[htbp]
    \centering
    \includegraphics[width=\linewidth]{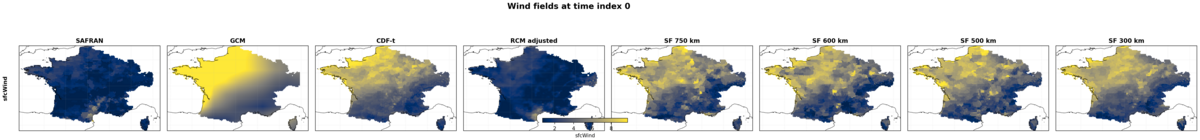}
    \caption{Wind maps for the first time step, for each method. The reanalysis values are only available for mainland France and Corsica. The reanalysis and downscaling methods (with the exception of the RCM) present small spatial patches, due to the design of the reanalysis itself. The data between SAFRAN and el GCM are not aligned in time, which explains the large difference between the two. The RCM represents a climate that is still different from the GCM, which explains the differences in the maps. We can see that the SerpentFlow (SF) versions are all slightly different. The lower the cutoff (300 km or 500 km), the closer we get to the CDF-t}
    \label{fig:wind_maps_safran}
\end{figure}

% Mean / std
\begin{figure}[htbp]
    \centering
    \includegraphics[width=\linewidth]{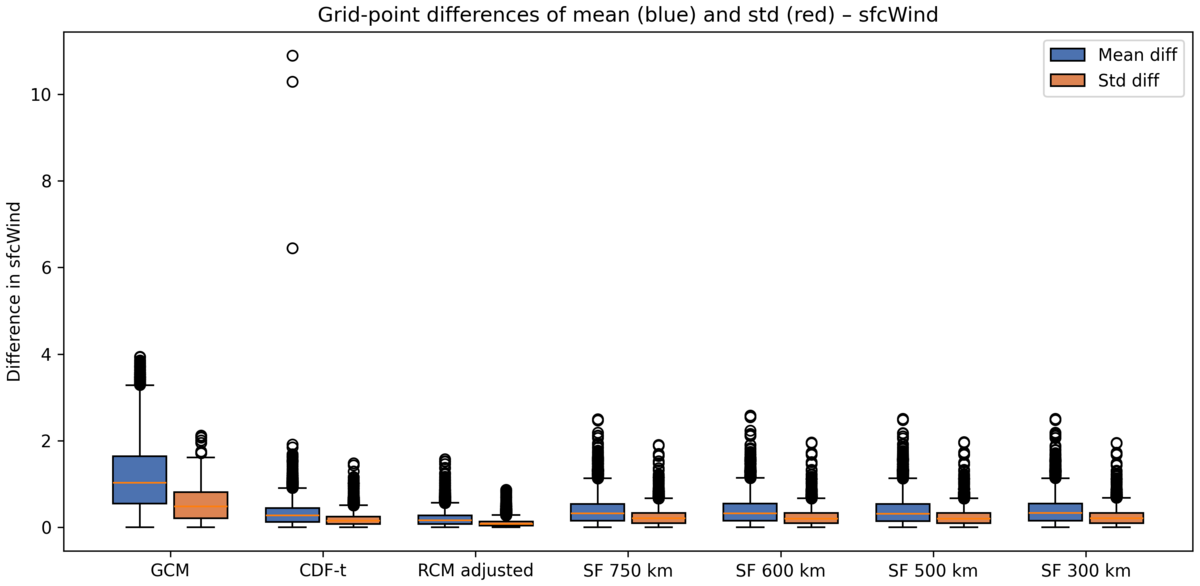}
    \caption{Distribution of mean and standard deviations differences w.r.t SAFRAN per grid point. Compared to the bias reduction between ACCESS and ERA5 (see Figure~\ref{fig:mean_std_ERA5}), this reduction is less pronounced, even for the CDF-t, which also shows a few rare large deviations from the mean. The performance of the different versions of SerpentFlow is fairly similar}
    \label{fig:mean_std_SAFRAN}
\end{figure}
% CDFs Global
\begin{figure}[htbp]
    \centering
    \includegraphics[width=\linewidth]{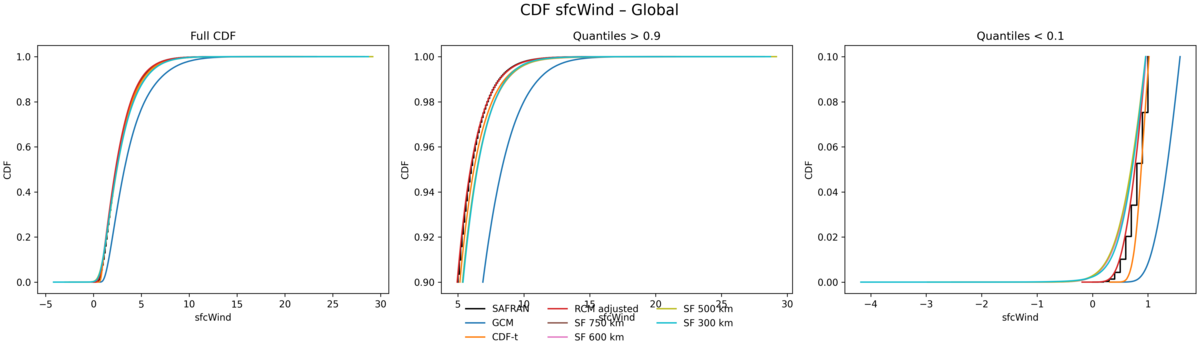}
     \caption{Cumulative Distribution Functions (CDFs) for each method over the validation period. For each sublot a zoom is done on the extreme quantiles ($q\leq 0.1$ and $q\geq 0.9$). SAFRAN's CDF is discontinuous in places due to the rounding used for its values, which may also explain the discrepancies seen in Figure~\ref{fig:mean_std_SAFRAN}. Apart from the RCM, which is very close to the reanalysis, the other methods tend to overestimate wind speed (still much less than the GCM). SF 300 km produces some negative values}
    \label{fig:cdf_global_SAFRAN}
\end{figure}
% CDFs Alps
\begin{figure}[htbp]
    \centering
    \includegraphics[width=\linewidth]{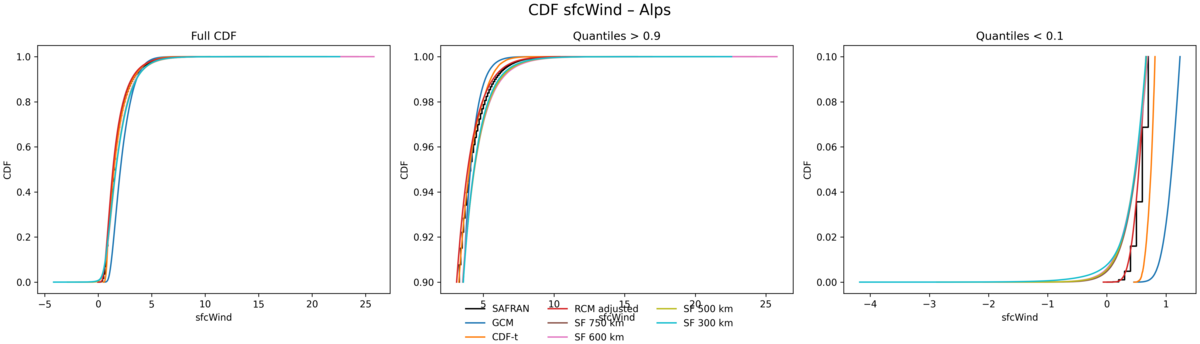}
    \caption{Cumulative distribution functions (CDFs) for each method over a small region in the Alps ($\mathrm{latitude} \in [44,47]$, $\mathrm{longitude} \in [5,8]$). For each subplot, a zoom is applied to the extreme quantiles ($q \leq 0.1$ and $q \geq 0.9$). The distribution correction is also correct on a small regional scale. In the high quantiles, the t-CDF tends to follow the GCM and underestimate the highest values. }

    \label{fig:cdf_alps_SAFRAN}
\end{figure}
% CDFs Med
\begin{figure}[htbp]
    \centering
    \includegraphics[width=\linewidth]{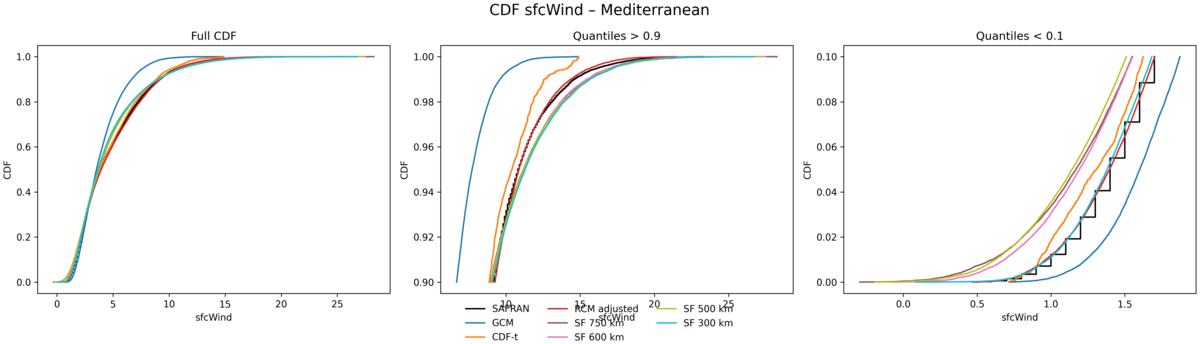}
    \caption{Cumulative distribution functions (CDFs) for each method over a small region in the Mediterranean ($\mathrm{latitude} \in [41,43]$, $\mathrm{longitude} \in [3,6]$). For each subplot, a zoom is applied to the extreme quantiles ($q \leq 0.1$ and $q \geq 0.9$). In this region, bias correction is slightly less effective for all methods except RCM. CDF-t is unable to reproduce the correct values at all for the high quantiles}
    \label{fig:cdf_med_SAFRAN}
\end{figure}

% Spatial Spearman correlation
\begin{figure}[htbp]
    \centering
    \includegraphics[width=\linewidth]{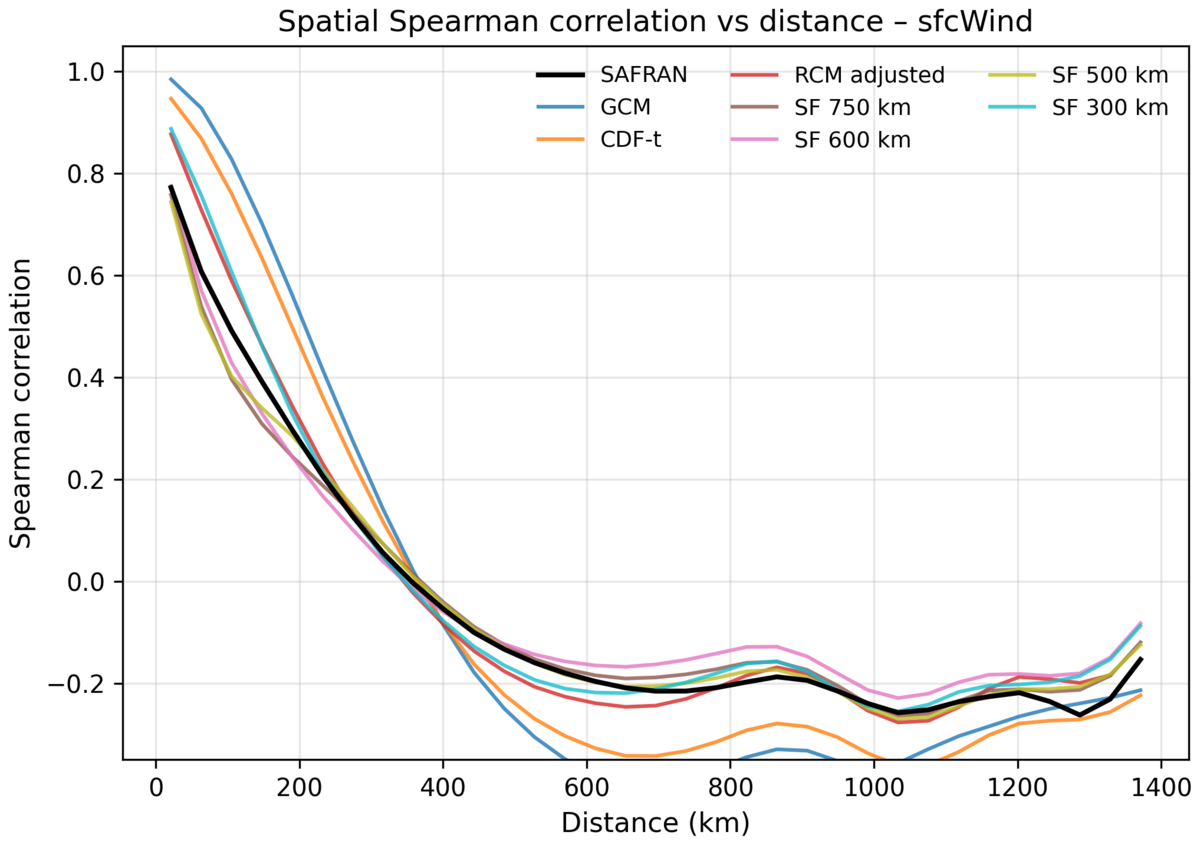}
    \caption{Mean spatial Spearman correlation between pairs of locations as a function of their separation distance (distance bins in km) for each climate variable.Correlations are spatially averaged within each distance bin. It is interesting to note here that SerpentFlow (and in particular the 500 km version) is better than RCM at modeling the spatial dynamics of SAFRAN}
    \label{fig:cdf_spearman_corr_SAFRAN}
\end{figure}

\subsection{Plots vs GCM}
% Global Annual Anomaly
\begin{figure}[htbp]
    \centering
    \includegraphics[width=\linewidth]{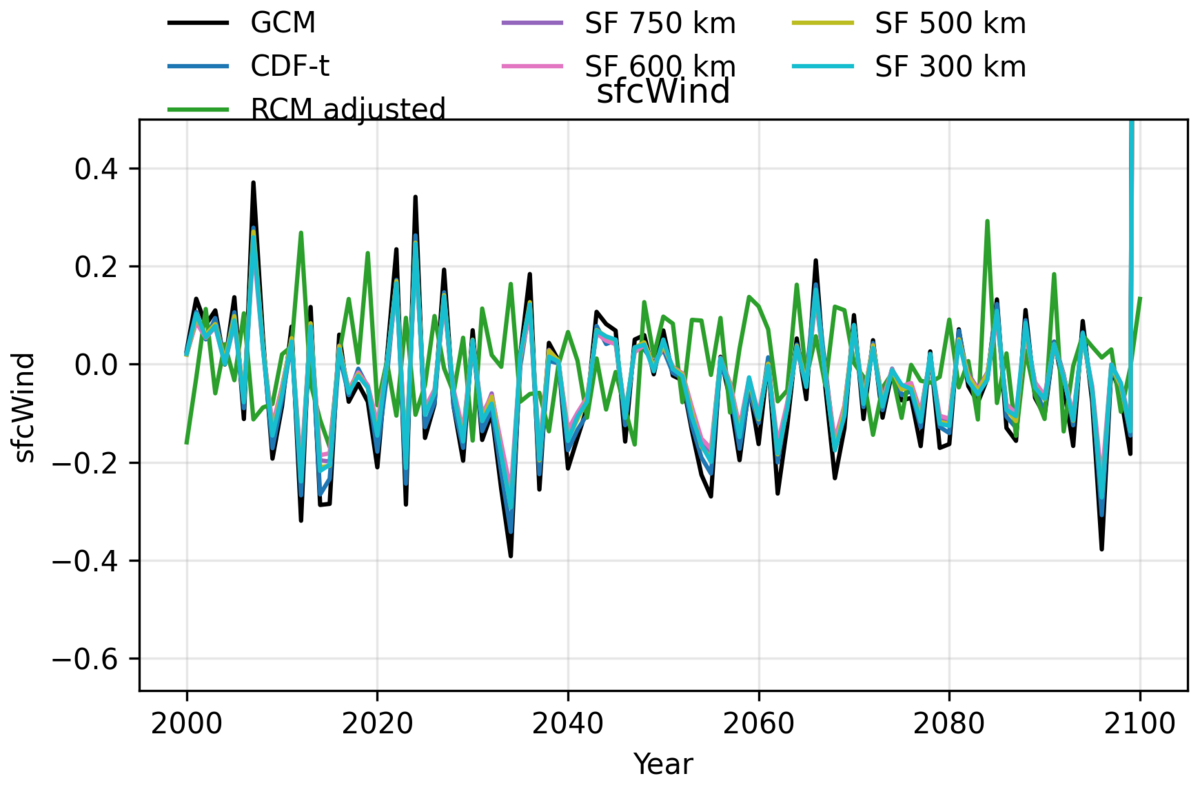}
    \caption{Global annual anomalies. Since the RCM models a different climate than the GCM, it is quite normal for the interannual dynamics to be different. As for the other methods, while they generally follow the GCM signal very closely, they tend not to reach the most extreme values}
    \label{fig:anomaly_safran}
\end{figure}

% Temporal correlation
\begin{figure}[htbp]
    \centering
    \includegraphics[width=\linewidth]{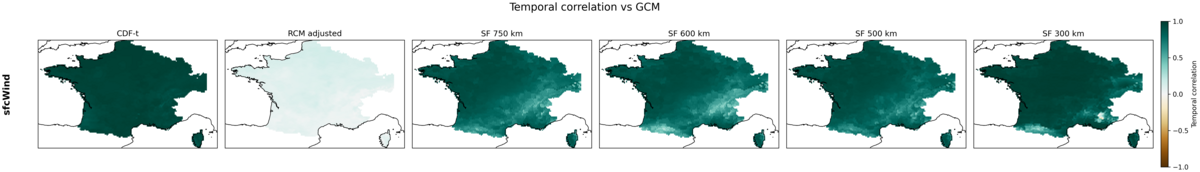}
    \caption{Temporal correlation between each method and the reference GCM. Each column shows one method. The color scale indicates the correlation coefficient at each grid point, ranging from ranging from $-1$ (perfect anti-correlation) to $1$ (perfect correlation). As shown in Figure~\ref{fig:anomaly_safran}, the RCM does not follow the dynamics of the GCM. For SerpentFlow, as seen in the ERA5 example (see Figure~\ref{fig:temp_corr_era5}), the more we cut at low frequencies, the further we move away from the dynamics of the GCM, particularly over mountainous terrain (the Alps and Pyrenees)}
    \label{fig:temp_corr_safran}
\end{figure}

% Full delta
\begin{figure}[htbp]
    \centering
    \includegraphics[width=\linewidth]{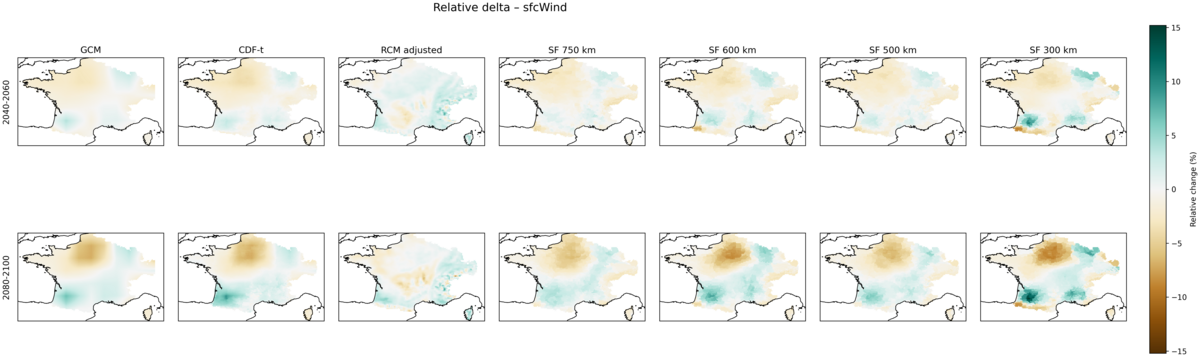}
     \caption{Relative change (delta, in \%) for multiple methods and future periods. Each row corresponds to a future period, and each column shows one method, including the reference GCM in the first column. Values represent the relative change with respect to the historical baseline, computed as \% difference. It is interesting to note that even for these deltas, the signal proposed by the RCM is quite different from that of the GCM. While we would expect the signal to be quite different from year to year, we might have expected a similar overall trend, as both models simulate the same scenario. However, we can see that the decrease in northwestern France, for example, is much less pronounced for the RCM than for the GCM. Otherwise, the dynamics of the SF versions are fairly similar to those of the GCM, with more pronounced variations for the SF 300 km version.}
    \label{fig:full_delta_safran}
\end{figure}
% Seasonal delta
\begin{figure}[htbp]
    \centering
    \includegraphics[width=\linewidth]{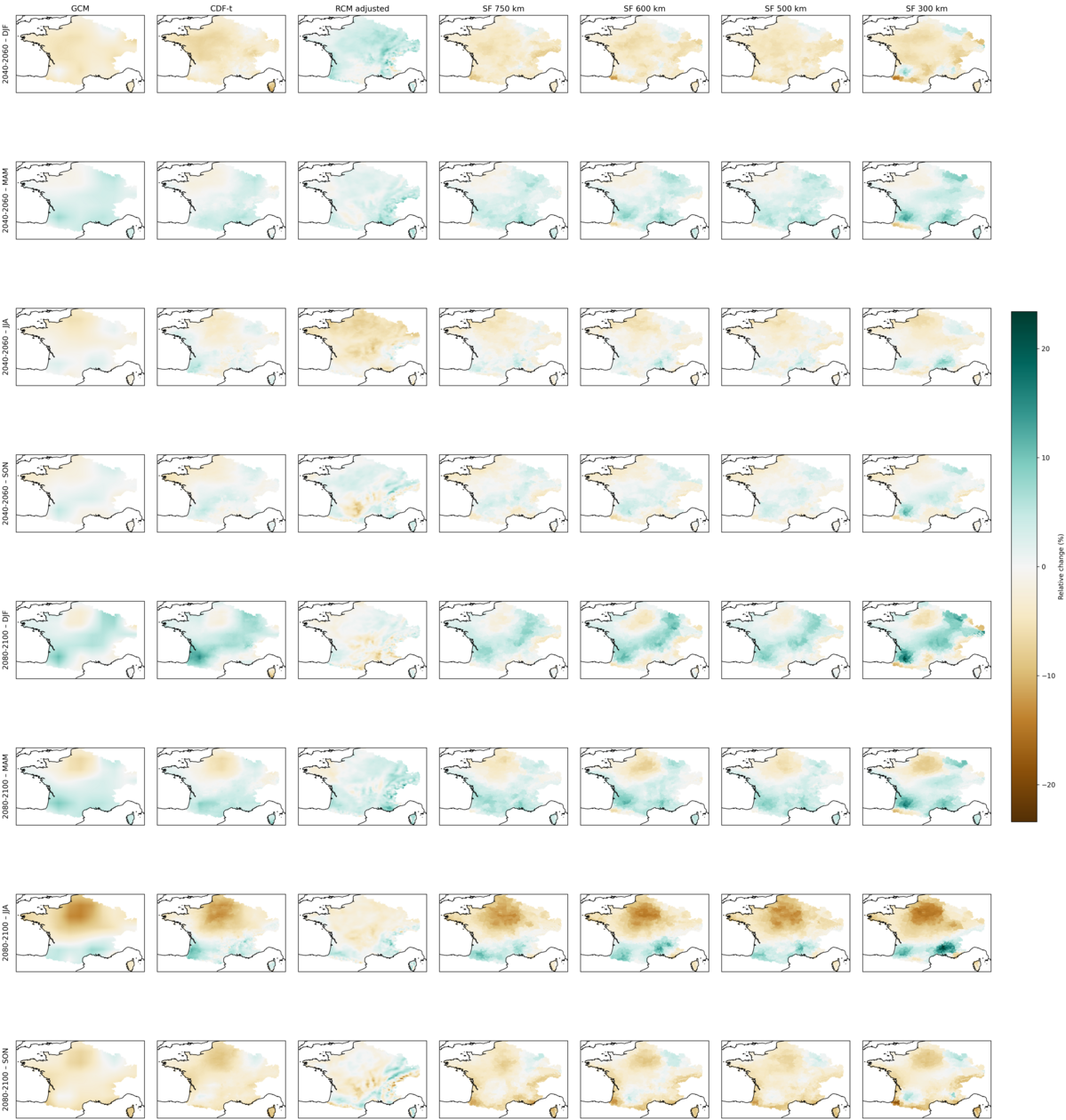}
    \caption{Relative seasonal change (delta, in \%) for multiple methods and future periods. Each row corresponds to a future period and a season, and each column shows one method, including the reference GCM in the first column. Values represent the relative change with respect to the historical baseline, computed as \% difference.}
    \label{fig:full_seasonal_delta_safran}
\end{figure}

\end{document}

%% file: diagram.tex
\tikzset{every picture/.style={line width=0.75pt}}  

\begin{figure}[htbp]
  \centering
  \resizebox{\textwidth}{!}{%

\begin{tikzpicture}[x=0.75pt,y=0.75pt,yscale=-1,xscale=1]
\draw (139.5,167.67) node  {\includegraphics[width=52.5pt,height=52.5pt]{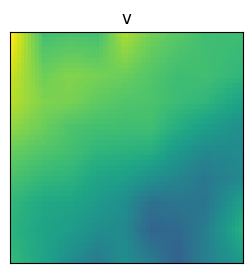}};
\draw (129.67,181.58) node  {\includegraphics[width=52.5pt,height=52.5pt]{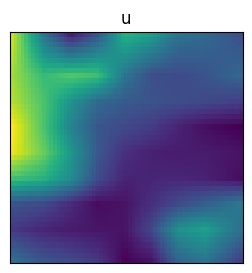}};
\draw (540,36.33) node  {\includegraphics[width=52.5pt,height=52.5pt]{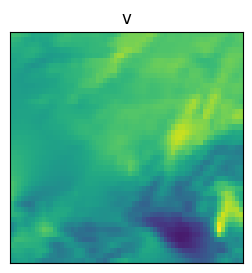}};
\draw (120.83,197.67) node  {\includegraphics[width=52.5pt,height=52.5pt]{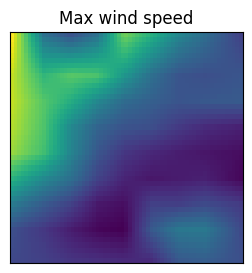}};
\draw (110.5,211.58) node  {\includegraphics[width=52.5pt,height=52.5pt]{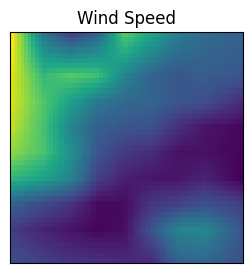}};
\draw (529.83,51) node  {\includegraphics[width=52.5pt,height=52.5pt]{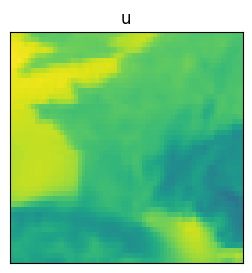}};
\draw (520.33,66.67) node  {\includegraphics[width=52.5pt,height=52.5pt]{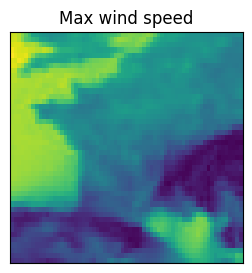}};
\draw (510.5,81.67) node  {\includegraphics[width=52.5pt,height=52.5pt]{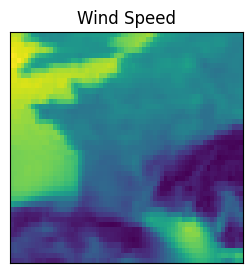}};
\draw (540,167.67) node  {\includegraphics[width=52.5pt,height=52.5pt]{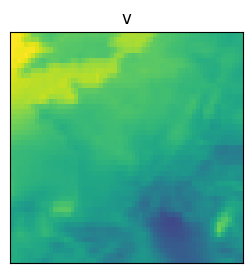}};
\draw (529.83,181.58) node  {\includegraphics[width=52.5pt,height=52.5pt]{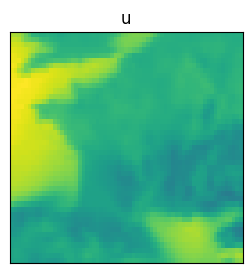}};
\draw (520.33,197.67) node  {\includegraphics[width=52.5pt,height=52.5pt]{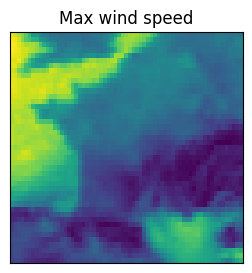}};
\draw (510.5,211.58) node  {\includegraphics[width=52.5pt,height=52.5pt]{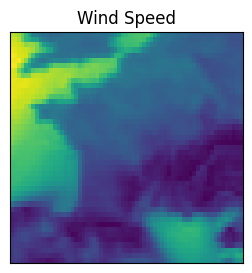}};
\draw (140.17,36.33) node  {\includegraphics[width=52.5pt,height=52.5pt]{data/era5_v.png}};
\draw (129.67,51) node  {\includegraphics[width=52.5pt,height=52.5pt]{data/era5_u.png}};
\draw (120.83,66.67) node  {\includegraphics[width=52.5pt,height=52.5pt]{data/era5_max.png}};
\draw (110.5,81.67) node  {\includegraphics[width=52.5pt,height=52.5pt]{data/era5_mean.png}};
\draw (340.09,35.67) node  {\includegraphics[width=52.5pt,height=52.5pt]{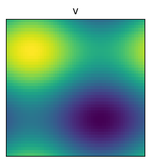}};
\draw (329.75,51.67) node  {\includegraphics[width=52.5pt,height=52.5pt]{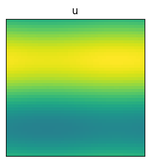}};
\draw (320.58,66.67) node  {\includegraphics[width=52.5pt,height=52.5pt]{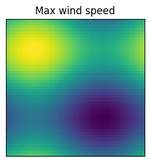}};
\draw (310.5,81.67) node  {\includegraphics[width=52.5pt,height=52.5pt]{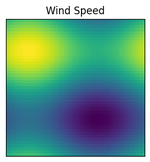}};
\draw (339.75,167.67) node  {\includegraphics[width=52.5pt,height=52.5pt]{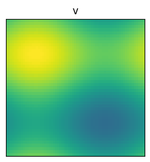}};
\draw (329.75,181.58) node  {\includegraphics[width=52.5pt,height=52.5pt]{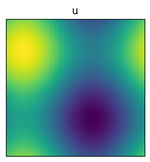}};
\draw (320.58,197.67) node  {\includegraphics[width=52.5pt,height=52.5pt]{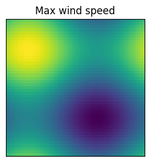}};
\draw (310.5,211.58) node  {\includegraphics[width=52.5pt,height=52.5pt]{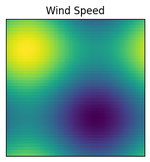}};
\draw  [dash pattern={on 4.5pt off 4.5pt}] (263.67,27.27) .. controls (263.67,13.86) and (274.53,3) .. (287.93,3) -- (360.73,3) .. controls (374.14,3) and (385,13.86) .. (385,27.27) -- (385,226.73) .. controls (385,240.14) and (374.14,251) .. (360.73,251) -- (287.93,251) .. controls (274.53,251) and (263.67,240.14) .. (263.67,226.73) -- cycle ;
%Straight Lines [id:da4287463628958349] 
\draw    (185,71) -- (252,70.36) ;
\draw [shift={(255,70.33)}, rotate = 179.45] [fill={rgb, 255:red, 0; green, 0; blue, 0 }  ][line width=0.08]  [draw opacity=0] (8.93,-4.29) -- (0,0) -- (8.93,4.29) -- cycle    ;
%Straight Lines [id:da8041335196210513] 
\draw    (185,202.67) -- (252,202.03) ;
\draw [shift={(255,202)}, rotate = 179.45] [fill={rgb, 255:red, 0; green, 0; blue, 0 }  ][line width=0.08]  [draw opacity=0] (8.93,-4.29) -- (0,0) -- (8.93,4.29) -- cycle    ;
%Straight Lines [id:da9790550892847215] 
\draw    (396.67,71) -- (463.67,70.36) ;
\draw [shift={(466.67,70.33)}, rotate = 179.45] [fill={rgb, 255:red, 0; green, 0; blue, 0 }  ][line width=0.08]  [draw opacity=0] (8.93,-4.29) -- (0,0) -- (8.93,4.29) -- cycle    ;
%Straight Lines [id:da7518832641719306] 
\draw    (396.67,202.67) -- (463.67,202.03) ;
\draw [shift={(466.67,202)}, rotate = 179.45] [fill={rgb, 255:red, 0; green, 0; blue, 0 }  ][line width=0.08]  [draw opacity=0] (8.93,-4.29) -- (0,0) -- (8.93,4.29) -- cycle    ;

% Text Node
\draw (385,40) node [anchor=north west][inner sep=0.75pt]   [align=left] {\begin{minipage}[lt]{63.65pt}\setlength\topsep{0pt}
\begin{center}
{\footnotesize Train generative }\\{\footnotesize approach $\displaystyle f_{\theta }$}
\end{center}

\end{minipage}};
% Text Node
\draw (22.67,4) node [anchor=north west][inner sep=0.75pt]  [font=\large] [align=left] {{\footnotesize \textbf{Training}}};
% Text Node
\draw (19.67,140.78) node [anchor=north west][inner sep=0.75pt]  [font=\large] [align=left] {{\footnotesize \textbf{Inference}}};
% Text Node
\draw (17.67,67.89) node [anchor=north west][inner sep=0.75pt]  [font=\normalsize] [align=left] {\begin{minipage}[lt]{42.64pt}\setlength\topsep{0pt}
\begin{center}
{\footnotesize Reference }\\{\footnotesize (obs/rea)}
\end{center}

\end{minipage}};
% Text Node
\draw (29.67,202.11) node [anchor=north west][inner sep=0.75pt]  [font=\normalsize] [align=left] {\begin{minipage}[lt]{25.4pt}\setlength\topsep{0pt}
\begin{center}
{\footnotesize GCM }\\{\footnotesize interp.}
\end{center}

\end{minipage}};
% Text Node
\draw (289.93,254) node [anchor=north west][inner sep=0.75pt]   [align=left] {{\scriptsize Joint distribution}};
% Text Node
\draw (176.83,76.33) node [anchor=north west][inner sep=0.75pt]   [align=left] {\begin{minipage}[lt]{54.88pt}\setlength\topsep{0pt}
\begin{center}
{\footnotesize Low pass filter}\\{\footnotesize $\displaystyle \omega _{cut}$}
\end{center}

\end{minipage}};
% Text Node
\draw (176.83,207.67) node [anchor=north west][inner sep=0.75pt]   [align=left] {\begin{minipage}[lt]{54.88pt}\setlength\topsep{0pt}
\begin{center}
{\footnotesize Low pass filter}\\{\footnotesize $\displaystyle \omega _{cut}$}
\end{center}

\end{minipage}};
% Text Node
\draw (409,190) node [anchor=north west][inner sep=0.75pt]   [align=left] {\begin{minipage}[lt]{31.22pt}\setlength\topsep{0pt}
\begin{center}
{\footnotesize Inference using $\displaystyle f_{\theta }$}
\end{center}

\end{minipage}};
% Text Node
\draw (580.33,67.89) node [anchor=north west][inner sep=0.75pt]  [font=\normalsize] [align=left] {\begin{minipage}[lt]{42.64pt}\setlength\topsep{0pt}
\begin{center}
{\footnotesize Reference }\\{\footnotesize (obs/rea)}
\end{center}

\end{minipage}};
\draw (578.33,202.11) node [anchor=north west][inner sep=0.75pt]  [font=\normalsize] [align=left] {\begin{minipage}[lt]{45.82pt}\setlength\topsep{0pt}
\begin{center}
{\footnotesize GCM}\\{\footnotesize downscaled}
\end{center}

\end{minipage}};
\end{tikzpicture}
 }
  \caption{Overview of the training and inference pipeline of SerpentFlow, a generative domain-adaptation method for statistical downscaling via shared-structure}
  \label{fig:pipeline}
\end{figure}

%% file: Sample.bbl
\begin{thebibliography}{32}
\providecommand{\natexlab}[1]{#1}
\providecommand{\url}[1]{\texttt{#1}}
\expandafter\ifx\csname urlstyle\endcsname\relax
  \providecommand{\doi}[1]{doi: #1}\else
  \providecommand{\doi}{doi: \begingroup \urlstyle{rm}\Url}\fi

\bibitem[Abdalla(2013)]{abdalla2013effective}
S~Abdalla.
\newblock Effective spectral resolution of ecmwf atmospheric forecast models.
\newblock \emph{ECMWF Newsletter}, 137:\penalty0 19, 2013.

\bibitem[Allard et~al.(2025)]{allard2025assessing}
Denis Allard et~al.
\newblock Assessing multivariate bias corrections of climate simulations on
  various impact models under climate change.
\newblock \emph{Hydrology and Earth System Sciences}, 29:\penalty0 4711--4729,
  2025.
\newblock \doi{10.5194/hess-29-4711-2025}.

\bibitem[Bart{\'o}k et~al.(2019)Bart{\'o}k, Tobin, Vautard, Vrac, Jin,
  Levavasseur, Denvil, Dubus, Parey, Michelangeli, et~al.]{bartok2019climate}
Blanka Bart{\'o}k, Isabelle Tobin, Robert Vautard, Mathieu Vrac, Xia Jin,
  Guillaume Levavasseur, S{\'e}bastien Denvil, Laurent Dubus, Sylvie Parey,
  Paul-Antoine Michelangeli, et~al.
\newblock A climate projection dataset tailored for the european energy sector.
\newblock \emph{Climate services}, 16:\penalty0 100138, 2019.

\bibitem[Bischoff and Deck(2024)]{bischoff2024unpaired}
Tobias Bischoff and Katherine Deck.
\newblock Unpaired downscaling of fluid flows with diffusion bridges.
\newblock \emph{Artificial Intelligence for the Earth Systems}, 3\penalty0
  (2):\penalty0 e230039, 2024.

\bibitem[Bo{\'e} et~al.(2020)Bo{\'e}, Somot, Corre, and Nabat]{boe}
Julien Bo{\'e}, Samuel Somot, Lola Corre, and Pierre Nabat.
\newblock Large discrepancies in summer climate change over europe as projected
  by global and regional climate models: causes and consequences.
\newblock \emph{Climate Dynamics}, 54\penalty0 (5):\penalty0 2981--3002, 2020.

\bibitem[Buzzicotti(2023)]{buzzicotti2023data}
Michele Buzzicotti.
\newblock Data reconstruction for complex flows using ai: Recent progress,
  obstacles, and perspectives.
\newblock \emph{Europhysics Letters}, 142\penalty0 (2):\penalty0 23001, 2023.

\bibitem[Cannon(2018)]{cannon2018mbcn}
Alex~J Cannon.
\newblock Multivariate quantile mapping bias correction: An n-dimensional
  probability density function transform for climate model simulations.
\newblock \emph{Journal of Climate}, 31\penalty0 (7):\penalty0 2649--2668,
  2018.

\bibitem[Colin et~al.(2010)Colin, D{\'e}qu{\'e}, Radu, and Somot]{rcm}
Jeanne Colin, Michel D{\'e}qu{\'e}, Raluca Radu, and Samuel Somot.
\newblock Sensitivity study of heavy precipitation in limited area model
  climate simulations: influence of the size of the domain and the use of the
  spectral nudging technique.
\newblock \emph{Tellus A: Dynamic Meteorology and Oceanography}, 62\penalty0
  (5):\penalty0 591--604, 2010.

\bibitem[Fran{\c{c}}ois et~al.(2020)Fran{\c{c}}ois, Vrac, Cannon, Robin, and
  Allard]{francois2020multivariate}
Bastien Fran{\c{c}}ois, Mathieu Vrac, Alex~J. Cannon, Yoann Robin, and Denis
  Allard.
\newblock Multivariate bias corrections of climate simulations: which benefits
  for which losses?
\newblock \emph{Earth System Dynamics}, 11:\penalty0 537--562, 2020.
\newblock \doi{10.5194/esd-11-537-2020}.

\bibitem[Groenke et~al.(2020)Groenke, Madaus, and
  Monteleoni]{groenke2020climalign}
Brian Groenke, Luke Madaus, and Claire Monteleoni.
\newblock Climalign: Unsupervised statistical downscaling of climate variables
  via normalizing flows.
\newblock In \emph{Proceedings of the 10th International Conference on Climate
  Informatics}, pages 60--66, 2020.

\bibitem[Hersbach et~al.(2020)Hersbach, Bell, Berrisford, Hirahara,
  Hor{\'a}nyi, Mu{\~n}oz-Sabater, Nicolas, Peubey, Radu, Schepers,
  et~al.]{era5}
Hans Hersbach, Bill Bell, Paul Berrisford, Shoji Hirahara, Andr{\'a}s
  Hor{\'a}nyi, Joaqu{\'\i}n Mu{\~n}oz-Sabater, Julien Nicolas, Carole Peubey,
  Raluca Radu, Dinand Schepers, et~al.
\newblock The era5 global reanalysis.
\newblock \emph{Quarterly journal of the royal meteorological society},
  146\penalty0 (730):\penalty0 1999--2049, 2020.

\bibitem[Hess et~al.(2025)Hess, Aich, Pan, and Boers]{hess2025fast}
Philipp Hess, Michael Aich, Baoxiang Pan, and Niklas Boers.
\newblock Fast, scale-adaptive and uncertainty-aware downscaling of earth
  system model fields with generative machine learning.
\newblock \emph{Nature Machine Intelligence}, 7\penalty0 (3):\penalty0
  363--373, 2025.

\bibitem[Keisler et~al.(2026)Keisler, Charantonis, Goude, Oueslati, and
  Monteleoni]{serpentflow2024}
Julie Keisler, Anastase~Alexandre Charantonis, Yannig Goude, Boutheina
  Oueslati, and Claire Monteleoni.
\newblock Serpentflow: Generative unpaired domain alignment via
  shared-structure decomposition.
\newblock \emph{arXiv preprint arXiv:2601.01979}, 2026.

\bibitem[Lipman et~al.(2023)Lipman, Chen, Ben-Hamu, Nickel, and
  Le]{flowmatching}
Yaron Lipman, Ricky~TQ Chen, Heli Ben-Hamu, Maximilian Nickel, and Matthew Le.
\newblock Flow matching for generative modeling.
\newblock In \emph{The Eleventh International Conference on Learning
  Representations}, 2023.

\bibitem[Maraun et~al.(2010)Maraun, Wetterhall, Ireson, Chandler, Kendon,
  Widmann, Brienen, Rust, Sauter, Theme{\ss}l, et~al.]{maraun2010}
Douglas Maraun, Fredrik Wetterhall, Andrew~M Ireson, Richard~E Chandler,
  Elizabeth~J Kendon, Martin Widmann, Stephan Brienen, Hyun-Goo Rust, Tobias
  Sauter, Michael Theme{\ss}l, et~al.
\newblock Precipitation downscaling under climate change: Recent developments
  to bridge the gap between dynamical models and the end user.
\newblock \emph{Reviews of Geophysics}, 48\penalty0 (3), 2010.

\bibitem[Mehrotra and Sharma(2020)]{mehrotra2020r2d2}
Rakesh Mehrotra and Ashish Sharma.
\newblock A stochastic multivariate bias correction approach for climate model
  simulations.
\newblock \emph{Water Resources Research}, 56\penalty0 (7), 2020.

\bibitem[Michelangeli et~al.(2009)Michelangeli, Vrac, and Loukos]{cdft}
P-A Michelangeli, Matthieu Vrac, and Harilaos Loukos.
\newblock Probabilistic downscaling approaches: Application to wind cumulative
  distribution functions.
\newblock \emph{Geophysical Research Letters}, 36\penalty0 (11), 2009.

\bibitem[Nabat et~al.(2025)Nabat, Somot, Bo{\'e}, Corre, Katragkou, Li, Mallet,
  van Meijgaard, Pavlidis, Pietik{\"a}inen, et~al.]{nabat}
Pierre Nabat, Samuel Somot, Julien Bo{\'e}, Lola Corre, Eleni Katragkou,
  Shuping Li, Marc Mallet, Erik van Meijgaard, Vasileios Pavlidis, J-P
  Pietik{\"a}inen, et~al.
\newblock Multi-model assessment of the role of anthropogenic aerosols in
  summertime climate change in europe.
\newblock \emph{Geophysical Research Letters}, 52\penalty0 (6):\penalty0
  e2024GL112474, 2025.

\bibitem[O’Neill et~al.(2016)O’Neill, Tebaldi, van Vuuren, Eyring,
  Friedlingstein, Hurtt, Knutti, Kriegler, Lamarque, Lowe, et~al.]{cmip6}
Brian~C O’Neill, Claudia Tebaldi, Detlef~P van Vuuren, Veronika Eyring,
  Pierre Friedlingstein, George Hurtt, Reto Knutti, Elmar Kriegler,
  Jean-Francois Lamarque, Jason Lowe, et~al.
\newblock The scenario model intercomparison project (scenariomip) for cmip6.
\newblock 2016.

\bibitem[Reynolds(1895)]{10.1098/rsta.1895.0004}
Osborne Reynolds.
\newblock Iv. on the dynamical theory of incompressible viscous fluids and the
  determination of the criterion.
\newblock \emph{Philosophical Transactions of the Royal Society of London,
  Series A: Containing Papers of a Mathematical or Physical Character},
  \penalty0 (186):\penalty0 123--164, 12 1895.
\newblock ISSN 0264-3952.
\newblock \doi{10.1098/rsta.1895.0004}.
\newblock URL \url{https://doi.org/10.1098/rsta.1895.0004}.

\bibitem[Robin et~al.(2019)Robin, Vrac, Naveau, and Yiou]{dotc}
Yoann Robin, Mathieu Vrac, Philippe Naveau, and Pascal Yiou.
\newblock Multivariate stochastic bias corrections with optimal transport.
\newblock \emph{Hydrology and Earth System Sciences}, 23\penalty0 (2):\penalty0
  773--786, 2019.

\bibitem[Soares et~al.(2024)]{soares2024cmip6dl}
Pedro M.~M. Soares et~al.
\newblock Deep learning–based statistical downscaling of cmip6 climate
  projections.
\newblock \emph{Geoscientific Model Development}, 17:\penalty0 229--247, 2024.
\newblock \doi{10.5194/gmd-17-229-2024}.

\bibitem[Suresh~Babu et~al.(2026)Suresh~Babu, Sadam, and
  Lermusiaux]{suresh2026guided}
Anantha~Narayanan Suresh~Babu, Akhil Sadam, and Pierre~FJ Lermusiaux.
\newblock Guided unconditional and conditional generative models for
  super-resolution and inference of quasi-geostrophic turbulence.
\newblock \emph{Journal of Advances in Modeling Earth Systems}, 18\penalty0
  (3):\penalty0 e2025MS005324, 2026.

\bibitem[Teutschbein and Seibert(2012)]{teutschbein2012}
Claudia Teutschbein and Jan Seibert.
\newblock Bias correction of regional climate model simulations for
  hydrological climate-change impact studies: Review and evaluation of
  different methods.
\newblock \emph{Journal of Hydrology}, 456:\penalty0 12--29, 2012.

\bibitem[Vandal et~al.(2017)Vandal, Kodra, Ganguly, Michaelis, Nemani, and
  Ganguly]{vandal2017deepsd}
Thomas Vandal, Evan Kodra, Sangram Ganguly, Andrew Michaelis, Ramakrishna
  Nemani, and Auroop~R. Ganguly.
\newblock Deepsd: Generating high resolution climate change projections through
  single image super-resolution.
\newblock In \emph{Proceedings of the 23rd ACM SIGKDD International Conference
  on Knowledge Discovery and Data Mining}, pages 1663--1672, 2017.
\newblock \doi{10.1145/3097983.3098004}.

\bibitem[Vidal et~al.(2010)Vidal, Martin, Franchist{\'e}guy, Baillon, and
  Soubeyroux]{safran}
Jean-Philippe Vidal, Eric Martin, Laurent Franchist{\'e}guy, Martine Baillon,
  and Jean-Michel Soubeyroux.
\newblock A 50-year high-resolution atmospheric reanalysis over france with the
  safran system.
\newblock \emph{International journal of climatology}, 30\penalty0
  (11):\penalty0 P--1627, 2010.

\bibitem[Voldoire et~al.(2019)Voldoire, Saint-Martin, S{\'e}n{\'e}si, Decharme,
  Alias, Chevallier, Colin, Gu{\'e}r{\'e}my, Michou, Moine, et~al.]{cnrm}
Aurore Voldoire, David Saint-Martin, St{\'e}phane S{\'e}n{\'e}si, B~Decharme,
  A~Alias, Matthieu Chevallier, Jeanne Colin, J-F Gu{\'e}r{\'e}my, Martine
  Michou, M-P Moine, et~al.
\newblock Evaluation of cmip6 deck experiments with cnrm-cm6-1.
\newblock \emph{Journal of Advances in Modeling Earth Systems}, 11\penalty0
  (7):\penalty0 2177--2213, 2019.

\bibitem[Vrac and Friederichs(2015)]{vrac2015cdf}
Mathieu Vrac and Petra Friederichs.
\newblock Nonstationary bias correction of climate simulations using a
  multivariate approach.
\newblock \emph{Journal of Climate}, 28\penalty0 (6):\penalty0 2189--2204,
  2015.

\bibitem[Wan et~al.(2023)Wan, Baptista, Chen, Anderson, Boral, Sha, and
  Zepeda-N{\'u}{\~n}ez]{wan2023debias}
Zhong~Yi Wan, Ricardo Baptista, Yi-Fan Chen, John~R. Anderson, Anudhyan Boral,
  Fei Sha, and Leonardo Zepeda-N{\'u}{\~n}ez.
\newblock Debias coarsely, sample conditionally: Statistical downscaling
  through optimal transport and probabilistic diffusion models.
\newblock \emph{arXiv preprint}, 2023.
\newblock arXiv:2305.15618.

\bibitem[Wohland(2022)]{wohland}
Jan Wohland.
\newblock Process-based climate change assessment for european winds using
  euro-cordex and global models.
\newblock \emph{Environmental Research Letters}, 17\penalty0 (12):\penalty0
  124047, 2022.

\bibitem[Zhang and Cannon(2021)]{zhang2021mrec}
Xuebin Zhang and Alex~J Cannon.
\newblock A multivariate bias correction method preserving rank structure and
  temporal dependence.
\newblock \emph{Climate Dynamics}, 56:\penalty0 1--19, 2021.

\bibitem[Ziehn et~al.(2020)Ziehn, Chamberlain, Law, Lenton, Bodman, Dix,
  Stevens, Wang, and Srbinovsky]{access}
Tilo Ziehn, Matthew~A Chamberlain, Rachel~M Law, Andrew Lenton, Roger~W Bodman,
  Martin Dix, Lauren Stevens, Ying-Ping Wang, and Jhan Srbinovsky.
\newblock The australian earth system model: Access-esm1. 5.
\newblock \emph{Journal of Southern Hemisphere Earth Systems Science},
  70\penalty0 (1):\penalty0 193--214, 2020.

\end{thebibliography}
